\documentclass[10pt,aps,prd,twocolumn,showpacs,amsmath,amssymb,nofootinbib,eqsecnum,preprintnumbers,longbibliography]{revtex4-1}

\usepackage[a-2u]{pdfx}
\usepackage[english]{babel}				
\usepackage[T3,T1]{fontenc}				
\usepackage[mathscr]{eucal} 			
\usepackage{accents}					
\usepackage{mathtools}					
\usepackage{graphicx}					

\usepackage[ddmmyyyy]{datetime}    		
\usepackage[dvipsnames]{xcolor}			
\usepackage[disable]{todonotes}        

\usepackage[shortcuts]{extdash}         

\newcommand\bs[1]{\boldsymbol{#1}}

\newcommand\tb[1]{\bs{#1}}

\newcommand\dd{\mathrm{d}}
\newcommand\pp{\partial}

\newcommand\imag{i}

\newcommand\e{e}

\newcommand\lieb[2]{\left[ #1\,{,}\,#2 \right]}

\newcommand\feq{\mathrel{\phantom{=}}}

\DeclareSymbolFont{tipa}{T3}{cmr}{m}{n}				
\DeclareMathAccent{\invbreve}{\mathalpha}{tipa}{16}

\newcommand\cir[1]{\protect\accentset{\circ}{#1}}

\newcommand\bul[1]{\protect\accentset{\bullet}{#1}}

\newcommand\st[1]{\protect\accentset{\star}{#1}}

\begin{document}


\title{Symmetry axes of Kerr--NUT--(A)dS spacetimes}

\author{Ivan Kol\'a\v{r}}
\email{Ivan.Kolar@utf.mff.cuni.cz}
\affiliation{Institute of Theoretical Physics,
Faculty of Mathematics and Physics, Charles University,
V~Hole\v{s}ovi\v{c}k\'ach 2, 180 00 Prague, Czech Republic}

\author{Pavel Krtou\v{s}}
\email{Pavel.Krtous@utf.mff.cuni.cz}
\affiliation{Institute of Theoretical Physics,
Faculty of Mathematics and Physics, Charles University,
V~Hole\v{s}ovi\v{c}k\'ach 2, 180 00 Prague, Czech Republic}

\date{May 16, 2019}

\begin{abstract}
We study fixed points of isometries of the higher-dimensional Kerr--NUT--(A)dS spacetimes that form generalizations of symmetry axes. It turns out that, in the presence of nonzero NUT charges, some parts of the symmetry axes are necessarily singular and their intersections are surrounded by regions with closed timelike curves. Motivated by similarities with the spacetime of a spinning cosmic string, we introduce geometric quantities that characterize various types of singularities on symmetry axes. Expanding the Kerr--NUT--(A)dS spacetimes around candidates for possible fixed points, we find the Killing vectors associated with generalized symmetry axes. By means of these Killing vectors we calculate the introduced geometric quantities describing axial singularities and show their relation to the parameters of the Kerr--NUT--(A)dS spacetimes. In addition, we identify the Killing coordinates that may be regarded as generalization of the Boyer--Lindquist coordinates.
\end{abstract}


\maketitle


\section{Introduction}
\label{sc:I}
The Kerr solution is one of the most interesting solutions of Einstein's equations \cite{Kerr:1963}. It describes a gravitational field of rotating compact objects such as black holes, which play an important role in modern astrophysics. In four dimensions, this geometry belongs to a wide class of spacetimes of type D called the Pleba\'{n}ski--Demia\'{n}ski family \cite{PlebanskiDemianski:1976,GriffithsPodolsky:2006b}. This class contains seven parameters: mass, rotation, acceleration, electric charge, magnetic charge, NUT charge, and cosmological constant.

Apart from the black hole spacetimes in four dimensions, there exist many solutions of higher-dimensional Einstein's equations representing gravitational fields of various black objects. Spacetimes describing black holes with spherical horizon topology rotating with respect to different planes of rotations are called the Myers--Perry metrics \cite{MyersPerry:1986}. Their best generalizations of type D known so far are the Kerr--NUT--(A)dS spacetimes \cite{ChenLuPope:2006} (for a recent review on their properties, see \cite{FrolovKrtousKubiznak:2017}). These solutions of Einstein's equations also incorporate NUT charges and a cosmological constant, but they are neither accelerated nor electrically charged.

Higher-dimensional spacetimes are interesting not only because of the ongoing development of the quantum gravity in the context of the string theory and the AdS/CFT correspondence, but also from the pure mathematical point of view. The study of various properties of the Kerr--NUT--(A)dS spacetimes led, for example, to a discovery of new general structures of hidden symmetries. Although these spacetimes have been studied extensively in the literature (particularly because of their integrability and separability properties), not much attention has been paid to the physical interpretation especially for non-vanishing NUT charges. It is because the meaning of the Kerr--NUT--(A)dS parameters is clear only in certain subcases, but not in the general case. The purpose of this paper is to identify the parameters related to different types of singularities on the symmetry axes of the Kerr--NUT--(A)dS spacetimes. We concentrate on a general case with nonzero NUT charges, but with a more specific choice of coordinate ranges than it is usually assumed in the literature.

The NUT charge (named after Neumann, Unti, and Tamburino) comes from the so called Taub--NUT spacetime \cite{Taub:1951,NewmanTamburinoUnti:1963}, which is known to have many unwanted properties such as the existence of closed timelike curves, the topological defects caused by the semi-infinite singularity on the symmetry axis, and others. In contrast to the Kerr spacetime, the Taub--NUT spacetime does not have a satisfactory interpretation (see, e.g., \cite{GriffithsPodolsky:2009book}). In the analogy with the theory of magnetic monopoles \cite{LyndenBellNouriZonos:1998}, the semi-infinite axial singularity resembles the Dirac's string \cite{DemianskiNewman:1966}. Contrary to the Dirac's string, which has no effect at all, the singular semi-axis of the Taub--NUT spacetime affects the spacetime geometry, and it can be seen as a thin massless spinning rod injecting angular momentum into the spacetime \cite{Bonnor:1969,Bonnor:1992,Bonnor:2001}. In the Kerr--NUT--(A)dS spacetimes, the NUT charges exhibit similar properties. Unfortunately, since these effects mix up with the rotations with respect to different planes, it is problematic to distinguish between the rotational parameters and the parameters responsible for NUT-like behavior.

The typical features of the four-dimensional NUT-charged spacetimes are similar to the properties of a spinning cosmic string. The exterior gravitational field of this object can be modelled by a very simple spacetime constructed as an identification of the Minkowski spacetime \cite{DeserJackiwtHooft:1984,Mazur:1986} (for an interior solution, see \cite{JensenSoleng:1992,MenaNatarioTod:2008}). This spacetime features a singularity on the symmetry axis that causes similar topological defects to those caused by the NUT charge. However, this axial singularity is equally distributed along the whole axis, unlike the axial singularities in the NUT-charged spacetimes, which can be different on different parts of the axis. As in the NUT-charged spacetimes, the singularity on the symmetry axis is surrounded by a region with closed timelike curves. Inspired by these similarities, we introduce geometric definitions of quantities describing various types of axial singularities for a sufficiently general class of higher-dimensional spacetimes that also contains the Kerr--NUT--(A)dS spacetimes.

The Kerr--NUT--(A)dS metrics are most often written in coordinates which arise from the hidden symmetries of the metrics and allow for separation of variables of many physical equations (Hamilton--Jacobi, Klein--Gordon \cite{FrolovEtal:2007}, Dirac \cite{OotaYasui:2008a}, Maxwell \cite{KrtousFrolovKubiznak:2008}, etc.). They were originally discovered by Carter \cite{Carter:1968b} (republished in \cite{Carter:2009}) and later generalized to higher dimensions by Chen, L\"{u}, and Pope \cite{ChenLuPope:2006}. Although these coordinates have proven to be extremely useful for many mathematical theorems, they are not very convenient for physical interpretations of the spacetimes. In \cite{KolarKrtous:2017}, we addressed this problem for non-Killing coordinates, where it was necessary to choose appropriate ranges of these coordinates to study various limiting procedures. Here, we follow up by introducing new Killing coordinates adjusted to particular isometries which in some cases lead to a generalization of the Boyer--Lindquist coordinates.

The paper is organized as follows: Sec.~\ref{sc:SSA} is devoted to a general description of singularities on the symmetry axes. In particular, we define quantities that capture different types of singularities. The definitions are inspired by the spinning cosmic string spacetime, which is studied in detail. In Sec.~\ref{sc:KNA}, we introduce the metrics of the Kerr--NUT--(A)dS spacetimes. We review the discussion of ranges of non-Killing coordinates from \cite{KolarKrtous:2017} and also introduce new Killing coordinates, whose meaning is explored further in the subsequent sections. This provides a more complete definition of the Kerr--NUT--(A)dS spacetimes. In Sec.~\ref{sc:SAKNA}, we explore the overall structure of the symmetry axes and deal with singularities located on them. In particular, we study the weak-field limit of the Kerr--NUT--(A)dS spacetimes and find the isometries with fixed points corresponding to the symmetry axes of the Kerr--NUT--(A)dS spacetimes. Also, we identify the coordinates generalizing the Boyer--Lindquist coordinates. Furthermore, we find some examples of the regions with closed timelike curves. Finally, the paper is concluded with a brief summary in Sec.~\ref{sc:C}. The ambiguities in definitions of the geometric quantities characterizing axial singularities are discussed separately in Appx.~\ref{ap:TKTW} where we find the corresponding transformation relations. Appx.~\ref{ap:FJAU} contains an overview of the notation and useful identities.


\section{Singularities on symmetry axes}
\label{sc:SSA}

\textit{Regular fixed points} of an isometry are points where the Killing vector generating the isometry vanishes. Unfortunately, this standard definition is not convenient for geometries with isometries whose fixed points are not regular points of spacetimes. This means that either the manifold is not smooth or the metric tensor degenerates. Since we would like to deal with such singular geometries, we introduce the notion of the \textit{generalized fixed points}.\footnote{For the sake of brevity we omit the words `regular' and `generalized' when their meaning is clear from the context.} Their definition only depends on their neighborhood and not on the points themselves. Strictly speaking, these points are not part of the manifold, but rather express the singular behavior of the spacetime. We say that a point is a generalized fixed point if all components of the Killing vector generating the isometry are small in the neighborhood of this point with respect to a `well-behaved' frame. 

A definition of these frames that would suit our purpose seems to be quite complicated. For this reason, we specify such frames for each spacetime with singular fixed points. Therefore, they can be regarded as a part of definitions of the singular spacetimes themselves. We refer to these frames as the \textit{semi-regular frames}, because we also require them to lead to regular frames in the limit of vanishing singularities. They define implicit maps between the spacetimes with singular fixed points and their regular limits. Typically, we choose such frames to be the Cartesian-type coordinate frame.

Important examples of singular fixed points are the singular symmetry axes which occur in many solutions of Einstein's equations with \textit{stationary rotational symmetries}. A symmetry axis is a well-defined concept in a general number of dimensions that can be also generalized to the presence of singularities. Let us consider a $D$-dimensional manifold equipped with a metric which is invariant under the action of the 1-parametric cyclic group $\mathrm{SO(2)}$. This property requires that the Killing vector generating the isometry of the metric has closed orbits. Such a Killing vector is often referred to as the \textit{cyclic Killing vector}. The set of fixed points of the cyclic Killing vector intuitively corresponds to the symmetry axis. Specifically, if the regular fixed points of the isometry generated by the cyclic Killing vector form a regular ${(D-2)}$-dimensional submanifold, such a manifold is said to be the \textit{regular symmetry axis}.

In the singular case, the situation is quite different, because the fixed points are not a part of the manifold and the cyclic Killing vector may not have fixed points at all. For this reason, the only condition we require for our metric apart from the $\mathrm{SO(2)}$ symmetry is the existence of a nontrivial Killing vector with the generalized fixed points. We refer to a set of such points as the \textit{generalized symmetry axis}. To distinguish it from the cyclic Killing vector, we call the Killing vector that vanishes towards the axis the \textit{axial Killing vector}.

\subsection{Minkowski spacetime}

A particular example of a regular axis is the symmetry axis of the four-dimensional Minkowski spacetime. The metric of the Minkowski spacetime ${(\mathcal{M}, \tb{g})}$ in the Cartesian coordinates~${(T,X,Y,Z)}$ reads
\begin{equation}\label{eq:metricflatcar}
	\tb{g} = -\tb{\dd}T^{\bs{2}}+\tb{\dd}X^{\bs{2}}+\tb{\dd}Y^{\bs{2}}+\tb{\dd}Z^{\bs{2}}\;.
\end{equation}
We assume that these coordinates are smooth and cover the complete manifold $\mathcal{M}$, so the spacetime is well defined everywhere.

Isometries of the Minkowski spacetime are described by the ten-parametric Poincar\'e group. Their algebra of generators is given by 10 independent Killing vectors, which consists of 4 translational, 3 rotational, and 3 boost symmetries. Without loss of generality, we concentrate on the rotation in the $X$-$Y$ plane, so we introduce more convenient cylindrical coordinates~${(t,\rho,\phi,z)}$,
\begin{equation}\label{eq:transfcarcyl}
\begin{gathered}
T=t\;,
\quad
X+\imag Y=\rho\e^{\imag\phi}\;,
\quad
Z=z\;,
\\
t=T\;,
\quad
\rho^2=\xi^2+\zeta^2\;,
\quad
\tan{\phi}=Y/X\;,
\quad
z=Z\;.
\end{gathered}
\end{equation}
where ${t,z\in\mathbb{R}}$, ${\rho>0}$, ${\phi\in (0,2\pi)}$. Employing this transformation the metric \eqref{eq:metricflatcar} takes the form\footnote{We denote the second tensor power of a tensor $\tb{T}$ by adding the bold superscript, ${{\tb{T}}^{\tb{2}}}$. In the abstract index notation, it reads \begin{equation*}
(\tb{T}^{\tb{2}})^{ab\dots}_{\tb{cd\dots}}=\tb{T}^{a\dots}_{c\dots}\tb{T}^{b\dots}_{d\dots}\;.
\end{equation*}}
\begin{equation}\label{eq:metricflatcyl}
	\tb{g} = -\tb{\dd}t^{\bs{2}}+\tb{\dd}\rho^{\bs{2}}+\rho^2\tb{\dd}\phi^{\bs{2}}+\tb{\dd}z^{\bs{2}}\;.
\end{equation}
Such a coordinate system is nicely adjusted to the time-translation symmetry ${\tb{\pp}_{t}=\tb{\pp}_T}$ and the rotational symmetry ${\tb{\pp}_{\phi}=X\tb{\pp}_Y{-}Y\tb{\pp}_X}$. Unfortunately, it has two obvious drawbacks: It does not cover the points ${X\neq0}$, ${Y=0}$ (${\phi= 0}$ or ${\phi= 2\pi}$ for ${\rho\not= 0}$) and the $Z$-axis, ${X=Y=0}$ (${\rho= 0}$), despite the fact that the manifold is smooth here. The first issue can be easily overcome by introducing a different coordinate $\tilde\phi$ which is shifted by a constant ${\tilde\phi_0\neq0}$,
\begin{equation}
X+\imag Y=\rho\e^{\imag(\tilde\phi+\tilde\phi_0)}\;,
\end{equation}
and therefore we can simply ignore it. However, the second one is more severe, since the determinant of the transformation \eqref{eq:transfcarcyl} vanishes at the $Z$-axis. Consequently, the Cartesian frame $\tb{\pp}_T$, $\tb{\pp}_X$, $\tb{\pp}_Y$, $\tb{\pp}_Z$ is more practical here, because it is regular everywhere.

Clearly, the Minkowski spacetime is $\mathrm{SO(2)}$ symmetric, because the Killing vector $\tb{\pp}_{\phi}$ is by definition cyclic. This Killing vector is axial, as its components with respect to the Cartesian frame smoothly vanish towards ${\rho= 0}$. Such points form a 2-dimensional manifold, i.e., the regular axis.

\subsection{Cosmic string spacetime}

\begin{figure}
    \centering
	\includegraphics[width=\columnwidth]{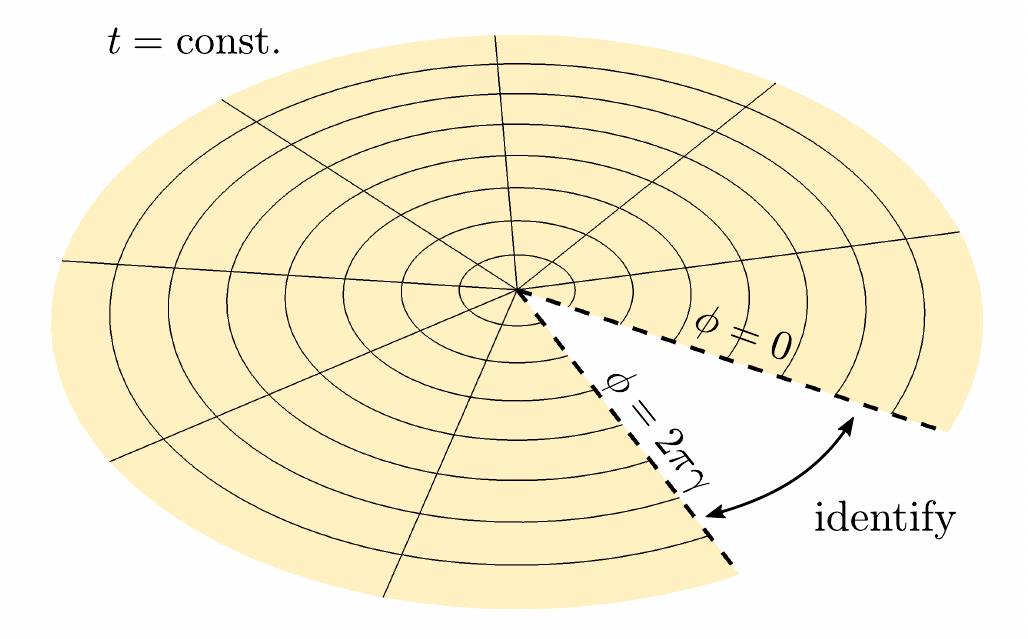}
    \caption{Construction of the cosmic string spacetime from the Minkowski spacetime.} \label{fig:cs}
\end{figure}

Consider a spacetime ${(\mathcal{M}, \tb{g})}$ with the metric $\tb{g}$ given by \eqref{eq:metricflatcyl}, but with ${\phi\in(0,2\pi\gamma)}$, ${\gamma>0}$, ${\gamma\neq 1}$. Such a spacetime is equivalent to the Minkowski spacetime where a wedge of angle $2\pi(1-\gamma)$ is artificially removed or added from all spatial sections of constant coordinates $t$ and $z$. The smoothness of the sections at ${\phi= 0}$ and ${\phi= 2\pi\gamma}$ can be regained by gluing these two edges together. It means that we identify the points with the values ${(t,\rho,\phi=0,z)}$ and ${(t,\rho,\phi=2\pi\gamma,z)}$. The construction is schematically illustrated in Fig.~\ref{fig:cs}.

Alternatively, one can define new coordinates $\tau$ and $\varphi$,
\begin{equation}
\tau = t\;,
\quad
\varphi = \frac{\phi}{\gamma}\;
\end{equation}
and bring the flat metric \eqref{eq:metricflatcyl} to the form
\begin{equation}\label{eq:cosmstr}
  \tb{g} = -\tb{\dd}\tau^{\bs{2}}+\tb{\dd}\rho^{\bs{2}}+\gamma^2\rho^2\tb{\dd}\varphi^{\bs{2}}+\tb{\dd}z^{\bs{2}}\;.
\end{equation}
Then, the identification can be reformulated simply as $2\pi$-periodicity in the coordinate $\varphi$. The resulting spacetime is often considered as a simple model of the infinite \textit{cosmic string/strut}, where the deficit and excess angles, given by the parameter $\gamma$, correspond to the tension in the string and compression of the strut, respectively.

We can also introduce the Cartesian-type coordinates ${(T,X,Y,Z)}$ by means of the transformation \eqref{eq:transfcarcyl} everywhere except for ${X=Y=0}$ (${\rho=0}$), as these points are singular, and thus do not belong to the spacetime. It is natural to regard the basis $\tb{\pp}_T$, $\tb{\pp}_X$, $\tb{\pp}_Y$, $\tb{\pp}_Z$ as the semi-regular frame defined above, as it reduces to the regular frame for ${\gamma\to1}$. We may also introduce frames $\tb{\pp}_{\tilde{T}}$, $\tb{\pp}_{\tilde{X}}$, $\tb{\pp}_{\tilde{Y}}$, $\tb{\pp}_{\tilde{Z}}$ rotated by an angle ${\phi_0\neq0}$, where 
\begin{equation}
    \tilde{T}=t\;,
    \quad
    \tilde{X}+i\tilde{Y}=\rho\e^{\imag(\phi+\phi_0)}\;,
    \quad
    \tilde{Z}=z\;,
\end{equation}
and consider these frames to be the semi-regular frames as well. The coordinates of two such frames cover the complete manifold and are smooth in the overlapping regions. Such coordinates represent `well-behaved' charts of an atlas and characterize this particular type of singularity.

Expressing the cyclic Killing vector $\tb{\pp}_{\varphi}$ with respect to this frame, ${\tb{\pp}_{\varphi}=\gamma\big(X\tb{\pp}_Y{-}Y\tb{\pp}_X\big)}$, we find that it has fixed points at ${\rho=0}$. These points form again a symmetry axis, however, because $\mathcal{M}$ is not smooth there, the symmetry axis is not regular. Thus, the cyclic vector ${\tb{\pp}_{\varphi}}$ is axial as well. In fact, all vectors proportional to ${\tb{\pp}_\phi=\gamma^{{-}1}\tb{\pp}_\varphi}$ with a constant coefficient share the same property. 

Please note that although the metric has the form \eqref{eq:metricflatcar}, it differs significantly from the Minkowski spacetime by the presence of the conical singularity on the axis which is encoded in rather complicated coordinate ranges of the Cartesian-type coordinates. The fact that the integral curves of $\tb{\pp}_{\varphi}$ are closed can be used to calculate the \textit{conicity} $\mathcal{C}$ of the 2-surface of constant coordinates $t$ and $z$ around the axis ${\rho=0}$. The conicity is a ratio of the circumference of a small circle with constant geodesic distance from the tip to the length along these radial geodesics (multiplied by $2\pi$). By using the metric form \eqref{eq:cosmstr}, we can verify that the conicity in this case is exactly the deficit/excess parameter $\gamma$,
\begin{equation}\label{eq:conicity}
	\mathcal{C}=\lim_{\rho\to 0}\frac{\int_0^{2\pi}|\tb{\pp}_\varphi|\,\dd\varphi}{2\pi\int_0^{\rho}|\tb{\pp}_{\rho}|\,\dd\rho}=\gamma\;.
\end{equation}

\subsection{Spinning cosmic string spacetime}\label{ssc:SCS}
\begin{figure}
    \centering
	\includegraphics[width=\columnwidth]{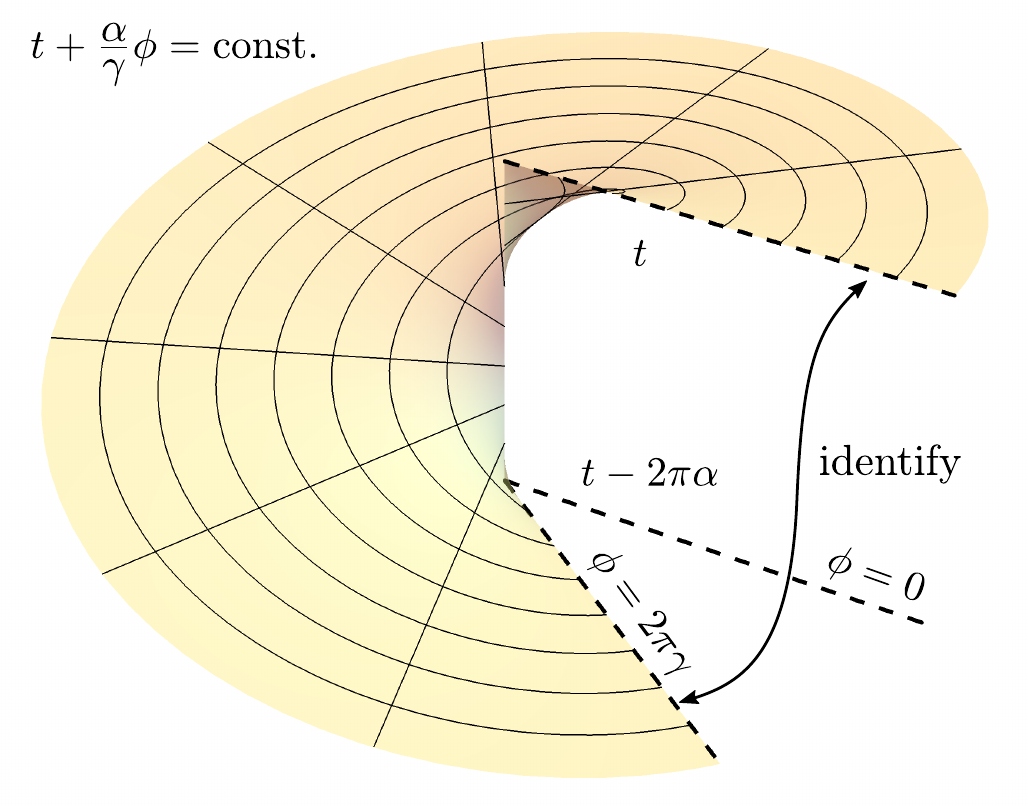}
    \caption{Construction of the spinning cosmic string spacetime from the Minkowski spacetime.}
    \label{fig:scs}
\end{figure}

The flat metric \eqref{eq:metricflatcyl} can be used to construct a spacetime with a more complicated axial singularity. As before, we again consider ${(\mathcal{M},\tb{g})}$, where ${\phi\in(0,2\pi\gamma)}$, ${\gamma>0}$ ${\gamma\neq 1}$. However, here we assume that the surfaces ${\phi=0}$ and ${\phi=2\pi\gamma}$ are shifted in time and joined together in a way that the points with the values ${(t,\rho,\phi=0,z)}$ and ${(t-2\pi\alpha,\rho,\phi=2\pi\gamma,z)}$, ${\alpha\neq 0}$, are identified. This gluing procedure is depicted in Fig.~\ref{fig:scs}.

Again, we can reformulate the identification by introducing new coordinates $\tau$ and $\varphi$,
\begin{equation}
\tau = t+\frac{\alpha}{\gamma}\phi\;,
\quad
\varphi = \frac{\phi}{\gamma}\;,
\end{equation}
and describing the identification in terms of $2\pi$\=/periodicity in the coordinate $\varphi$. In these coordinates, the flat metric \eqref{eq:metricflatcyl} takes the form
\begin{equation}\label{eq:scsmetric}
  \tb{g} = -\big(\tb{\dd}\tau-\alpha\tb{\dd}\varphi\big)^{\!\bs{2}}+\tb{\dd}\rho^{\bs{2}}+\gamma^2\rho^2\tb{\dd}\varphi^{\bs{2}}+\tb{\dd}z^{\bs{2}}\;.
\end{equation}
Such a metric is often interpreted as describing an external gravitational field around a general \textit{spinning cosmic string/strut} with the tension/compression given by~$\gamma$ and the angular momentum proportional to~$\alpha$.

\begin{figure}
    \centering
	\includegraphics[width=\columnwidth/\real{1.3}]{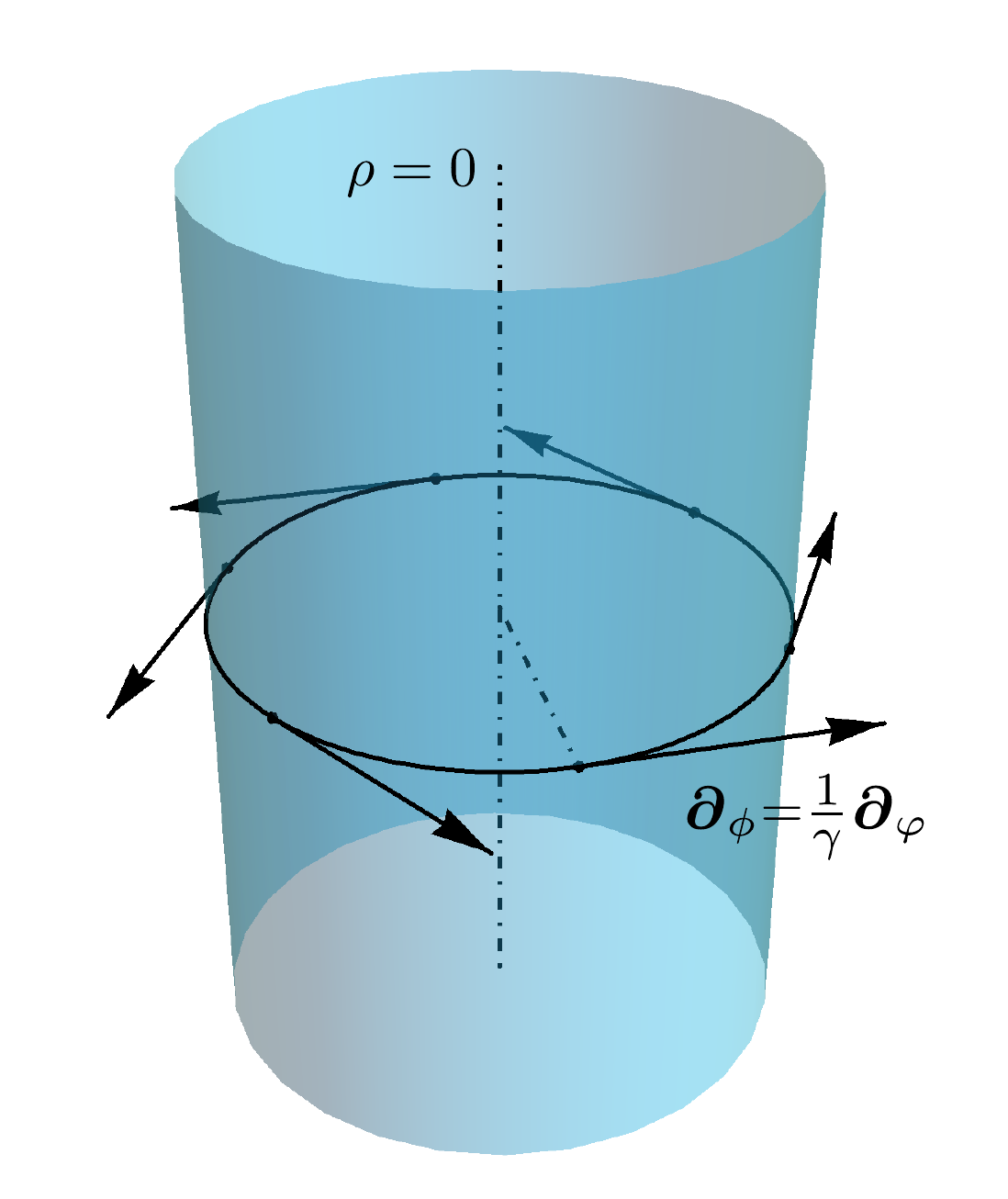}
	\includegraphics[width=\columnwidth/\real{1.3}]{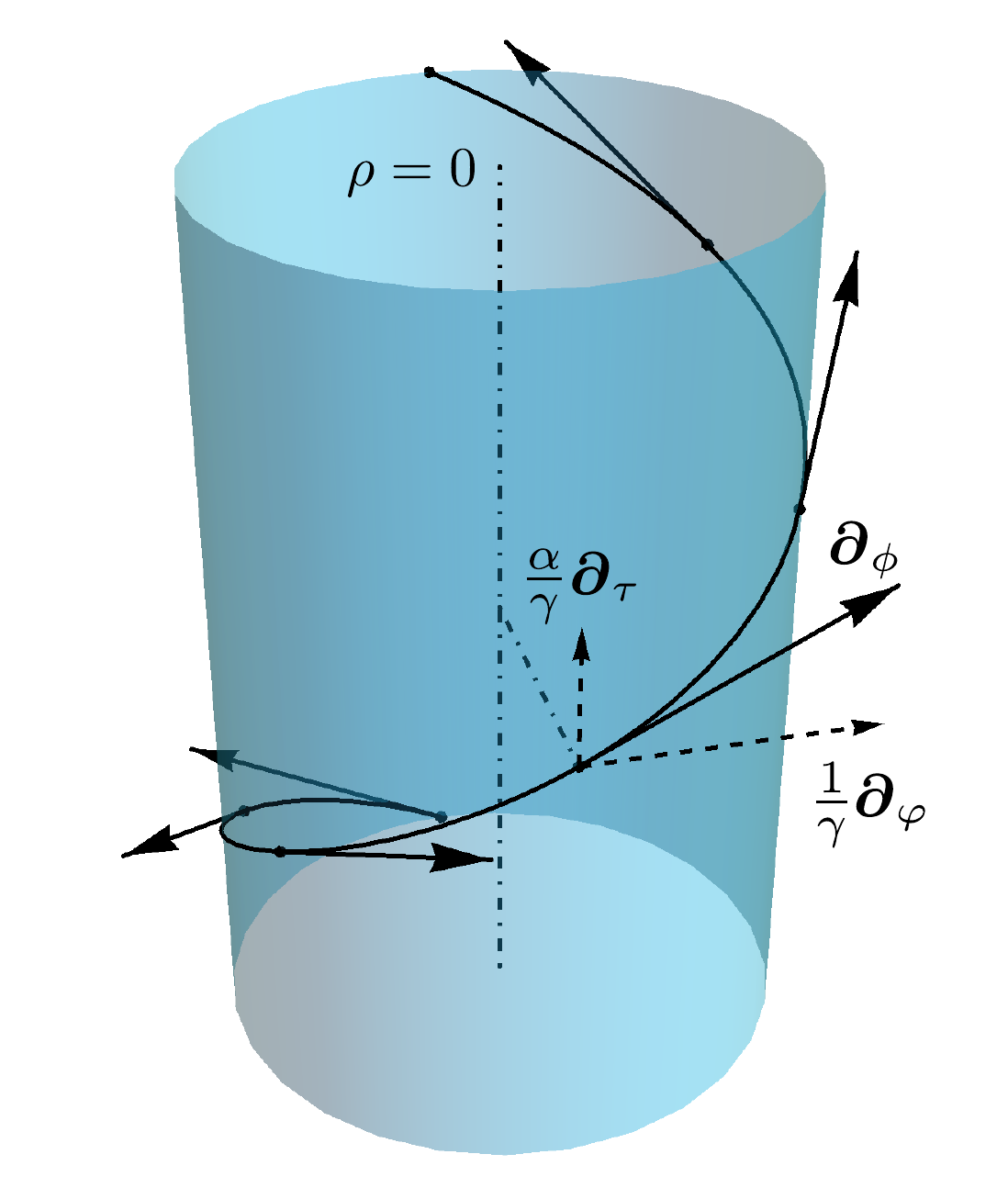}
    \caption{Orbits of the axial Killing vector~$\tb{\pp}_\phi$ that give rise to the symmetry axis at ${\rho=0}$. Top: the closed orbit of $\tb{\pp}_\phi$ in the non-spinning spacetime --- $\tb{\pp}_\phi$ is aligned with the cyclic Killing vector $\tb{\pp}_\varphi$. Bottom: the open orbit of $\tb{\pp}_\phi$ in the spacetime of the spinning cosmic string --- $\tb{\pp}_\phi$ differs from the cyclic Killing vector~$\tb{\pp}_\varphi$. Pictures are drawn after the smooth periodic gluing (indicated in Fig.~\ref{fig:cs} and Fig.~\ref{fig:scs}), i.e., $\rho$, $\varphi$, and $\tau$ are represented as cylindrical coordinates.}
    \label{fig:fp}
\end{figure}

As before, we can introduce the Cartesian-type coordinates~${(T,X,Y,Z)}$ by means of the transformation \eqref{eq:transfcarcyl} and assume that $\tb{\pp}_T$, $\tb{\pp}_X$, $\tb{\pp}_Y$, $\tb{\pp}_Z$ is the semi-regular frame. It clearly leads to a regular frame of the Minkowski spacetime if ${\alpha\to0}$ and ${\gamma\to1}$. We can follow the same procedure as that for the non-spinning case. We find that the cyclic Killing vector $\tb{\pp}_\varphi$ has no fixed points at ${\rho=0}$  (${X=Y=0}$), which can be verified by expressing $\tb{\pp}_\varphi$ with respect to the Cartesian-type frame, ${\tb{\pp}_{\varphi}=\gamma\big(X\tb{\pp}_Y{-}Y\tb{\pp}_X\big)-\alpha\tb{\pp}_T}$. The axial Killing vector corresponding to this axis~is
\begin{equation}\label{eq:phiphialphatau}
    \tb{\pp}_\phi=\frac{1}{\gamma}\big(\tb{\pp}_\varphi+\alpha\tb{\pp}_\tau\big)\;,
\end{equation}
as it vanishes towards ${\rho=0}$, ${\tb{\pp}_{\phi}=X\tb{\pp}_Y{-}Y\tb{\pp}_X}$. The misalignment of the cyclic and axial Killing vectors has interesting consequences for the orbits of the axial Killing vectors. While these orbits are closed for the non-spinning case, and thus coincide with the orbits of the cyclic Killing vectors, they are open for the spacetime of the spinning cosmic string as shown in Fig.~\ref{fig:fp}.

The cancellation of the two vectors on the right-hand side of \eqref{eq:phiphialphatau} in the limit ${\rho \to 0}$ may be used to introduce a new geometric quantity. This quantity, which is properly defined in the next section, is called the \textit{time shift} $\mathcal{T}$, because it is created by a time-shifted identification of the spacetime. In this case, it reduces to the proportionality constant between the cyclic Killing vector $\tb{\pp}_\varphi$ and the temporal Killing vector $\tb{\pp}_\tau$ that becomes aligned with it at the axis,
\begin{equation}\label{eq:timeshiftscs}
    \mathcal{T}=\alpha\;.
\end{equation}
Apart from the cyclicity of $\tb{\pp}_\varphi$ and ${2\pi}$-periodicity of $\varphi$, the definition also employs that $\tb{\pp}_\tau$ is timelike far from the axis ${\rho\to+\infty}$ with the normalization ${\tb{\pp}_\tau^2=-1}$.\footnote{For a tensor quantity $\tb{T}$, we introduce the second power $\tb{T}^2$ by the expression (in the abstract index notation) \begin{equation*}
\tb{T}^{2}=\tb{T}^{a\dots}_{c\dots}\tb{T}^{b\dots}_{d\dots}\dots\tb{g}_{ab}\tb{g}^{cd}\dots\;.
\end{equation*}}

\begin{figure}
    \centering
	\includegraphics[width=\columnwidth]{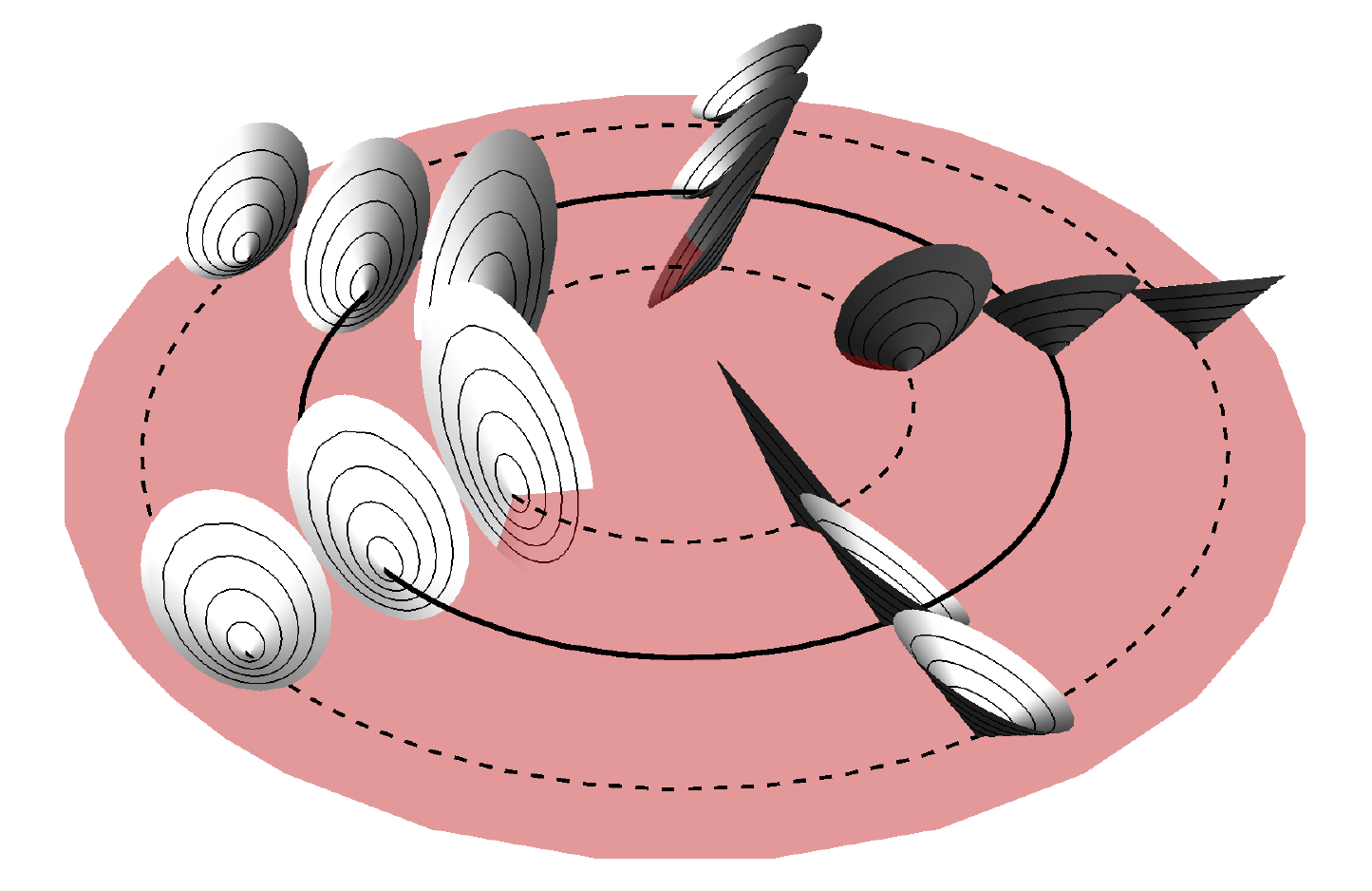}
    \caption{Future null cones of the spinning cosmic string spacetime on the surface of constant $\tau$. The circle denotes the position of the ergosurface, ${\rho=|\alpha|/\gamma}$. Outside the region, the cones lie completely above the surface, while inside the region the cones intersect the surface.}
    \label{fig:scsergo}
\end{figure}

The presence of the time-shift singularity manifests itself in the fact that the closed orbits of Killing vectors may have different causal characters in different regions of the spacetime. The surface where the causal character changes from spacelike to timelike is called the \textit{ergosurface}. We can find it by studying the norm of the cyclic Killing vector,
\begin{equation}
	\tb{\pp}_\varphi^2=\gamma^2\rho^2-\alpha^2\;.
\end{equation}
In this spacetime, the closed curves in direction $\tb{\pp}_\varphi$ are spacelike only for ${\rho>|\alpha|/\gamma}$, but for ${\rho<|\alpha|/\gamma}$ they are timelike. This behavior is visualized in terms of the future null cones in Fig.~\ref{fig:scsergo}.

So far we have investigated the fixed points at ${\rho=0}$, however, we could ask whether there exist fixed points for ${\rho>0}$ as well. Since the metric \eqref{eq:scsmetric} is regular in this coordinate patch, we can use the frame $\tb{\pp}_\tau$, $\tb{\pp}_\rho$, $\tb{\pp}_\varphi$, $\tb{\pp}_z$ for seeking axial Killing vectors. Given that a Killing vector is a linear combination of $\tb{\pp}_t$ and $\tb{\pp}_\varphi$ with constant coefficients, we conclude that the spacetime has no axial Killing vectors with fixed points at ${\rho>0}$.

\subsection{Conicities, time shifts, and twists}\label{ssc:CTST}

In the previous subsections, we encountered two types of axial singularities and associated quantities that measure these topological defects: the conicity in the non-spinning cosmic string and the time shift in the spinning cosmic string. Here, we elaborate on this by introducing proper definitions suitable for more general cases. Apart from generalizing the quantities to a general number of dimensions, we also introduce topological defects similar to time shifts called twists and extend the definitions of the conicities to the presence of other axial singularities. 

Let us consider a general higher-dimensional spacetime ${(\mathcal{M},\tb{g})}$ of dimension $D$ which admits $n$, ${2n+1\leq D}$ cyclic Killing vectors $\tb{u}_a$ and one additional non-cyclic Killing vector $\tb{t}$. The Latin indices take values
\begin{equation}
	a,b,\,\ldots =1,\,\dots,\,n\;.
\end{equation}
From now on, we do not use the Einstein summation convention for any indices, but we write sums and products explicitly as, for example, $\sum_a$ and $\prod_a$. Due to the $\mathrm{{SO(2)}}^n$ symmetry, it is natural to expect that there exist $n$ symmetry axes labelled by index $a$. Intuitively, the Killing vector $\tb{t}$ describes time symmetry, i.e., homogeneity in time. Therefore we assume that the Killing vector $\tb{t}$ is timelike in some region near the axes (for instance above the black hole horizon) and that all the Killing vectors $\tb{u}_a$ and $\tb{t}$ mutually Lie-commute,
\begin{equation}\label{eq:Liecomut}
    \lieb{\tb{u}_a}{\tb{u}_b}=0\;,
    \quad
    \lieb{\tb{u}_a}{\tb{t}}=0\;.
\end{equation}
The first condition alone implies that the set of Killing vectors $\tb{u}_a$ generate the foliation of $\mathcal{M}$ by $n$-dimensional surfaces~$\Sigma$. Each such surface is the $n$-dimensional torus $\mathbb{T}^n$. In addition we assume that the chosen cyclic Killing vectors $\tb{u}_a$ are tangent vectors of some $2\pi$-periodic coordinates which smoothly cover the whole $\mathbb{T}^n$ (except for the endpoints of coordinate ranges). The first and second conditions in \eqref{eq:Liecomut} together give rise to the foliation of $\mathcal{M}$ by ${(n+1)}$-dimensional surfaces~$\Gamma$ which have the topology of ${\mathbb{T}^n\times\mathbb{R}}$. The induced metric on each surface of both foliations $\Sigma$ and $\Gamma$ is flat, because the components of the metrics with respect to the coordinates given by the set of Lie-commuting Killing vectors are constant. The metric signature as well as the scaling in various directions may change from surface to surface within the foliations.

It turns out that different parts of the symmetry axis may be associated with different axial Killing vectors. Therefore, we introduce the index $p$ and denote these axial vectors by $\tb{v}_a^p$. Let us consider non-Killing radial vector fields $\tb{r}_a^p$ perpendicular to $\tb{t}$ and $\tb{u}_b$, at the axis,\footnote{More specifically, the perpendicularity means that ${\tb{r}_a^p\cdot\tb{g}\cdot\tb{t}}$ and ${\tb{r}_a^p\cdot\tb{g}\cdot\tb{u}_b}$, vanish towards the axes.} whose integral curves approximate geodesics near the axes. Furthermore, we assume that $\tb{v}_a^p$ and $\tb{r}_a^p$ generate the foliation of $\mathcal{M}$ by 2-surfaces $\Upsilon_a^p$. To write down the formula for the conicities of these surfaces in the presence of other axial singularities, we employ the integral curves of the axial Killing vectors $\tb{v}_a^p$ instead of the cyclic ones $\tb{u}_a$. The reason for this is that the lengths of the orbits of $\tb{v}_a^p$ shrink to zero as we approach the axes, although they may not be cyclic. 

\begin{figure}
    \centering
	\includegraphics[width=\columnwidth/\real{1.4}]{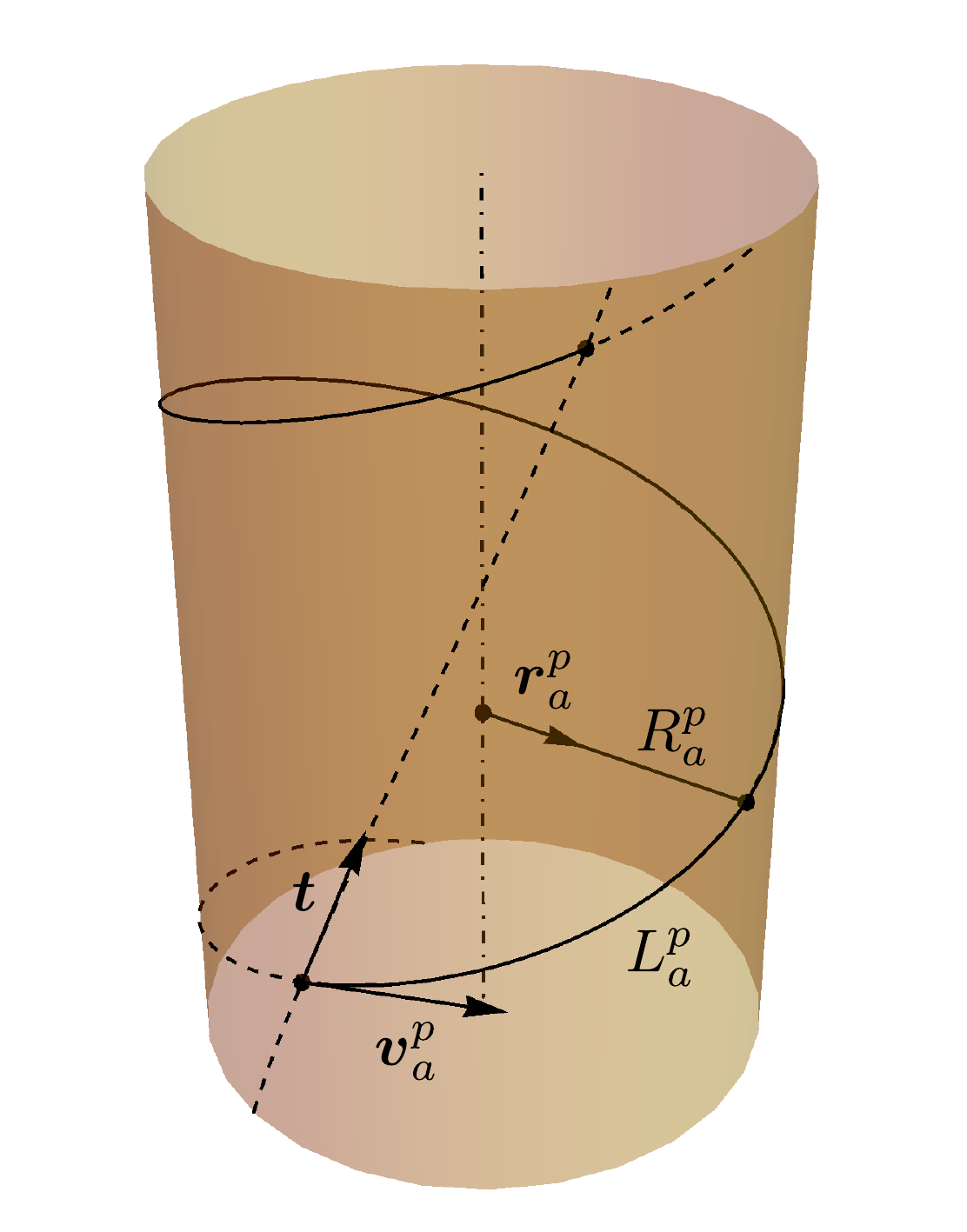}
    \caption{Generalized definition of the conicity in the presence of other axial singularities.}
    \label{fig:c}
\end{figure}

The conicity $\mathcal{C}_a^p$ of the surface of foliation $\Upsilon_a^p$ can then be defined~by the expression
\begin{equation}\label{eq:conicities}
    \mathcal{C}_a^p=\lim_{R_a^p\to0}\frac{L_a^p}{2\pi R_a^p}\;,
\end{equation}
where $L_a^p$ stands for the length of the integral curve of $\tb{v}_a^p$ which starts at the intersection with the integral curve of the temporal Killing vector $\tb{t}$ winds around the axes and ends at the next intersection. The symbol $R_a^p$ denotes the distance from the axis along the vector field $\tb{r}_a^p$ to a point at the orbit of $\tb{v}_a^p$. The conicity is well defined if $L_a^p$ is of the order $R_a^p$ and $R_a^p$ is independent of the point at the orbit of $\tb{v}_a^p$. The described geometric relation is illustrated in Fig.~\ref{fig:c}.

The axial Killing vectors $\tb{v}_a^p$ are not independent of $\tb{u}_a$ and $\tb{t}$. They can be expressed as a linear combination with constant coefficients,
\begin{equation}\label{eq:genlincom}
    \tb{v}_a^p=\mathcal{A}_a^p\,\tb{t}+\sum_b\mathcal{B}_{ab}^p\,\tb{u}_b\;.
\end{equation}
However, by analogy with the spinning cosmic string, we assume that $\tb{v}_a^p$ takes a particular form resembling \eqref{eq:phiphialphatau} and \eqref{eq:timeshiftscs},
\begin{equation}\label{eq:timeshifttwist}
    \tb{v}_a^p=\frac{1}{\mathcal{K}_a^p}\Bigg[\tb{u}_a+\mathcal{T}_a^p\,\tb{t}+\sum_{\substack{b \\ b\neq a}}\mathcal{W}_{ab}^p\,\tb{u}_b\Bigg]\;.
\end{equation}
Here, the constants $\mathcal{T}_a^p$ and $\mathcal{W}_{ab}^p$, ${b
\neq a}$, are referred to as the \textit{time shifts} and \textit{twists}, respectively. The cyclic and temporal Killing vectors become linearly dependent as we approach the axis, because the axial Killing vector $\tb{v}_a^p$ vanishes towards it. Thus, the quantities time shifts and twists are simply the coefficients of the $\tb{v}_a^p$ expressed in terms of the other Killing vectors at the axis (or infinitely close to it, when the axis is singular). As we will see on a particular example of the Kerr--NUT--(A)dS spacetimes, the constants $\mathcal{K}_a^p$ are equal to the conicities $\mathcal{C}_a^p$ if the vectors $\tb{v}_a^p$ are normalized so that the induced metric on $\Upsilon_a^p$ near the respective part of the axis is proportional to ${\tb{\dd}\rho^2+\rho^2\tb{\dd}\phi^2}$, where ${\tb{v}_a^p=\tb{\pp}_\phi}$ and ${\tb{r}_a^p=\tb{\pp}_\rho}$. After the separation of the vector $\tb{u}_a$ (with the same index $a$ as~$\tb{v}_a^p$) the coefficients $\mathcal{K}_a^p$, $\mathcal{T}_a^p$, and $\mathcal{W}_{ab}^p$, ${b\neq a}$, still carry the same information as the original coefficients $\mathcal{A}_a^p$ and~$\mathcal{B}_{ab}^p$.

The quantities $\mathcal{C}_a^p$ (or equivalently $\mathcal{K}_a^p$), $\mathcal{T}_a^p$, $\mathcal{W}_{ab}^p$ described above depend on the choice of the cyclic and temporal directions $\tb{u}_a$ and $\tb{t}$ on $\mathbb{T}^n\times\mathbb{R}$. Therefore, these quantities express the conicities, time shifts, and twists with respect to a given Killing frame. However, there often exist criteria that justify some choices in certain cases. The freedom in the change of the Killing directions, the corresponding transformations, and the some particular choices are thoroughly described in Appx.~\ref{ap:TKTW}.


\section{Kerr--NUT--(A)dS spacetimes}
\label{sc:KNA}

In this section, we introduce a standard form of the Kerr--NUT--(A)dS metrics which was found in \cite{ChenLuPope:2006}. For simplicity we concentrate on the even-dimensional spacetimes, ${D=2N}$, however, we expect that our results can be simply translated to odd dimensions as well. Additionally, we review an introductory part of \cite{KolarKrtous:2017}, where we addressed the question of when such metrics represent gravitational fields of black holes.

\subsection{Canonical coordinates}
The higher-dimensional Kerr--NUT--(A)dS metrics are usually written in the \textit{canonical coordinates} $x_\mu$ and $\psi_k$,
\begin{equation}\label{eq:KerrNUTAdSmetric}
  \tb{g} = \sum_{\mu}\bigg[\frac{U_\mu}{X_\mu}\,\tb{\dd}x_\mu^{\bs{2}}
  +\frac{X_\mu}{U_\mu}\Big(\sum_{k}A_{\mu}^{(k)}\,\tb{\dd}\psi_k\Big)^{\!\bs{2}}\bigg]\;,
\end{equation}
where the Greek and Latin indices take values
\begin{equation}
\begin{split}
	\mu,\nu,\,\dots &=1,\,\dots,\,N\;,
	\\
	k,l,\,\dots &=0,\,\dots,\,N-1\;.
\end{split}
\end{equation}
The functions $U_\mu$ and $A_\mu^{(k)}$ are polynomial expressions
\begin{equation}\label{eq:AUexp}
    A_\mu^{(k)}=\sum_{\mathclap{\substack{\nu_1,\,\dots,\,\nu_k\\ \nu_1<\dots<\nu_k\\ \nu_j\neq\mu}}}x_{\nu_1}^2\dots x_{\nu_k}^2\;,\quad
   U_\mu=\prod_{\substack{\nu\\ \nu\neq\mu}}(x_\nu^2-x_\mu^2)\;,
\end{equation}
and ${X_\mu= X_\mu(x_\mu)}$ are single-variable functions. The Einstein's equations are satisfied if and only if $X_\mu$ take form of a particular polynomials in $x_\mu$, 
\begin{equation}
 X_\mu = \lambda\mathcal{J}(x_\mu^2)-2b_\mu x_\mu\;,
\end{equation}
where ${\mathcal{J}(x^2)}$ is an even polynomial of degree ${2N}$ in ${x}$, which can be parametrized using its roots $a_\mu$,
\begin{equation}
    \mathcal{J}(x^2) = \prod_\mu (a_\mu^2-x^2)\;.
\end{equation}
The parameters $a_\mu$ may acquire complex values, provided that the polynomials $X_\mu$ remain real. The parameter $\lambda$ is the rescaled cosmological constant,
\begin{equation}
	\Lambda=(2N-1)(N-1)\lambda\;.
\end{equation}
The metrics \eqref{eq:KerrNUTAdSmetric} have many interesting properties that hold true even for $X_\mu$ being completely general functions ${X_\mu= X_\mu(x_\mu)}$. Such geometries are called the \textit{off-shell Kerr--NUT--(A)dS spacetimes}.


\subsection{Black holes}\label{ssc:BH}
Although \eqref{eq:KerrNUTAdSmetric} are remarkable geometries from the mathematical point of view, which possess a high degree of symmetry that gives rise to many interesting properties, they are not always physically significant. The reason for this is that their common definition given in the literature lacks information about ranges of coordinates and possible values of the parameters. In order for these geometries to describe black hole spacetimes, we have to specify all important details. Unfortunately, this problem is quite complex and we do not deliver its complete solution in this paper. Instead, we focus on some particular cases, however, most results of this paper should remain true even in more general settings. After we deal with a proper choice of ranges of coordinates $x_\mu$ \cite{KolarKrtous:2017}, we encounter an issue when we try to specify the ranges of coordinates $\psi_k$. This problem will be solved by introducing a new set of Killing coordinates.

\subsubsection{Non-Killing coordinates}

A metric of a physically meaningful spacetime should have the Lorentzian signature. For the metrics \eqref{eq:KerrNUTAdSmetric}, this can be achieved, for example, by taking the coordinate $x_{N}$ and the parameter $b_{N}$ to be imaginary,
\begin{equation}
	x_{N}=\imag r\;,
	\quad
	b_{N}=\imag m\;,
\end{equation}
and the coordinates $x_{\bar\mu}$ and parameters $b_{\bar\mu}$ to be real and chosen so that
\begin{equation}\label{eq:XU}
\frac{X_{\bar\mu}}{U_{\bar\mu}}>0\;.
\end{equation}
Here, the `barred' indices take values
\begin{equation}
\begin{split}
\bar\mu,\bar\nu,\dots &=1,2,\dots,\bar N\;,
\\
\bar k,\bar l,\dots &=0,1,\dots,\bar N-1\;,
\end{split}
\end{equation}
where ${\bar N=N-1}$. The condition \eqref{eq:XU} can be met by restricting $x_{\bar\mu}$ between adjacent roots ${}^\pm x_{\bar\mu}$ of $X_{\bar\mu}$,
\begin{equation}\label{eq:coorranges}
	{}^- x_{\bar\mu} <x_{\bar\mu}<{}^+ x_{\bar\mu}\;,
\end{equation}
in a way that the intervals for different $x_{\bar\mu}$ do not overlap. Without loss of generality, we assume that 
\begin{equation}\label{eq:coorarr}
x_1^2<x_2^2<\dots<x_{\bar N}^2\;.
\end{equation}

The coordinates $x_{\bar\mu}$ have the meaning of the \textit{latitudinal directions}, so one may introduce the angular-type coordinates ${\vartheta_{\bar\mu}\in(0,\pi)}$,
\begin{equation}\label{eq:thetatrans}
\begin{split}
\cos{\vartheta_{\bar\mu}} &=\frac{2}{{}^{+\!}x_{\bar\mu}-{}^{-\!}x_{\bar\mu}}\bigg(x_{\bar\mu}-\frac{{}^{+\!}x_{\bar\mu}+{}^{-\!}x_{\bar\mu}}{2}\bigg)\;.
\\
\end{split}
\end{equation}
Applying this transformation, the metrics can be rewritten as
\begin{equation}\label{eq:gtheta}
\begin{split}
  \tb{g} &= \sum_{\bar\mu}\bigg[\frac{U_{\bar\mu}}{Y_{\bar\mu}}\,\tb{\dd}\vartheta_{\bar\mu}^{\bs{2}}
  +\frac{Y_{\bar\mu}}{U_{\bar\mu}}\sin^2{\vartheta_{\bar\mu}}\Big(\sum_{k}A_{\bar\mu}^{(k)}\,\tb{\dd}\psi_k\Big)^{\!\bs{2}}\bigg]
  \\
  &\feq+\frac{J(-r^2)}{\Delta}\,\tb{\dd}r^{\bs{2}}-\frac{\Delta}{J(-r^2)}\Big(\sum_{k}A_{N}^{(k)}\,\tb{\dd}\psi_k\Big)^{\!\bs{2}}\;,
\end{split}
\end{equation}
where all $x_{\bar\mu}$ are replaced by $x_{\bar\mu}(\vartheta_{\bar\mu})$ according to \eqref{eq:thetatrans}. The polynomials $Y_{\bar\mu}$ are defined by factoring out the roots~${}^{\pm\!}x_{\bar\mu}$,
\begin{equation}
\begin{split}
X_{\bar\mu} &=({}^{+\!}x_{\bar\mu}-x_{\bar\mu})(x_{\bar\mu}-{}^{-\!}x_{\bar\mu})Y_{\bar\mu}
\\
&=\frac{{({}^{+\!}x_{\bar\mu}-{}^{-\!}x_{\bar\mu})}^2}{4}Y_{\bar\mu}\big(x_{\bar\mu}(\vartheta_{\bar\mu})\big)\sin^2{\vartheta_{\bar\mu}}\;.
\end{split}
\end{equation}
The resulting metrics \eqref{eq:gtheta} do not contain the divergences in angular part, but they still degenerate when ${\cos{\vartheta_{\bar\mu}}\to\pm 1}$. As we will see later, the properties of the endpoints of ${x_{\bar\mu}}$ depend on the proper identifications of the spacetime in Killing directions, so we postpone the discussion for now. We also introduced the polynomial ${\Delta=-X_N}$.

In contrast with $x_{\bar\mu}$, the coordinate $r$ is not bounded by any values in general, since it has the meaning of the \textit{radial direction}.\footnote{In special subcases, the coordinate $r$ may be restricted to the positive values only. This is because for some parameter values the endpoint ${r=0}$ may correspond to the singularity or the origin.} Its domain of definition contains all real numbers except for the roots of $\Delta$, which correspond to the positions of horizons. 

The metrics \eqref{eq:KerrNUTAdSmetric} contain a gauge freedom which allows us to set one parameter $a_\mu$ to an arbitrary value. After fixing the parameter $a_{N}$, 
\begin{equation}\label{eq:gaugechlamb}
	a_{N}^2=-\frac{1}{\lambda}\;,
\end{equation}
we are left with mass $m$, rotations $a_{\bar{\mu}}$, NUT charges $b_{\bar{\mu}}$, and the cosmological constant $\lambda$. Even though this is a common naming of the parameters, their physical interpretation might be different. The relationship to actual physical quantities have been verified so far only for certain subcases such as the Myers--Perry black holes (${b_{\bar\mu}=0}$ and ${\lambda=0}$). The meaning of the NUT charges in various limits of the Kerr--NUT--(A)dS also differs considerably \cite{KrtousEtal:2016,KolarKrtous:2017}.

\begin{figure}
    \centering
	\includegraphics[width=\columnwidth]{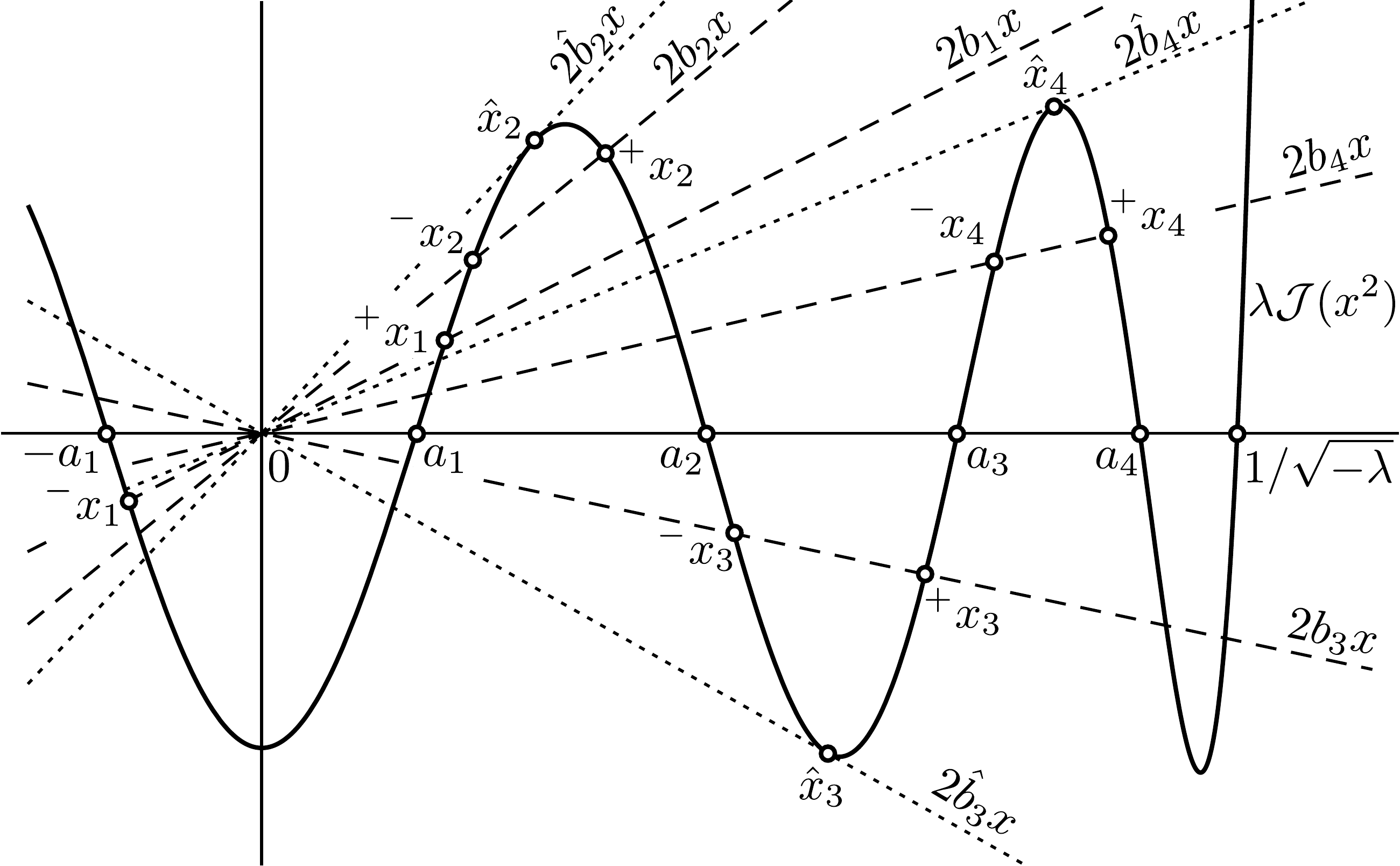}
    \caption{Graph of the polynomial $\lambda\mathcal{J}(x^2)$. Intersections of the polynomial and the lines $2b_{\bar{\mu}}x$ passing through the origin determine the ranges of coordinates $x_{\bar{\mu}}$, see \eqref{eq:coorranges}, \eqref{eq:xrootsineq1}.}
    \label{fig:rangex1}
\end{figure}

\begin{figure}
    \centering
	\includegraphics[width=\columnwidth]{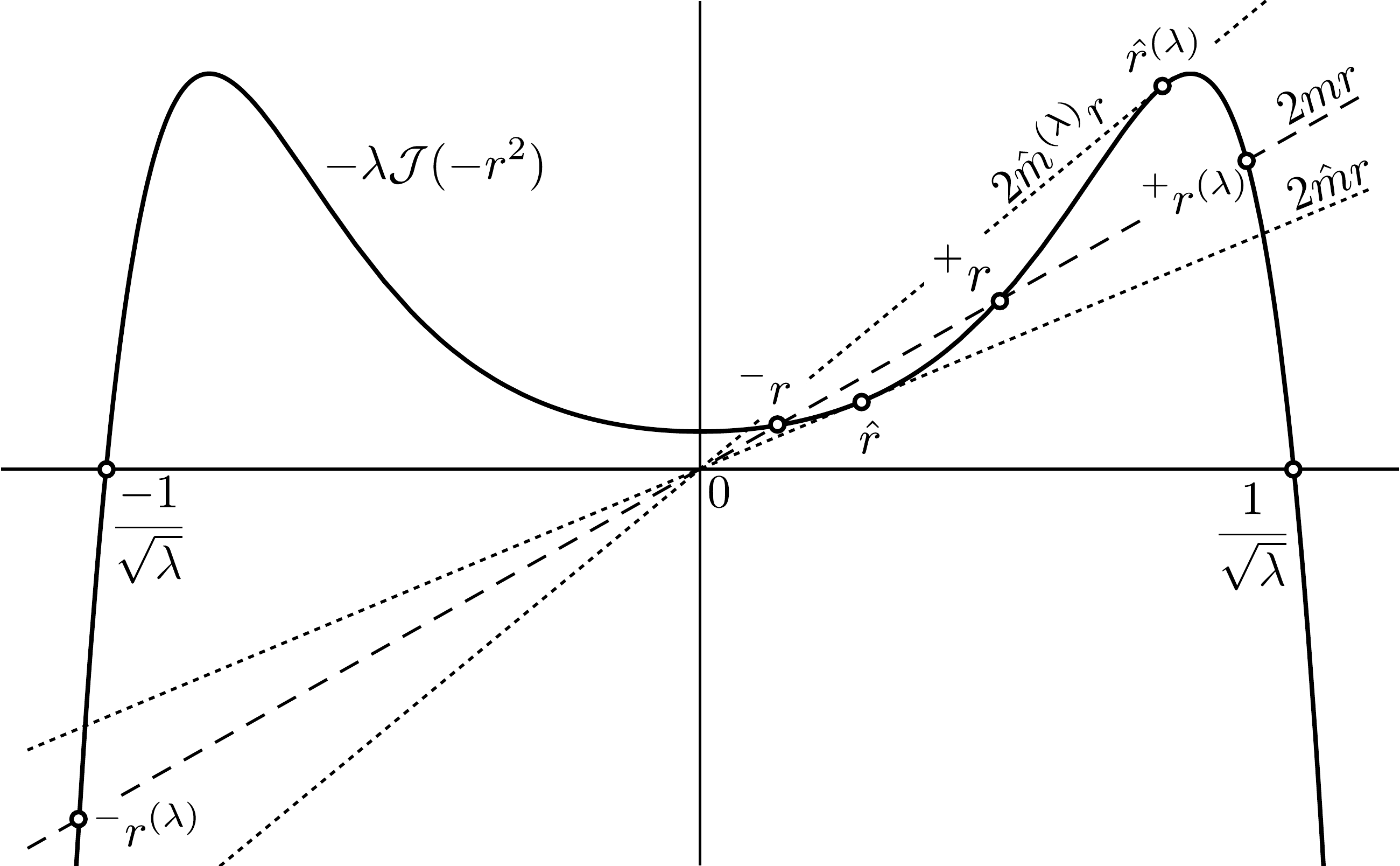}
    \caption{Graph of the polynomial $-\lambda\mathcal{J}(-r^2)$. Intersections of the polynomial and the line $2mr$ passing through the origin represent the horizons ${}^{\pm}r$, ${}^{\pm}r^{(\lambda)}$.}
    \label{fig:ranger1}
\end{figure}

Let us consider that the parameters $a_{\bar\mu}$ are positive and well ordered,
\begin{equation}
	0<a_1<a_2<\dots<a_{\bar{N}}<\frac{1}{|\lambda|}\;,
\end{equation}
and search for the possible values of the parameters $b_{\bar\mu}$ such that the metrics have the Lorentzian signature. Unfortunately, this problem cannot be solved this way in full generality, because it requires algebraic solutions of high-degree polynomial equations. Still, we can deduce a few basic properties from graphs of intersections of the polynomial ${\lambda\mathcal{J}(x^2)}$ with the lines $2b_{\mu}x$. A typical example of this graph for the ranges of coordinates $x_{\bar\mu}$ is shown in Fig.~\ref{fig:rangex1}. An analogous graph for the radial coordinate $r$ is depicted in Fig.~\ref{fig:ranger1}.

In Fig.~\ref{fig:rangex1}, we see that further from the origin there are just alternating `hills' and `valleys' of $\lambda\mathcal{J}(x^2)$. Thus, to meet the requirements \eqref{eq:coorranges} and \eqref{eq:coorarr}, we can assume that the parameters $b_{\bar\mu}$ are of alternating signs and satisfy
\begin{equation}
\begin{split}
	0 &<b_2 <\hat{b}_2\;,
	\\
	0 &<b_4 <\hat{b}_4\;,
	\\
	&\;\;\;\;\;\;\vdots
\end{split}
\quad
\begin{split}
	\hat{b}_3 &<b_3 <0\;,
	\\
	\hat{b}_5 &<b_5 <0\;,
	\\
	&\;\;\;\;\;\;\vdots
\end{split}
\end{equation}
The constants $\hat{b}_{\bar\mu}$ are critical values of parameters $\hat{b}_{\bar\mu}$ for which the roots ${}^{\pm} x_{\bar\mu}$ merge into double roots $\hat{x}_{\bar\mu}$ of $X_{\bar\mu}$, which also label the positions of the tangent points in Fig.~\ref{fig:rangex1}. Closer to the origin the situation differs significantly, because the roots ${}^{\pm\!}x_1$ have different signs and cannot merge, 
\begin{equation}\label{eq:xrootsineq1}
	-a_1<{}^{-\!}x_1<0<a_1<{}^{+\!}x_1<{}^{-\!}x_2\;.
\end{equation}
For this reason, the parameter $b_{1}$ is bounded by $b_2$,
\begin{equation}
   0<b_1<b_2.\;
\end{equation}
Such a choice of parameters $b_{\mu}$ always exists if the parameters $b_{\bar\mu}$ are sufficiently small, 
\begin{equation}
|b_{\bar\mu}|\lesssim 1\;.
\end{equation}
We refer to this choice as to the \textit{Kerr-like choice}, because it corresponds to a rotating black hole with small NUT charges. Another option is the \textit{NUT-like choice}, which is interpreted as NUT-like black hole with small rotations and requires the value of ${a_1}$ to be imaginary \cite{KolarKrtous:2017}.

The geometries \eqref{eq:KerrNUTAdSmetric} with the Kerr-like choice admit a similar causal structure to the four-dimensional Kerr spacetime. The metric signature remains Lorentzian when we cross the root of the function $\Delta$, but the coordinate $r$ changes its character. It is spatial if ${\Delta>0}$ and temporal if ${\Delta<0}$. Therefore, the surfaces ${\Delta=0}$ really correspond to the horizons, because they separate the stationary and nonstationary regions of the spacetime. Fig.~\ref{fig:ranger1} shows that depending on the value of $m$, the spacetime has inner ${}^{-}r$, outer ${}^{+}r$, and two cosmological ${}^{\pm}r^{(\lambda)}$ horizons. The cosmological horizons are present above and below ${r=0}$ for ${\lambda>0}$ only. Values $\hat{m}$, $\hat{m}^{(\lambda)}$ are critical masses for which the horizons ${}^{-}r$, ${}^{+}r$ or ${}^{+}r$, ${}^{+}r^{(\lambda)}$ merge into a single extreme horizon, which are denoted by $\hat{r}$, $\hat{r}^{(\lambda)}$ respectively.

\subsubsection{Killing coordinates}

In order to give a complete definition of the Kerr--NUT--(A)dS spacetimes, we still need to specify the restrictions on the Killing coordinates $\psi_k$. Since we want the spacetimes to describe black holes that arbitrarily rotate in $\bar{N}$ independent planes of rotations, the Killing coordinates should be adjusted to the \textit{longitudinal} and \textit{temporal directions}. Naturally, we would like the longitudinal coordinates to be $2\pi$-periodic with the tangent vectors being cyclic Killing vectors. Unfortunately, the canonical coordinates $\psi_k$ are not of this type in contrast to, for instance, the Boyer--Lindquist coordinates $\tau$, $\varphi$ in the four-dimensional Kerr spacetime (${N=2}$, ${b_{\bar{\mu}}=0}$, ${\lambda=0}$). By comparing the canonical coordinates to the Boyer--Lindquist coordinates,
\begin{equation}
	\tau=\psi_0+a_1^2\psi_1\;,
	\quad
	\varphi=a_1\psi_1\;,
\end{equation}
we see that the vector ${\tb{\pp}_{\psi_1}=a_1(\tb{\pp}_{\varphi}+a_1\tb{\pp}_{\tau})}$ is not aligned with the cyclic Killing vector $\tb{\pp}_{\varphi}$. 

Thus, the coordinates $\psi_k$ are not very convenient and the original metrics of the form \eqref{eq:KerrNUTAdSmetric} do not provide necessary information for the physical interpretation. For this reason, we consider the coordinate transformation (the inverse transformation follows from  \eqref{eq:AmuU})
\begin{equation}\label{eq:psiphitran}
\psi_k = \sum_{\mu}\frac{(-\cir{x}_{\mu}^2)^{N-k-1}}{\cir{c}_{\mu}}\varphi_{\mu}\;,
\quad
\varphi_{\mu} = \frac{\cir{c}_{\mu}}{\cir{U}_{\mu}}\sum_k \cir{A}_{\mu}^{(k)}\psi_k\;,
\end{equation}
where we introduced new coordinates $\varphi_{\mu}$ as well as new arbitrary parameters $\cir{x}_{\mu}$ and $\cir{c}_{\mu}$, such that ${\cir{x}_{\mu}^2\neq\cir{x}_{\nu}^2}$ for ${\mu\neq\nu}$ and ${\cir{c}_{\mu}\neq0}$. The circle above the quantities such as $\cir{A}_{\mu}^{(k)}$ or $\cir{U}_{\mu}$ are defined by the same expressions as $A_{\mu}^{(k)}$ or $U_{\mu}$, but involving $\cir{x}_{\bar\mu}$ instead of $x_{\bar\mu}$. We can ignore the signs of $\cir{x}_{\mu}$, because only $\cir{x}_{\mu}^2$ occur in \eqref{eq:psiphitran}. In order to avoid expressions with confusing signs we supplement the definition of the Kerr--NUT--(A)dS spacetimes with several additional conditions on parameters $\cir{x}_{\mu}$ and $\cir{c}_{\mu}$, see \eqref{eq:xcircond1}, \eqref{eq:xcircond2}, and \eqref{eq:ccond} below. The corresponding transformation of the Killing vectors reads
\begin{equation}\label{eq:phipsiKVtrans}
    \tb{\pp}_{\psi_k} {=}\sum_\mu \cir{c}_{\mu}\frac{\cir{A}_{\mu}^{(k)}}{\cir{U}_\mu}\tb{\pp}_{\varphi_\mu}\;,
    \;\;\;
    \tb{\pp}_{\varphi_\mu} {=}\frac{1}{\cir{c}_{\mu}}\sum_k (-\cir{x}_{\mu}^2)^{N-k-1}\tb{\pp}_{\psi_k}\;.
\end{equation}

Applying the coordinate transformation \eqref{eq:psiphitran}, the metrics \eqref{eq:KerrNUTAdSmetric} can be cast to the form
\begin{equation}\label{eq:KerrNUTAdSmetricALT}
  \tb{g} = \sum_{\mu}\bigg[\frac{U_\mu}{X_\mu}\,\tb{\dd}x_\mu^{\bs{2}}
  +\frac{X_\mu}{U_\mu}\Big(\sum_{\nu}\frac{J_{\mu}(\cir{x}_{\nu}^2)}{\cir{c}_{\nu}}\,\tb{\dd}\varphi_{\nu}\Big)^{\!\bs{2}}\bigg]\;.
\end{equation}
Since we want the Killing coordinates $\varphi_{\bar\mu}$ to represent the longitudinal coordinates, we assume that the points with values ${\varphi_{\bar\mu}= 0}$ and ${\varphi_{\bar\mu}= 2\pi}$ are identified so that the Killing vectors $\tb{\pp}_{\varphi_{\bar\mu}}$ are cyclic and generate the $\bar{N}$-di\-mensional torus. We leave the temporal coordinate $\varphi_N$ and the vector $\tb{\pp}_{\varphi_N}$ unspecified.

Our definition of the Kerr--NUT--(A)dS spacetimes is not only mathematically complete, but also it allows for a study of symmetry axes, which we deal with in the rest of the paper. For instance, we show that the parameters $\cir{x}_{\bar\mu}$ and $\cir{c}_{\bar\mu}$ have a nice geometric meaning while $\cir{x}_N$ and $\cir{c}_N$ are related to a gauge freedom in a coordinate transformation. Altogether, the Kerr--NUT--(A)dS spacetimes \eqref{eq:KerrNUTAdSmetricALT} with the gauge choices for $a_N$ \eqref{eq:gaugechlamb}, $\cir{c}_N$, and $\cir{x}_N$ (\eqref{eq:paramxcgauge} below) contain ${4N{-}3}$ physical parameters (excluding the parameter of the theory $\lambda$). Besides ${2N{-}1}$ parameters $a_{\bar\mu}$, $b_{\bar\mu}$, and $m$, there are additional ${2N{-}2}$ parameters $\cir{x}_{\bar\mu}$ and $\cir{c}_{\bar\mu}$, which are connected with the conicities, time shifts, and twists.

\section{Symmetry axes of Kerr--NUT--(A)dS}
\label{sc:SAKNA}

Before we begin studying the symmetry axes in the general Kerr--NUT--(A)dS spacetimes \eqref{eq:KerrNUTAdSmetricALT}, we first examine the weak-field limits. This provides us with some insight into the structure of the symmetry axes in higher dimensions, particularly their coordinate description. If the NUT charges are non-vanishing, the symmetry axes are necessarily singular. We find the axial Killing vectors and demonstrate the presence of the conical, time-shift, and twist singularities by calculating the appropriate quantities defined in Sec.~\ref{ssc:CTST}.


\subsection{Weak-field limit}

\begin{figure}
    \centering
	\includegraphics[width=\columnwidth]{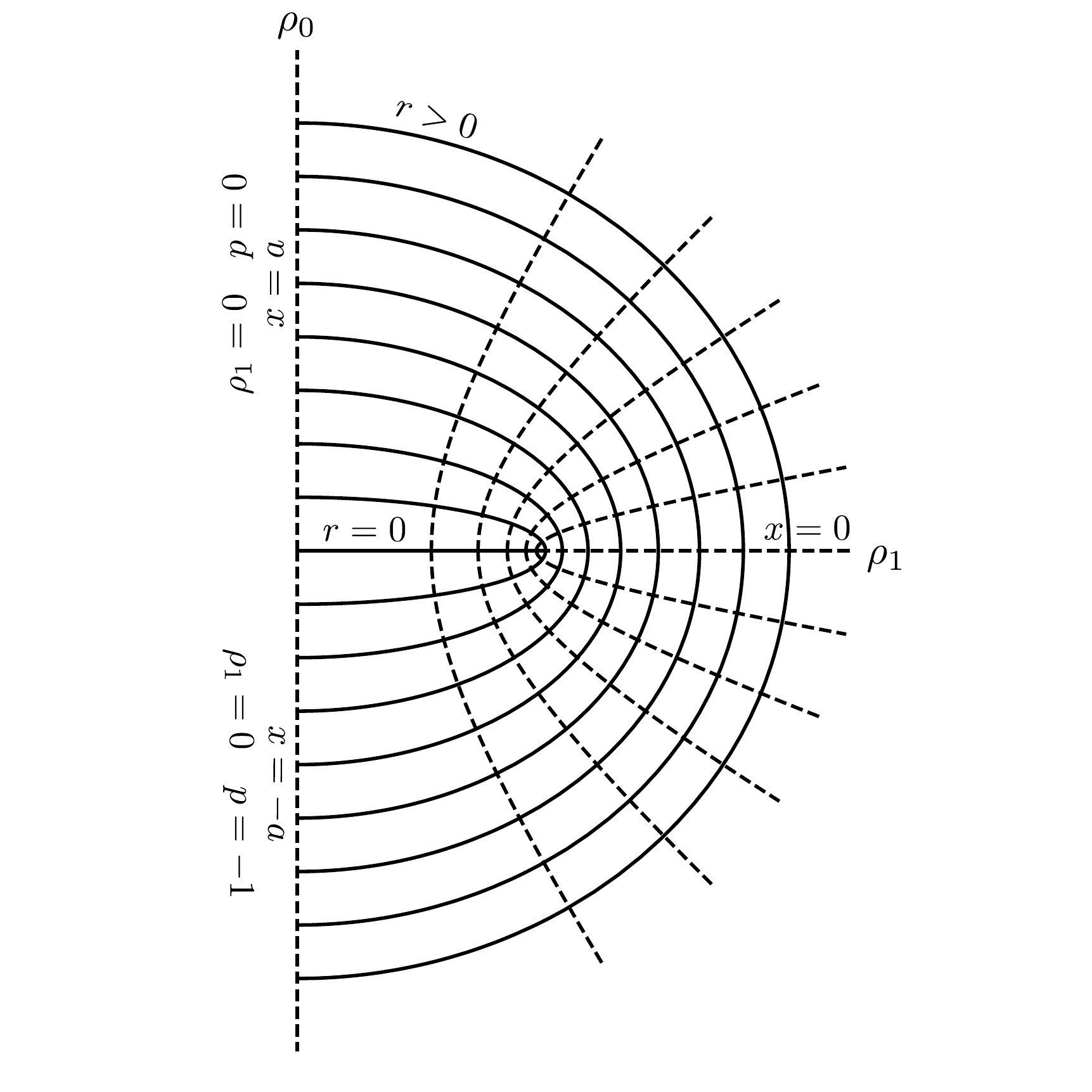}
    \caption{Coordinates ${x_1=x}$, ${x_2=\imag r}$ with respect to $\rho_0$, $\rho_1$ for ${D=4}$, ${\lambda=0}$. Coordinate values ${x=\pm a}$ correspond to the regular axis at ${\rho_1=0}$.}
    \label{fig:coorflat4}
\end{figure}

Let us first recall the weak-field limit of the four-dimensional Kerr spacetime. The geometry we obtain by taking the zero mass limit ${m\to0}$ is the Minkowski spacetime in the oblate spheroidal coordinates ${(\tau,r,x,\varphi)}$, where the rotational parameter $a$ plays a role of the spheroid equatorial radius. These coordinates are depicted for constant values of $\tau$ and $\varphi$ in Fig.~\ref{fig:coorflat4}. In the case of the NUT charge described by the parameter $b$, the situation is different. Not only does $b$ induce the axial singularity, but it also contributes to the curvature even if ${m=0}$. Therefore the flat spacetime is obtained only by taking both limits ${m\to0}$ and ${b\to0}$. The procedure can be also generalized to the spacetime with a nonzero cosmological constant, which leads to either the de Sitter or anti-de Sitter spacetimes.

Now we consider the higher-dimensional Kerr--NUT--(A)dS spacetimes ${(\mathcal{M},\tb{g})}$ with the metric $\tb{g}$ of the form \eqref{eq:KerrNUTAdSmetricALT}. In order to get the limit of maximally symmetric spacetimes, we set the parameters $\cir{x}_{\mu}$ and $\cir{c}_{\mu}$ to the specific values,
\begin{equation}\label{eq:xgcira}
\begin{gathered}
\cir{x}_{\bar\mu}^2=a_{\bar{\mu}}^2\;,
\quad
\cir{x}_N^2=a_{N}^2\;,
\\
\cir{c}_{\bar{\mu}} =a_{\bar{\mu}}|\lambda \mathcal{U}_{\bar{\mu}}|\;,
\quad
\cir{c}_N =\mathcal{U}_N \;,
\end{gathered}
\end{equation}
where $\mathcal{U}_{\mu}$ is defined by similar expressions as \eqref{eq:AUexp}, but involving $a$'s instead of $x$'s,
\begin{equation}
    \mathcal{U}_\mu=\prod_{\substack{\nu\\ \nu\neq\mu}}(a_\nu^2-a_\mu^2)\;.
\end{equation}
By turning off all NUT charges and mass,
\begin{equation}\label{eq:zeronutsmass}
    b_{\mu}=0,
\end{equation}
we specified all necessary parameters and the metric \eqref{eq:KerrNUTAdSmetricALT} simplifies to
\begin{equation}
	\tb{g}=\sum_\mu\frac{U_\mu}{\lambda\mathcal{J}(x_\mu^2)}\tb{\dd}x_\mu^{\bs{2}}{+}\sum_{\bar\mu}\frac{J(a_{\bar\mu}^2)}{-\lambda a_{\bar\mu}^2 \mathcal{U}_{\bar\mu}}\tb{\dd}\varphi_{\bar\mu}^{\bs{2}}{-}\lambda\frac{J(a_{N}^2)}{\mathcal{U}_{N}}\tb{\dd}\varphi_N^{\bs{2}}\;,
\end{equation}
by means of the identity \eqref{eq:idJJJ}.


To see that this geometry describes maximally symmetric spacetimes, we introduce new coordinates 
\begin{equation}\label{eq:multicyl}
   \rho_{0}^2=\frac{A^N}{\lambda \mathcal{A}^N}\;,
	\quad
	\rho_{\bar\mu}^2=\frac{J(a_{\bar\mu}^2)}{-\lambda a_{\bar\mu}^2 \mathcal{U}_{\bar\mu}}\;.
\end{equation}
By using the relations\footnote{Equations \eqref{eq:idfl} can be obtained from \eqref{eq:AN1} and \eqref{eq:idJJdJ}.}
\begin{equation}\label{eq:idfl}
\begin{split}
\frac{A^{N-2}_{\mu\nu}}{\mathcal{A}^N}+\sum_\kappa\frac{J_{\mu\nu}(a_\kappa^2)}{-a_\kappa^2\mathcal{U}_\kappa} &=0\;,
\\
\frac{A^{N-1}_{\mu}}{\mathcal{A}^N}+\frac{J_\mu(a_\kappa^2)}{-a_\kappa^2\mathcal{U}_\kappa}\frac{x_\mu^2}{x_\mu^2-a_\kappa^2} &=\frac{U_\mu}{\mathcal{J}(x_\mu^2)}\;,
\end{split}
\end{equation}
we can cast the metric to the form
\begin{equation}
\tb{g}=\tb{\dd}\rho_0^{\bs{2}} +\sum_{\bar\mu}\big(\tb{\dd}\rho_{\bar\mu}^{\bs{2}} +\rho_{\bar\mu}^2\tb{\dd}\varphi_{\bar\mu}^{\bs{2}}\big) +\tb{\dd}\rho_N^{\bs{2}} -\lambda \rho_N^2 \tb{\dd}\varphi_N^{\bs{2}}\;,
\end{equation}
where
\begin{equation}
	\rho_N^2=\frac{1}{\lambda}-\sum_{\mu=0}^{N-1}\rho_\mu^2\;.
\end{equation}
Here, the de Sitter spacetime (${\lambda>0}$) or anti-de Sitter spacetime (${\lambda<0}$) are expressed as a $2N$-dimensional surface embedded in the ${(2N{+}1)}$-dimensional flat space with the signature ${(-,+,\dots,+)}$ or ${(-,-,+,\dots,+)}$, respectively. To show the signature more explicitly, for ${1/\lambda<\sum_{\mu=0}^{N-1}\rho_\mu^2}$, when $\rho_N^2$ is negative, we may replace the function $\rho_N$ by a Wick-rotated one ${\rho_N=\imag s}$.

\begin{figure}
    \centering
	\includegraphics[width=\columnwidth/\real{1.205}]{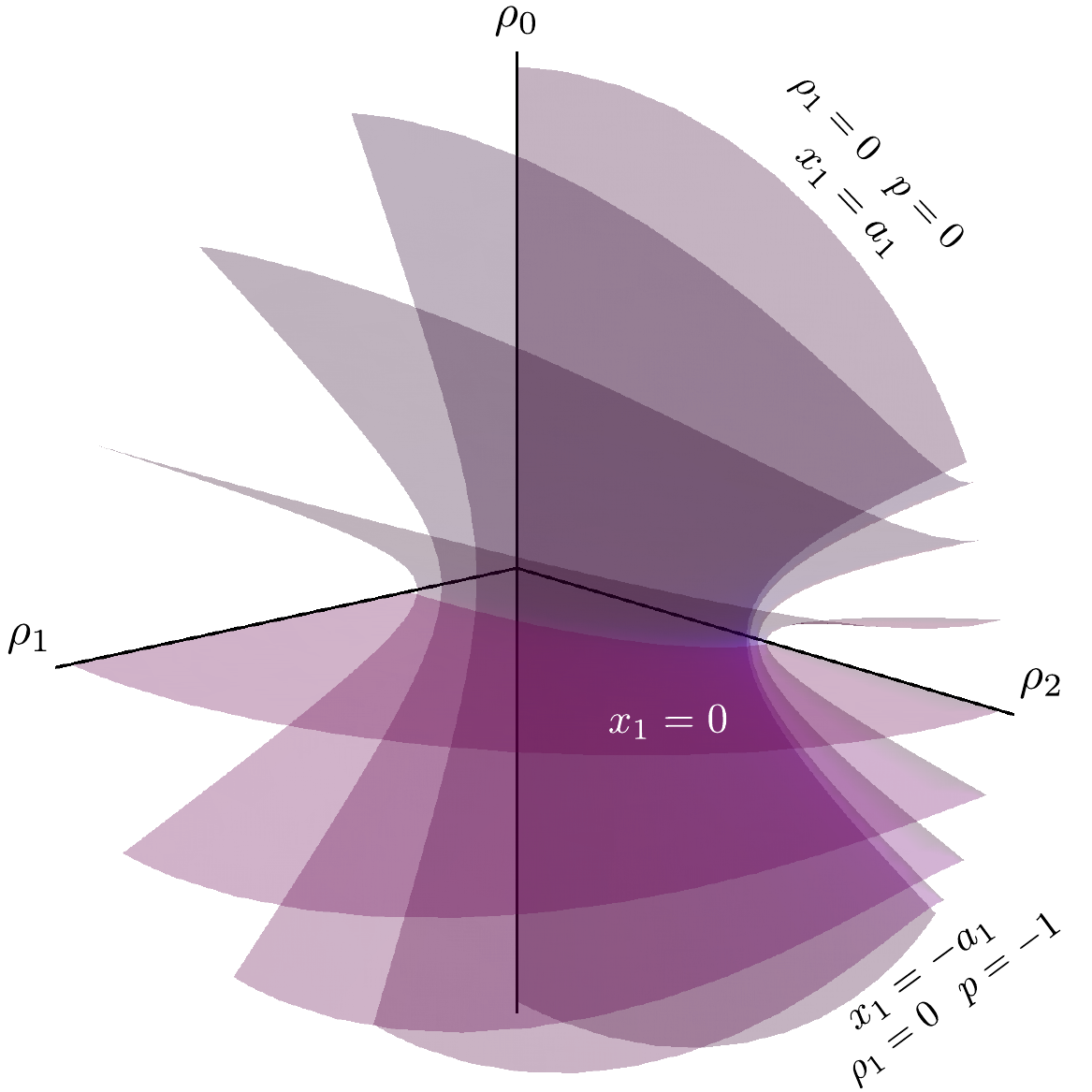}
	\includegraphics[width=\columnwidth/\real{1.205}]{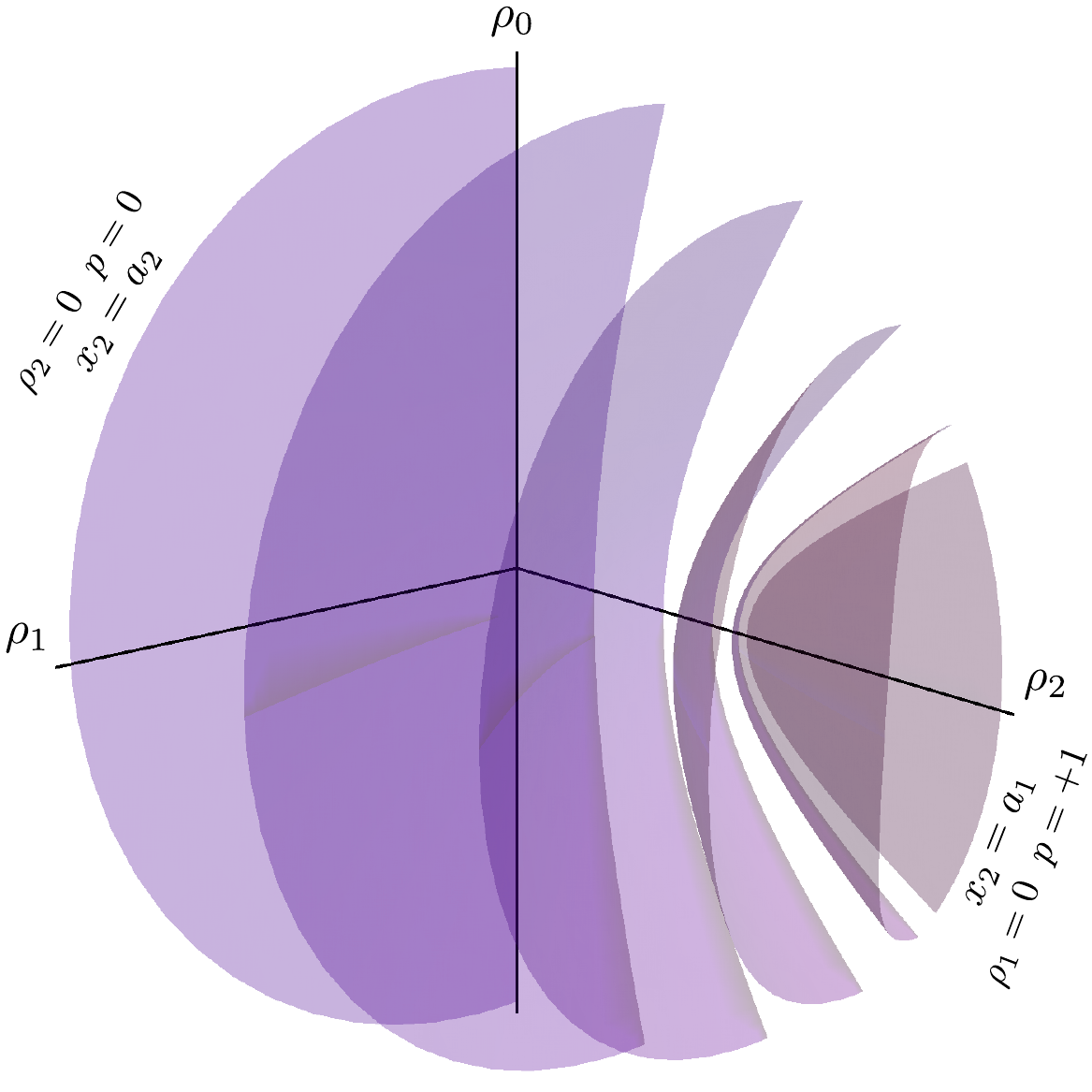}
	\includegraphics[width=\columnwidth/\real{1.205}]{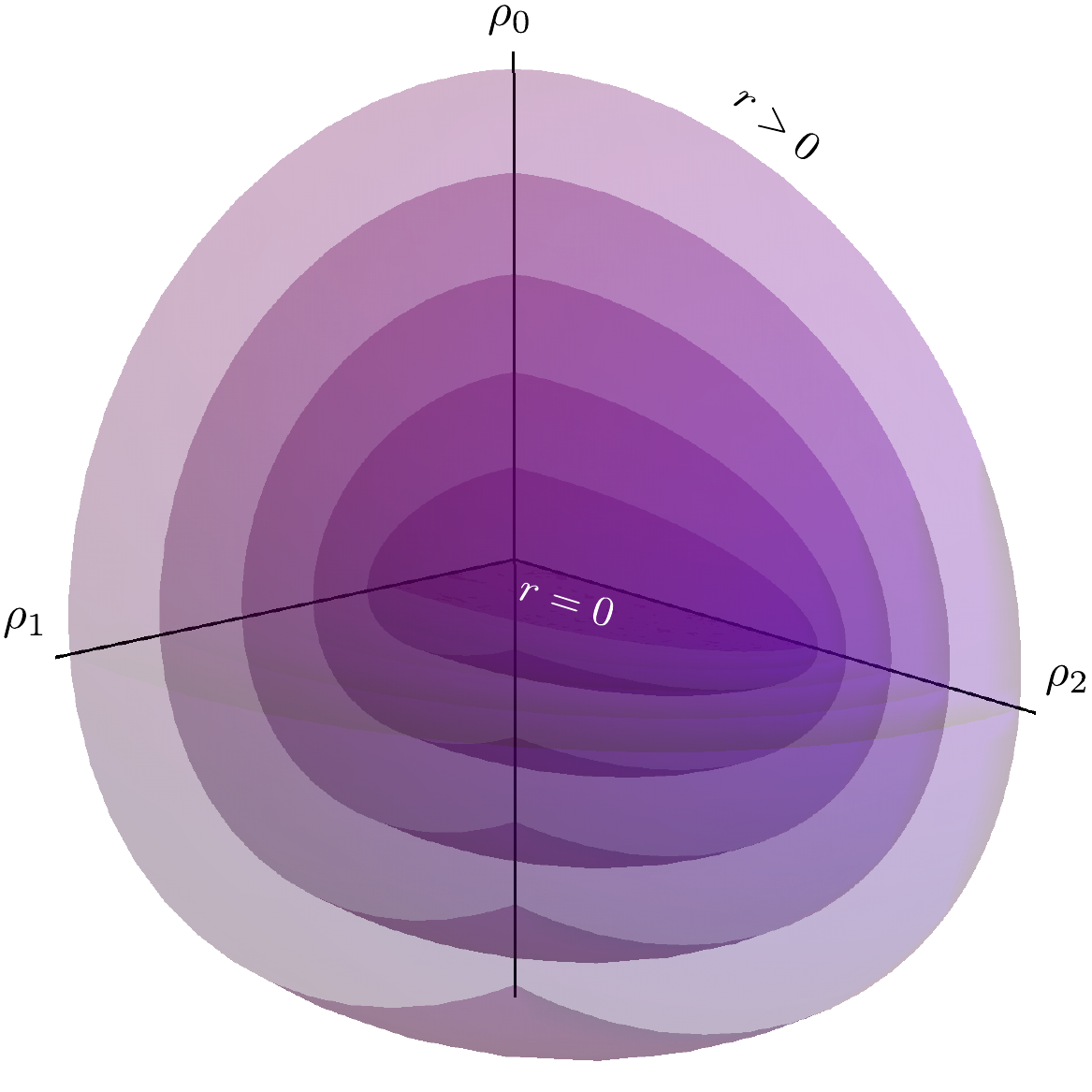}
    \caption{Coordinates ${x_1}$, ${x_2}$, ${x_3=\imag r}$ (top, center, bottom) with respect to $\rho_0$, $\rho_1$, $\rho_2$ for ${D=6}$, ${\lambda=0}$. Coordinate values ${x_1=\pm a_1}$, ${x_2=a_1}$ correspond to the regular axis at ${\rho_1=0}$ and values ${x_2= a_2}$ to the regular axis at ${\rho_2=0}$.}
    \label{fig:coorflat6}
\end{figure}

Alternatively, we can express $\rho_N$ in terms of the coordinates $\rho_0$, $\rho_{\bar\mu}$, so we obtain the metric
\begin{equation}\label{eq:maxsymsp}
\begin{split}
\tb{g} &=\tb{\dd}\rho_0^{\bs{2}} +\sum_{\bar\mu}\big(\tb{\dd}\rho_{\bar\mu}^{\bs{2}} +\rho_{\bar\mu}^2\tb{\dd}\varphi_{\bar\mu}^{\bs{2}}\big) 
\\
&\feq+\frac{\lambda}{1{-}\lambda \sum\limits_{\mu=0}^{N-1}\rho_{\mu}^2} \Big(\sum_{\mu=0}^{N-1}\rho_{\mu}\tb{\dd}\rho_{\mu}\Big)^{\!\!\bs{2}} {-}\Big(1{-}\lambda \sum_{\mu=0}^{N-1}\rho_{\mu}^2\Big)\tb{\dd}\varphi_N^{\bs{2}}\;.
\end{split}
\end{equation}
The obvious advantage of this form is that depending on the sign of $\lambda$ it represents the maximally symmetric spacetimes of positive, negative, as well as zero scalar curvature. The coordinates $\rho_0$, $\rho_{\bar\mu}$, $\phi_{\bar\mu}$, $\phi_N$ are generalizations of cylindrical coordinates in the four-dimensional Minkowski spacetime (${\lambda=0}$) to higher dimensions with an arbitrary cosmological constant ${\lambda}$. We call them the \textit{multi-cylindrical coordinates}. If we suppress the Killing directions, we can plot the relation between $\rho$'s and $x$'s for ${\lambda=0}$ as it is shown in Fig.~\ref{fig:coorflat4} and Fig.~\ref{fig:coorflat6} for ${D=4}$ and ${D=6}$, respectively. The choice of parameters \eqref{eq:xgcira} and \eqref{eq:zeronutsmass} guarantees that ${\rho_{\bar\mu}=0}$ correspond to regular axes with the axial (and cyclic) Killing vectors $\tb{\pp}_{\varphi_{\bar\mu}}$. 

In the canonical coordinates $x_{\bar\mu}$, the symmetry axes ${\rho_{\bar\mu}=0}$ are described by the endpoints $a_{\bar\mu}$ of the ranges of $x_{\bar\mu}$, cf. \eqref{eq:multicyl}. However, the particular expressions are rather involved, since different endpoints correspond to different parts of the symmetry axes. Specifically, the symmetry axes decompose into a one, two, or three parts depending on the index $\bar\mu$ of the axis (and the dimension $D$). The particular expressions of the symmetry axes in canonical coordinates are shown in Tab.~\ref{tab:axes}. For example, in the six-dimensional case depicted in Fig.~\ref{fig:coorflat6}, the axis ${\rho_1=0}$ consists of three parts ${x_1=-a_1}$, ${x_1=a_1}$, and ${x_2=a_1}$, while the axis ${\rho_2=0}$ comprises a single part ${x_2=a_2}$. In the table, we use a notation that reflects this decomposition. Apart from the index $\bar\mu$ denoting individual axes, we also introduce another index ${p=-1,0,1}$ which labels different parts of the axes. Please note that not all pairs of indices $(\bar\mu,p)$ are realized.

\begin{table}\centering
\begin{tabular}{c||c|c}
${D=4}$ ${\bar{N}=1}$ &${p=-1}$ &${p=0}$\\\hline
       ${\rho_1=0}$  & ${x_1=-a_1}$ & ${x_1=a_1}$
\end{tabular}
\begin{tabular}{c||c|c|c}
${D\geq 6}$ ${\bar{N}\geq 2}$ &${p=-1}$ &${p=0}$ &${p=+1}$\\\hline
       ${\rho_1=0}$  & ${x_1=-a_1}$ & ${x_1=a_1}$ & ${x_2=a_1}$ \\
       ${\rho_2=0}$  &  & ${x_2=a_2}$ & ${x_3=a_2}$ \\
       	\;\;$\vdots$ & &$\vdots$ &$\vdots$\\
       ${\rho_{\bar{N}{-}1}=0}$  &  & ${x_{\bar{N}{-}1}=a_{\bar{N}{-}1}}$ & ${x_{\bar{N}}=a_{\bar{N}{-}1}}$ \\
       ${\rho_{\bar{N}}=0}$  &  & ${x_{\bar{N}}=a_{\bar{N}}}$ & 
\end{tabular}
    \caption{Symmetry axes ${\rho_{\bar\mu}=0}$ expressed in canonical coordinates $x_{\bar\mu}$. Different parts of symmetry axes are distinguished by different values of $p$.}
    \label{tab:axes}
\end{table}

By turning on the NUT charges $b_{\bar\mu}$, the ranges of coordinates $x_{\bar\mu}$ change in accordance with \eqref{eq:coorranges}, \eqref{eq:xrootsineq1}, see also Fig.~\ref{fig:rangex1}. Therefore, the endpoints do not coincide with the roots $a_{\bar\mu}$ of a single polynomial ${\lambda\mathcal{J}(x_{\bar\mu}^2)}$, but they are given by roots ${}^{\pm}x_{\bar\mu}$ of different polynomials $X_{\bar\mu}$ instead. As mentioned above, the coordinate $r$ runs over all real values. Since $\Delta$ switches signs in the Lorentzian part only, its roots correspond to the horizons and do not restrict the coordinate $r$.

Switching to general values of parameters $\cir{x}_\mu$ and $\cir{c}_\mu$ is connected with introducing axial singularities as we discuss in greater generality in the next subsections. 

For a general Kerr--NUT--(A)dS spacetimes, we may introduce coordinates $\tilde{x}_{\bar\mu}$ that are given by rescaling the coordinates in the latitudinal directions $\cos{\vartheta_{\bar\mu}}$ (introduced in \eqref{eq:thetatrans}),
\begin{equation}\label{eq:thetatranssym}
\begin{split}
\cos{\vartheta_{\bar\mu}} &=\frac{2}{a_{\bar\mu}-a_{\bar\mu-1}}\bigg(\tilde{x}_{\bar\mu}-\frac{a_{\bar\mu}+a_{\bar\mu-1}}{2}\bigg)\;,
\\
\tilde{x}_{\bar\mu} &=\frac{a_{\bar\mu}-a_{\bar\mu-1}}{2}\cos{\vartheta_{\bar\mu}}+\frac{a_{\bar\mu}+a_{\bar\mu-1}}{2}\;,
\end{split}
\end{equation}
where for convenience, we set the parameter ${a_0=-a_1}$. Combining \eqref{eq:thetatrans} and \eqref{eq:thetatranssym} we find the relations
\begin{equation}\label{eq:xxprimetrans}
\begin{split}
x_{\bar\mu} &= \frac{{}^+x_{\bar\mu}-{}^-x_{\bar\mu}} {a_{\bar\mu}-a_{\bar\mu-1}}\bigg(\tilde{x}_{\bar\mu}-\frac{a_{\bar\mu}+a_{\bar\mu-1}}{2}\bigg)+\frac{{}^+x_{\bar\mu}+{}^-x_{\bar\mu}}{2}\;,
\\
\tilde{x}_{\bar\mu} &= \frac{a_{\bar\mu}-a_{\bar\mu-1}}{{}^+x_{\bar\mu}-{}^-x_{\bar\mu}}\bigg(x_{\bar\mu}-\frac{{}^+x_{\bar\mu}+{}^-x_{\bar\mu}}{2}\bigg)+\frac{a_{\bar\mu}+a_{\bar\mu-1}}{2}\;.
\end{split} 	
\end{equation}
This transformation maps the intervals ${x_{\bar\mu}\in ({}^-x_{\bar\mu}, {}^+x_{\bar\mu})}$ to ${\tilde{x}_{\bar\mu}\in (a_{\bar\mu-1}, a_{\bar\mu})}$. This change effectively shifts the dependence on the NUT charges from the ranges of the latitudinal coordinates exclusively to the metric coefficients, which are, however, more complicated. Such ranges are unchanged in the weak-field limit similar to the four-dimensional Kerr spacetime. Motivated by the maximally symmetric spacetimes, we introduce the coordinates $\tilde{\rho}_0$, $\tilde{\rho}_{\bar\mu}$ of the multi-cylindrical type by the same relations as \eqref{eq:multicyl},
\begin{equation}\label{eq:rhoprime}
   \tilde{\rho}_{0}^2=\frac{\tilde{A}^N}{\lambda \mathcal{A}^N}\;,
	\quad
	\tilde{\rho}_{\bar\mu}^2=\frac{\tilde{J}(a_{\bar\mu}^2)}{-\lambda a_{\bar\mu}^2 \mathcal{U}_{\bar\mu}}\;,
\end{equation}
but involving the coordinates $\tilde{x}_{\bar\mu}$ and ${\tilde{x}_N=x_N}$. We are interested in the singular points ${\tilde{\rho}_{\bar\mu}=0}$ since they reduce to the regular symmetry axes ${\rho_{\bar\mu}=0}$ in the weak-field limit \eqref{eq:xgcira}, \eqref{eq:zeronutsmass}. By means of \eqref{eq:xxprimetrans}, \eqref{eq:rhoprime}, and inspired by Tab.~\ref{tab:axes}, we can express the singular points ${\tilde{\rho}_{\bar\mu}=0}$ in terms of the canonical coordinates, see Tab.~\ref{tab:axesprimed}.

\begin{table}
    \begin{tabular}{c||c|c}
${D=4}$ ${\bar{N}=1}$ &${p=-1}$ &${p=0}$ \\\hline
       ${\tilde{\rho}_1=0}$  & ${x_1={}^{-\!}x_{1}}$ & ${x_1={}^{+\!}x_{1}}$\\
    \end{tabular}
    \begin{tabular}{c||c|c|c}
${D\geq 6}$ ${\bar{N}\geq 2}$ &${p=-1}$ &${p=0}$ &${p=+1}$\\\hline
       ${\tilde{\rho}_1=0}$  & ${x_1={}^{-\!}x_{1}}$ & ${x_1={}^{+\!}x_{1}}$ & ${x_2={}^{-\!}x_{2}}$ \\
       ${\tilde{\rho}_2=0}$  &  & ${x_2={}^{+\!}x_{2}}$ & ${x_3={}^{-\!}x_{3}}$ \\
       	\;\;$\vdots$ & &$\vdots$ &$\vdots$\\
       ${\tilde{\rho}_{\bar{N}{-}1}=0}$  &  & ${x_{\bar{N}{-}1}={}^{+\!}x_{\bar{N}{-}1}}$ & ${x_{\bar{N}}={}^{-\!}x_{\bar{N}}}$ \\
       ${\tilde{\rho}_{\bar{N}}=0}$  &  & ${x_{\bar{N}}={}^{+\!}x_{\bar{N}}}$ & 
    \end{tabular}
        \caption{The sets of singular points ${\tilde{\rho}_{\bar\mu}=0}$ expressed in canonical coordinates~$x_{\bar\mu}$.}
    \label{tab:axesprimed}
\end{table}

Let us now introduce two functions ${\bar\eta=\bar\eta(\bar\mu,p)}$ and ${\sigma=\sigma(p)}$ which relabel the endpoints ${}^{\pm\!}x_{\bar\mu}$ in a way that is explicitly shown in Tab.~\ref{tab:opt}, or equivalently,
\begin{equation}\label{eq:sigmaeta}
\begin{split}
\bar\eta &=
\left\{\begin{array}{lll}
1\;,  &p=-1\;, &\bar\mu=1\;,
\\
\bar\mu\;,  &p=0\;, &1\leq\bar\mu\leq\bar{N}\;,
\\
\bar\mu+1\;, &p=+1\;, &1\leq\bar\mu\leq\bar{N}{-}1\;,
\end{array}\right.
\\
\sigma &={(-1)}^{p}\;.
\end{split}
\end{equation}
These functions define a matching between pairs of indices ${(\bar\mu,p)\Longleftrightarrow (\bar\eta,\sigma)}$ so that the singular points ${\tilde{\rho}_{\bar\mu}=0}$ with a given $p$ are described by ${x_{\bar\eta}={}^{\sigma\!}x_{\bar\eta}}$ in the canonical coordinates.

\begin{table}
\begin{tabular}{c||c|c}
${D=4}$ ${\bar{N}=1}$ &${p=-1}$ &${p=0}$\\\hline
	${\bar\mu=1}$ &${}^{-\!}x_{1}$ &${}^{+\!}x_{1}$ \\
\end{tabular}
\begin{tabular}{c||c|c|c}
${D\geq 6}$ ${\bar{N}\geq 2}$ &${p=-1}$ &${p=0}$ &${p=+1}$\\\hline
	${\bar\mu=1}$ &${}^{-\!}x_{1}$ &${}^{+\!}x_{1}$ &${}^{-\!}x_{2}$ \\
	${\bar\mu=2}$	 & &${}^{+\!}x_{2}$ &${}^{-\!}x_{3}$ \\
	\;\;\;$\vdots$ & &$\vdots$ &$\vdots$\\
	${\bar\mu=\bar{N}{-}1}$	& &${}^{+\!}x_{\bar{N}{-}1}$ &${}^{-\!}x_{\bar{N}}$ \\
	${\bar\mu=\bar{N}}$	& &${}^{+\!}x_{\bar{N}}$ &\\
\end{tabular}
\caption{Introduction of the notation for ${}^{\sigma\!}x_{\bar\eta}$. The indices ${\bar\eta=\bar\eta(\bar\mu,p)}$ and ${\sigma=\sigma(p)}$ are such functions that ${}^{\sigma\!}x_{\bar\eta}$ corresponds to the particular values ${}^{\pm\!}x_{\bar\mu}$ in this table.}
\label{tab:opt}
\end{table}

For simplicity we impose a few conditions on the parameters $\cir{x}_{\mu}$ and $\cir{c}_{\mu}$ which ensure that the parameter values are not significantly different from the weak-field values. In particular, we assume that $\cir{x}_{\mu}$ are well ordered
\begin{equation}\label{eq:xcircond1}
    \cir{x}_1^2<\cir{x}_2^2<\dots<\cir{x}_{\bar N}^2<\cir{x}_N^2\;,
\end{equation}
and satisfy
\begin{equation}\label{eq:xcircond2}
    \cir{x}_{\bar\mu{-}1}^2<{}^{\sigma\!}x_{\bar\eta}^2<\cir{x}_{\bar\mu+1}^2\;,
\end{equation}
where the first inequality is used only if ${\bar\mu>1}$ and the last inequality with ${\bar\mu=\bar N}$ only if ${\cir{x}_N^2>0}$. This condition, for example, implies that the expression ${\cir{J}_{\bar\mu}({}^{\sigma\!}x_{\bar\eta}^2)/\cir{U}_{\bar\mu}}$, which occurs frequently in the next subsection, is always positive. Furthermore, we require
\begin{equation}\label{eq:ccond}
    \cir{c}_{\bar\mu}>0\;,
\end{equation}
which fixes the signs of $\varphi_{\bar\mu}$. This assumption does not affect the cyclicity of $\tb{\pp}_{\varphi_{\bar\mu}}$.

\subsection{Axial Killing vectors}

The singular points ${\tilde{\rho}_{\bar\mu}=0}$ are natural candidates for the symmetry axes not only because they give rise to the regular axes in the weak-field limit, but mainly because they describe all endpoints of $x_{\bar\mu}$ where the Kerr--NUT--(A)dS metrics degenerate. Therefore, it is reasonable to ask whether these singular points describe the singular axes of such spacetimes. For this purpose, we need to find a semi-regular frame at ${\tilde{\rho}_{\bar\mu}=0}$. We can try to introduce a semi-regular frame by imitating the reversed procedure of construction of the spinning cosmic string spacetime.

Our goal is to transform the Kerr--NUT--(A)dS metric to the coordinates resembling the coordinate system of the spacetimes with a partially regularized axis. Since the Kerr--NUT--(A)dS spacetimes are not flat like the spinning cosmic string spacetime, this can only be done locally in a close vicinity of an axis. Also, this applies to only one part of one axis at a time. Thus, for a particular part $p$ of an axis ${\tilde{\rho}_{\bar\mu}=0}$ (described by a pair of indices ${(\bar\mu,p)}$), we cast the metrics \eqref{eq:KerrNUTAdSmetricALT} to the form that resembles the Kerr--NUT--(A)dS spacetimes with the regular part of the axis ${(\bar\mu,p)}$. The difference from the regular case is in different ranges of newly introduced Killing coordinates which capture the axial singularities. The semi-regular frame is then given by Cartesian-type coordinates that are found by a simple conversion from polar-type coordinates involving one of the new Killing coordinates.

The impossibility of introducing the collective coordinates for all $p$'s at the same time suggests that the spacetimes are significantly different at various parts of the axes. This is a well-known characteristic of all four-dimensional NUT-charged spacetimes such as the Taub--NUT spacetime. Therefore, it is not unexpected in higher dimensions and more complicated NUT-charged spacetimes.

Let us fix a pair of indices ${(\bar\mu,p)}$ with the corresponding pair ${(\bar\eta,\sigma)}$, so the singular points ${\tilde{\rho}_{\bar\mu}=0}$ are described by ${x_{\bar\eta}={}^{\sigma\!}x_{\bar\eta}}$ as mentioned above. By employing the transformation of the Killing coordinates (the inverse transformation follows from \eqref{eq:JJ})
\begin{equation}\label{eq:varphiphidotbul}
    \varphi_\nu=\sum_\kappa\frac{\cir{c}_\nu}{\bul{c}_\kappa}\frac{\cir{J}_\nu(\bul{x}_\kappa^2)}{\cir{U}_\nu}\phi_\kappa\;,
    \quad
    \phi_\nu=\sum_\kappa\frac{\bul{c}_\nu}{\cir{c}_\kappa}\frac{\bul{J}_\nu(\cir{x}_\kappa^2)}{\bul{U}_\nu}\varphi_\kappa\;,
\end{equation}
we can bring the metrics \eqref{eq:KerrNUTAdSmetricALT} to the form
\begin{equation}\label{eq:KerrNUTAdSmetricALTBul}
  \tb{g} = \sum_{\nu}\bigg[\frac{U_\nu}{X_\nu}\,\tb{\dd}x_\nu^{\bs{2}}
  +\frac{X_\nu}{U_\nu}\Big(\sum_{\kappa}\frac{J_{\nu}(\bul{x}_{\kappa}^2)}{\bul{c}_{\kappa}}\,\tb{\dd}\phi_{\kappa}\Big)^{\!\bs{2}}\bigg]\;,
\end{equation}
which is given by the same expression as \eqref{eq:KerrNUTAdSmetricALT} with $\varphi_\nu$ replaced by $\phi_\nu$, and ${\cir{c}_\nu}$, ${\cir{x}_\nu}$ replaced by ${\bul{c}_\nu}$, ${\bul{x}_\nu}$ that are defined as follows:
\begin{equation}\label{eq:xcbul}
\begin{gathered}
    \bul{x}_{\bar{\mu}}^2 ={}^{\sigma\!}x_{\bar\eta}^2\;,
    \quad
    \bul{c}_{\bar{\mu}}=\frac12|X'_{\bar\eta}|\big|_{{}^{\sigma\!}x_{\bar\eta}}\;,
    \\
    \bul{x}_{\kappa}^2 =\cir{x}_{\kappa}^2\;, 
    \quad
    \bul{c}_{\kappa} =\cir{c}_{\kappa}\;, 
    \quad
    \kappa\neq\bar{\mu}\;.
\end{gathered}
\end{equation}
The notation $|_{{}^{\sigma\!}x_{\bar\eta}}$ means that the expression is evaluated at ${x_{\bar\eta}={}^{\sigma\!}x_{\bar\eta}}$. The quantities with the solid circles above are defined in the same manner as the ones with the hollow circles, but involving $\bul{x}$'s instead of $\cir{x}$'s. Employing \eqref{eq:varphiphidotbul} and the identity \eqref{eq:JJ}, we find the transformation of the Killing vectors,
\begin{equation}\label{eq:varphiphidotbulKV}
    \tb{\pp}_{\varphi_\nu}=\sum_\kappa\frac{\bul{c}_\kappa}{\cir{c}_\nu}\frac{\bul{J}_\kappa(\cir{x}_\nu^2)}{\bul{U}_\kappa}\tb{\pp}_{\phi_\kappa}\;,
    \quad
    \tb{\pp}_{\phi_\nu}=\sum_\kappa\frac{\cir{c}_\kappa}{\bul{c}_\nu}\frac{\cir{J}_\kappa(\bul{x}_\nu^2)}{\cir{U}_\kappa}\tb{\pp}_{\varphi_\kappa}\;.
\end{equation}
By analogy with \eqref{eq:phipsiKVtrans}, the transformation between $\tb{\pp}_{\phi_\nu}$ and $\tb{\pp}_{\psi_k}$ reads\footnote{It can be also directly checked that \eqref{eq:phipsiKVtransdot} together with \eqref{eq:varphiphidotbulKV} gives back \eqref{eq:phipsiKVtrans}.}
\begin{equation}\label{eq:phipsiKVtransdot}
    \tb{\pp}_{\psi_k} {=}\sum_\nu \bul{c}_{\nu}\frac{\bul{A}_{\nu}^{(k)}}{\bul{U}_\nu}\tb{\pp}_{\phi_\nu}\;,
\;\;\;
    \tb{\pp}_{\phi_\nu} {=}\frac{1}{\bul{c}_{\nu}}\sum_k (-\bul{x}_{\nu}^2)^{N-k-1}\tb{\pp}_{\psi_k}\;.
\end{equation}
The equations \eqref{eq:phipsiKVtrans} and \eqref{eq:phipsiKVtransdot} say that, from the viewpoint of the canonical coordinates $\psi_k$, the quantities ${\cir{c}_\nu}$ and ${\cir{x}_\nu}$ define the cyclic Killing vectors $\tb{\pp}_{\varphi_\nu}$, while ${\bul{c}_\nu}$ and ${\bul{x}_\nu}$ define the axial Killing vectors $\tb{\pp}_{\phi_\nu}$.

Before we proceed any further, let us briefly discuss the motivation behind the definition \eqref{eq:xcbul}. First, it should be noted that if we choose ${\cir{c}_{\bar{\mu}}=\bul{c}_{\bar{\mu}}}$, ${\cir{x}_{\bar{\mu}}=\bul{x}_{\bar{\mu}}}$, i.e.,
\begin{equation}\label{eq:cirxcchoice3}
    \cir{x}_{\bar{\mu}}^2={}^{\sigma\!}x_{\bar\eta}^2\;,
    \quad
    \cir{c}_{\bar{\mu}}=\frac12|X'_{\bar\eta}|\big|_{{}^{\sigma\!}x_{\bar\eta}}\;,
\end{equation}
the metrics \eqref{eq:KerrNUTAdSmetricALT} describe spacetimes with the regular part of the axis.\footnote{This could be seen by transforming the metrics directly to the Cartesian-type coordinates near the axis. We skip this step here, because it is technically the same as the derivation in a general case presented below. It leads to the metrics \eqref{eq:gaxis} where $\phi_\nu$ are replaced by $\varphi_\nu$. Since ${\xi,\zeta\in\mathbb{R}}$, cf. \eqref{eq:xizeta}, these metrics are smooth at origin ${\xi=\zeta=0}$.} The transformation \eqref{eq:varphiphidotbul} with \eqref{eq:xcbul} and general parameters $\cir{c}_{\bar{\mu}}$, $\cir{x}_{\bar{\mu}}$ is, thus, chosen so that the resulting metrics \eqref{eq:KerrNUTAdSmetricALTBul} with \eqref{eq:xcbul} have the same form as the metrics with the regular part of the axis \eqref{eq:KerrNUTAdSmetricALT} with \eqref{eq:cirxcchoice3}. These two cases only differ by the ranges of the Killing coordinates. While $\varphi_{\bar\nu}$ are $2\pi$-periodic coordinates and their tangent vectors are cyclic Killing vectors, none of the coordinates $\phi_{\nu}$ introduced in \eqref{eq:varphiphidotbul} have such properties. As we show below, the choice \eqref{eq:xcbul} also guaranties that the Killing vector $\tb{\pp}_{\phi_{\bar\mu}}$ is axial.

In Sec.~\ref{sc:KNA}, we mentioned that the latitudinal coordinate $x_{\bar\eta}$ is restricted by the two endpoints ${}^{\pm\!}x_{\bar\eta}$, which are the roots of $X_{\bar\eta}$. The polynomial $X_{\bar\eta}$ can be expanded in the neighborhood of one of the roots ${}^{\sigma\!}x_{\bar\eta}$, ${|x_{\bar\eta}-{}^{\sigma\!}x_{\bar\eta}|\ll 1}$,
\begin{equation}\label{eq:Xexpansion}
	X_{\bar\eta}=X'_{\bar\eta}\big|_{{}^{\sigma\!}x_{\bar\eta}}(x_{\bar\eta}{-}{}^{\sigma\!}x_{\bar\eta})+\mathcal{O}\big({(x_{\bar\eta}{-}{}^{\sigma\!}x_{\bar\eta})}^2\big)\;.
\end{equation}
Consider the polar--Cartesian-type transformation of the two coordinates $x_{\bar\eta}$ and $\phi_{\bar\mu}$ with the remaining coordinates $x_\nu$, $\phi_\kappa$, ${\nu\neq\bar\eta}$, ${\kappa\neq\bar\mu}$ unchanged,
\begin{equation}\label{eq:xizeta}
\begin{gathered}
\xi+\imag\zeta= \sqrt{|{}^{\sigma\!}x_{\bar\eta}-x_{\bar\eta}|}\exp(\imag\phi_{\bar\mu})\;,
	\\
	|{}^{\sigma\!}x_{\bar\eta}-x_{\bar\eta}|=\xi^2+\zeta^2\;,
	\quad
	\tan{\phi_{\bar\mu}}=\zeta/\xi\;.
\end{gathered}
\end{equation}
The gradients of the coordinates read
\begin{equation}\label{eq:xzdif}
    \begin{aligned}
    \tb{\dd}x_{\bar\eta}&=-2\sigma\big(\xi\tb{\dd}\xi+\zeta\tb{\dd}\zeta\big)\;,
    \\
    \tb{\dd}\phi_{\bar\mu}&=\frac{\xi\tb{\dd}\zeta-\zeta\tb{\dd}\xi}{\xi^2+\zeta^2}\;.
    \end{aligned}
\end{equation}

Applying the transformation \eqref{eq:xizeta} and expanding the metrics \eqref{eq:KerrNUTAdSmetricALTBul} around ${\xi=\zeta=0}$ (${x_{\bar\eta}={}^{\sigma\!}x_{\bar\eta}}$), we find the following expression for the Kerr--NUT--(A)dS metrics in the coordinates  ${(\xi,\zeta,x_\nu,\phi_\kappa)}$, ${\nu\neq\bar\eta}$, ${\kappa\neq\bar\mu}$,
\begin{equation}\label{eq:gaxis}
\begin{split}
\tb{g} &=
4\bigg|\frac{U_{\bar\eta}}{X'_{\bar\eta}}\bigg|\Bigg|_{{}^{\sigma\!}x_{\bar\eta}}
\!\!\big(\tb{\dd}{\xi}^{\bs{2}}+\tb{\dd}{\zeta}^{\bs{2}}\big)
\\
&\feq
{+}\sum_{\substack{\nu \\ \nu\neq{\bar\eta}}}\bigg[\frac{U_\nu\big|_{{}^{\sigma\!}x_{\bar\eta}}}{X_\nu}\,\tb{\dd}x_\nu^{\bs{2}}
  {+}\frac{X_\nu}{U_\nu\big|_{{}^{\sigma\!}x_{\bar\eta}}}\bigg(\!\sum_{\substack{\kappa \\ \kappa\neq\bar{\mu}}}\frac{J_{\nu}(\cir{x}_{\kappa}^2)\big|_{{}^{\sigma\!}x_{\bar\eta}}}{\cir{c}_{\kappa}}\,\tb{\dd}\phi_{\kappa}\!\bigg)^{\!\!\bs{2}}\bigg]
\\
&\feq
{-}\sum_{\substack{\kappa,\nu \\ \kappa\neq\bar\mu}}\frac{2\sigma\big[X_\nu J_\nu({}^{\sigma\!}x_{\bar\eta}^2)\big]_{,\bar\eta}J_\nu(\cir{x}_{\kappa}^2)}{\cir{c}_\kappa |X'_{\bar\eta}|U_\nu}\Bigg|_{{}^{\sigma\!}x_{\bar\eta}}\!\!\big(\xi\tb{\dd}{\zeta}-\zeta\tb{\dd}{\xi}\big)\vee\tb{\dd}{\phi_\kappa}
\\
&\feq{+}\mathcal{O}(\xi,\zeta;2)\;.
\end{split}
\end{equation}
Here, we used \eqref{eq:xzdif} and the relations
\begin{equation}\label{eq:metricfuncexp}
    \begin{aligned}
    X_{\bar\eta}&=-\sigma X'_{\bar\eta}\big|_{{}^{\sigma\!}x_{\bar\eta}}(\xi^2+\zeta^2)+\mathcal{O}(\xi,\zeta;4)\;,
    \\
        U_{\bar\eta}&= U_{\bar\eta}\big|_{{}^{\sigma\!}x_{\bar\eta}}+\mathcal{O}(\xi,\zeta;2)\;,
    \\
    U_\nu&= U_{\nu}\big|_{{}^{\sigma\!}x_{\bar\eta}}+\mathcal{O}(\xi,\zeta;2)\;,
    \\
    J_\nu({}^{\sigma\!}x_{\bar\eta}^2)&=-2\sigma\,{}^{\sigma\!}x_{\bar\eta}\,J_{\nu\bar\eta}({}^{\sigma\!}x_{\bar\eta}^2)\,(\xi^2+\zeta^2)+\mathcal{O}(\xi,\zeta;4)\;,
    \\
    J_\nu(\cir{x}_\kappa^2)&=J_\nu(\cir{x}_\kappa^2)\big|_{{}^{\sigma\!}x_{\bar\eta}}+\mathcal{O}(\xi,\zeta;2)\;,
    \end{aligned}
\end{equation}
which hold true for ${\nu\neq\bar\eta}$, ${\kappa\neq\bar\mu}$. The symbol ${\mathcal{O}(\xi,\zeta;n)}$ stands for the terms whose sum of the powers of $\xi$ and $\zeta$ is $n$ or higher. The third line of \eqref{eq:gaxis} is of the first order, ${\mathcal{O}(\xi,\zeta;1)}$, and vanishes towards ${\xi=\zeta=0}$ as well as the fourth line containing the second and higher order terms ${\mathcal{O}(\xi,\zeta;2)}$.

Let us focus on the part close to ${\xi=\zeta=0}$ and neglect the distortion caused by the terms ${\mathcal{O}(\xi,\zeta;1)}$. The sections of constant coordinates $x_\nu$ and $\phi_\kappa$, ${\nu\neq\bar\eta}$, ${\kappa\neq\bar\mu}$ are flat and the corresponding 2-metrics are proportional to the standard Cartesian form ${\tb{\dd}{\xi}^{\bs{2}}+\tb{\dd}{\zeta}^{\bs{2}}}$ in the first line of \eqref{eq:gaxis}. This suggests that the basis $\tb{\pp}_{\xi}$, $\tb{\pp}_{\zeta}$, $\tb{\pp}_{x_\nu}$, $\tb{\pp}_{\phi_\kappa}$, ${\nu\neq\bar\eta}$, ${\kappa\neq\bar\mu}$ may be regarded as the semi-regular frame. This frame becomes regular when we take the limit ${\cir{x}_{\bar\mu}\to\bul{x}_{\bar\mu}}$, ${\cir{c}_{\bar\mu}\to\bul{c}_{\bar\mu}}$, leading to \eqref{eq:cirxcchoice3}. This is because the transformation \eqref{eq:varphiphidotbul} is redundant in this case and all coordinates $\varphi_\kappa$ and $\phi_\kappa$ coincide. Therefore, $\phi_{\bar\mu}$ is a $2\pi$-periodic coordinate and $\tb{\pp}_{\phi_{\bar\mu}}$ is a cyclic Killing vector in this limit. As a result, the transformation \eqref{eq:xizeta} allows for a smooth inclusion of the regular points ${\xi=\zeta=0}$, so ${\xi,\zeta\in\mathbb{R}}$. This is impossible in a general case when the ranges of coordinates $\xi$, $\zeta$ are much more complicated and $\mathcal{M}$ cannot be smoothed out.

With our definition of the semi-regular frame, the axial Killing vector responsible for the symmetry axis ${x_{\bar\eta}={}^{\sigma\!}x_{\bar\eta}}$ (${\xi=\zeta=0}$) is indeed ${\tb{\pp}_{\phi_{\bar\mu}}=\xi\tb{\pp}_{\zeta}-\zeta\tb{\pp}_{\xi}}$, as its components vanish towards the axis. In fact, the axial Killing vector $\tb{\pp}_{\phi_{\bar\mu}}$ is associated with a pair of indices ${(\bar\mu,p)}$ and not just $\bar\mu$. Therefore, it generally does not admit fixed points for different values of $p$. In our terminology, adopted from the maximally symmetric spacetimes, the axes are composed of several parts formed by sets of fixed points of isometries corresponding to different axial Killing vectors. To stress the dependency on $p$, we often denote various axial Killing vectors by $\tb{\pp}_{\phi_{\bar\mu}^p}$. From the second line of \eqref{eq:gaxis}, we can read off the induced metric on the respective part of the axis (or in its close neighborhood if the axis is singular). Interestingly, it has a form of a ${(2N{-}2)}$-dimensional off-shell Kerr--NUT--(A)dS spacetimes \eqref{eq:KerrNUTAdSmetricALT} with functions $X_{\nu}$ replaced by  ${X_{\nu}/({}^{\sigma\!}x_{\bar\eta}^2-x_{\nu}^2)}$. These rational functions become polynomials in $x_{\nu}$ only if ${b_{\nu}=b_{\bar\eta}=0}$.

Employing \eqref{eq:varphiphidotbulKV}, we can find a useful expression for the set of axial Killing vectors $\tb{\pp}_{\phi_{\bar\mu}^p}$ in terms of the original coordinates $\varphi_\kappa$, cf. \eqref{eq:KerrNUTAdSmetricALT},
\begin{equation}\label{eq:axialKVKerrNUTAdS}
	\tb{\pp}_{\phi_{\bar\mu}^p}=\frac{2\cir{c}_{\bar\mu}}{|X'_{\bar\eta}|\big|_{{}^{\sigma\!}x_{\bar\eta}}\!\!}\frac{\cir{J}_{\bar\mu}({}^{\sigma\!}x_{\bar\eta}^2)}{\cir{U}_{\bar\mu}}\Bigg[\tb{\pp}_{\varphi_{\bar\mu}}+\sum_{\substack{\kappa \\ \kappa\neq\bar\mu}}\frac{\cir{c}_\kappa}{\cir{c}_{\bar\mu}}\frac{{}^{\sigma\!}x_{\bar\eta}^2-\cir{x}_{\bar\mu}^2}{{}^{\sigma\!}x_{\bar\eta}^2-\cir{x}_\kappa^2}\frac{\cir{U}_{\bar\mu}}{\cir{U}_\kappa}\tb{\pp}_{\varphi_\kappa}\Bigg]\;.
\end{equation}
Obviously these ${2N{-}2}$ Killing vectors are not independent, but a complete set of $N$ independent Killing vectors can be always constructed by selecting one vector $\tb{\pp}_{\phi_{\bar\mu}^p}$ for each index $\bar\mu$ and supplementing them with a temporal Killing vector, for example, one of $\tb{\pp}_{\varphi_N^q}$ defined in the next subsection. The factor in front of the square bracket is always positive, thanks to the supplementary conditions \eqref{eq:xcircond1}, \eqref{eq:xcircond2}, and \eqref{eq:ccond}.

From the expression \eqref{eq:axialKVKerrNUTAdS}, we see that the axial Killing vector $\tb{\pp}_{\phi_{\bar\mu}^p}$ becomes aligned with the cyclic Killing vector $\tb{\pp}_{\varphi_{\bar\mu}}$ if we choose
\begin{equation}\label{eq:cirxcchoice1}
    \cir{x}_{\bar{\mu}}^2 ={}^{\sigma\!}x_{\bar\eta}^2\;,
\end{equation}
and the prefactor in \eqref{eq:axialKVKerrNUTAdS} cancels out for
\begin{equation}\label{eq:cirxcchoice2}
    \cir{c}_{\bar{\mu}}=\frac12|X'_{\bar\eta}|\big|_{{}^{\sigma\!}x_{\bar\eta}}\frac{\cir{J}_{\bar\mu}({}^{\sigma\!}x_{\bar\eta}^2)}{\cir{U}_{\bar\mu}}\;.
\end{equation}
Together with \eqref{eq:cirxcchoice1}, this leads to the condition \eqref{eq:cirxcchoice3} under which ${\tb{\pp}_{\phi_{\bar\mu}^p}=\tb{\pp}_{\varphi_{\bar\mu}}}$. Due to the particular dependence of $\bar\eta$ and $\sigma$ on $p$ in \eqref{eq:cirxcchoice3}, it is impossible to set the parameters so that this condition holds true simultaneously for multiple parts of the axis for the nonzero NUT charge $b_{\bar\eta}$. 

By switching off the NUT charges, the dependence on $p$ effectively disappears from the formula \eqref{eq:axialKVKerrNUTAdS}, because ${{}^{\sigma\!}x_{\bar\eta}^2= a_{\bar\mu}^2}$. Therefore the vectors $\tb{\pp}_{\phi_{\bar\mu}^p}$ with different values of $p$ coincide. Altogether, if the mass as well as all NUT charges vanish, ${b_{\nu}=b_{\bar\eta}=0}$ (cf. \eqref{eq:zeronutsmass}), then the condition \eqref{eq:cirxcchoice3} reduces to \eqref{eq:xgcira} and the resulting geometries are the maximally symmetric spacetimes.

Thus far we have been dealing with the fixed points given by the roots of the polynomials $X_{\bar\mu}$, but there might exist other fixed points as well. First, we restrict ourselves to the regions where ${X_\mu}\neq0$ for all $\mu$. Since the metric \eqref{eq:KerrNUTAdSmetricALT} is regular in these regions, we may use, for example, the frame $\tb{\pp}_{x_\mu}$, $\tb{\pp}_{\varphi_\mu}$ to study fixed points of isometries. Considering that a general Killing vector compatible with stationary rotational symmetry is a linear combination of $\tb{\pp}_{\varphi_\mu}$ with constant coefficients, there exists no other Killing vector generating isometry with fixed points in these regions. Lastly, we should also mention the roots of the polynomial ${X_N}$. As these surfaces correspond to the horizons, we expect that the fixed points on them (for ${X_{\bar\mu}\neq0}$) are of a different nature and do not form the symmetry axes.


\subsection{Horizon-adjusted coordinates}\label{ssc:HAC}

Depending on the parameters $a_{\bar\mu}$, $\lambda$, and $m$, the polynomial $X_N$ can have various number of roots, which determine the positions of the horizons. Unlike the latitudinal coordinates ${x_{\bar\mu}}$, the radial-type coordinate ${x_N=\imag r}$ is not restricted by these roots. We denote these roots by ${}^{q\!}x_N$, where index $q$ labels particular horizons, for instance, ${}^{\pm}r$ and ${}^{\pm}r^{(\lambda)}$.

Let us consider a coordinate transformation similar to \eqref{eq:varphiphidotbul}, where $\bul{x}_{\nu}$, $\bul{c}_{\nu}$ are defined as:
\begin{equation}
\begin{gathered}
    \bul{x}_{\bar\mu}^2 =\cir{x}_{\bar\mu}^2\;, 
    \quad
    \bul{c}_{\bar\mu} =\cir{c}_{\bar\mu}\;,
    \\
    \bul{x}_N^2 =\st{x}_N^2\;,
    \quad
    \bul{c}_N=\st{c}_N\;.
\end{gathered}
\end{equation}
Such a transformation effectively sets the parameters $\cir{x}_N$, $\cir{c}_N$ to new arbitrary values $\st{x}_N$, $\st{c}_N$ and keep the remaining parameters $\cir{x}_{\bar\mu}$, $\cir{c}_{\bar\mu}$ unchanged (cf. \eqref{eq:KerrNUTAdSmetricALTBul}) by introducing new coordinates ${\phi_\nu}$. Applying this transformation, we find that the Killing vectors ${\tb{\pp}_{\phi_{\bar\mu}}}$ are equal to the cyclic Killing vectors, ${\tb{\pp}_{\phi_{\bar\mu}}=\tb{\pp}_{\varphi_{\bar\mu}}}$, while $\tb{\pp}_{\phi_N}$ is given by a linear combination of $\tb{\pp}_{\varphi_{\nu}}$, cf. \eqref{eq:varphiphidotbulKV},
\begin{equation}\label{eq:phiNxstar}
     \tb{\pp}_{\phi_N}=\sum_\nu\frac{\cir{c}_{\nu}}{\st{c}_N}\frac{\cir{J}_\nu(\st{x}_N^2)}{\cir{U}_\nu}\tb{\pp}_{\varphi_\nu}\;.
\end{equation}
This transformation is in accordance with the ambiguity in the choice of the temporal direction \eqref{eq:phiNambi}, where the coordinate $\phi_N$ is the new temporal coordinate ${\tilde\varphi}_N$. This means that both parameters $\cir{x}_N$ and $\cir{c}_N$ are actually redundant and we can choose them arbitrarily. They are not connected with any geometrical quantities, but correspond to a change of coordinates instead.

We can employ this freedom to find the coordinates adjusted to a particular horizon $q$. Consider the choice \begin{equation}\label{eq:paramxcgauge}
    \cir{x}_N^2={}^{q\!}x_N^2\;,
    \quad
    \cir{c}_N=\cir{J}_N({}^{q\!}x_N^2)\;,
\end{equation}
and label the coordinates $\varphi_\nu$ corresponding to a horizon $q$ by $\varphi_\nu^q$. Following the previous subsection modified to suit the Lorentzian part of the metric, one could also verify that the vectors tangent to the coordinates $\tb{\pp}_{\varphi_\nu^q}$ are the Killing vectors corresponding to the isometries with fixed points. The Killing vectors $\tb{\pp}_{\varphi_N^q}$ generate the (Killing) horizons ${x_N={}^{q\!}x_N}$, because they become null as we approach the horizons. The fixed points, however, form only the \textit{bifurcation surfaces} in the whole structure that is often referred to as the \textit{bifurcate Killing horizons}. The bifurcation surfaces are given by the limit ${x_N\to{}^{q\!}x_N}$ with the remaining coordinates fixed.

The transformation between the two coordinate systems $\varphi_\nu^q$ and  $\varphi_\nu^{q'}$ is
\begin{equation}
\varphi_N^{q'} = \varphi_N^q\;,
\quad
\varphi_{\bar\mu}^{q'} = \varphi_{\bar\mu}^q-\Omega_{\bar\mu}^{q'\!q}\,\varphi_N^q\;,
\end{equation}
where we introduced the symbol $\Omega_{\bar\mu}^{q'\!q}$,
\begin{equation}\label{eq:angularvel}
\Omega_{\bar\mu}^{q'\!q}=\frac{{}^{q'\!\!}x_N^2-{}^{q\!}x_N^2}{({}^{q'\!\!}x_N^2-\cir{x}_{\bar\mu}^2)({}^{q\!}x_N^2-\cir{x}_{\bar\mu}^2)}\frac{\cir{c}_{\bar\mu}}{\cir{J}_{\bar\mu N}(\cir{x}_{\bar\mu}^2)}\;,
\end{equation}
which is antisymmetric with respect to indices $q$ and $q'$, ${\Omega_{\bar\mu}^{q'\!q}=-\Omega_{\bar\mu}^{qq'}}$. Therefore, the transformation of the tangent vectors is
\begin{equation}\label{eq:horizonadjKVtrans}
	\tb{\pp}_{\varphi_{\bar\mu}^{q'}} =\tb{\pp}_{\varphi_{\bar\mu}^q}\;,
	\quad
	\tb{\pp}_{\varphi_N^{q'}} = \tb{\pp}_{\varphi_N^q}+\sum_{\bar\mu}\Omega_{\bar\mu}^{q'\!q}\,\tb{\pp}_{\varphi_{\bar\mu}^q}\;.
\end{equation}
The quantity $\Omega_{\bar\mu}^{q'\!q}$ can be interpreted as relative angular velocity of the horizon $q'$ with respect to the horizon $q$. The coordinates $\varphi_\nu^q$ thus represent a frame co-rotating with the horizon $q$. If $q$ is a cosmological horizon, then such coordinates may be considered as generalizations of the Boyer--Lindquist-type coordinates. This follows from the fact that in the limit ${\lambda\to+\infty}$ the Killing vector $\tb{\pp}_{\varphi_N^q}$ reduces to a Killing vector which is timelike far from the axes, ${r\to+\infty}$, and normalized to ${-1}$.


\subsection{Axial singularities}
\label{ssc:AS}

We have all necessary ingredients to examine the singularities on the axes of the Kerr--NUT--(A)dS spacetimes by means of the geometric quantities describing topological defects such as the time shifts, twists, and conicities. As before, consider a fixed pair of indices ${(\bar\mu,p)}$ (with the corresponding pair ${(\bar\eta,\sigma)}$, cf. Tab.~\ref{tab:opt}) and focus our attention on the part of the axis (${x_{\bar\eta}={}^{\sigma\!}x_{\bar\eta}}$) described by such indices. 

Since the axial Killing vector $\tb{\pp}_{\phi_{\bar\mu}^p}$ can be written in terms of the Lie-commuting Killing vectors $\tb{\pp}_{\varphi_\nu}$, we can read off the quantities $\mathcal{K}_{\bar\mu}^p$, $\mathcal{T}_{\bar\mu}^p$, and $\mathcal{W}_{\bar\mu\bar\nu}^p$ by comparing \eqref{eq:axialKVKerrNUTAdS} with \eqref{eq:timeshifttwist}. The overall prefactor is
\begin{equation}
\mathcal{K}_{\bar\mu}^p =\frac{|X'_{\bar\eta}|\big|_{{}^{\sigma\!}x_{\bar\eta}}\!\!}{2\cir{c}_{\bar\mu}}\frac{\cir{U}_{\bar\mu}}{\cir{J}_{\bar\mu}({}^{\sigma\!}x_{\bar\eta}^2)}\;.
\end{equation}
Since the axial Killing vector $\tb{\pp}_{\phi_{\bar\mu}^p}$ vanishes towards the part of the axis given by ${x_{\bar\eta}={}^{\sigma\!}x_{\bar\eta}}$, the expressions for the time shifts $\mathcal{T}_{\bar\mu}^p$ and twists $\mathcal{W}_{\bar\mu\bar\nu}^p$ are basically the coefficients of the $\tb{\pp}_{\varphi_{\bar\mu}^p}$ with respect to ${x_{\bar\eta}={}^{\sigma\!}x_{\bar\eta}}$,
\begin{equation}\label{eq:tstwistKN}
\begin{aligned}
    \mathcal{T}_{\bar\mu}^p &=\frac{\cir{c}_N}{\cir{c}_{\bar\mu}}\frac{{}^{\sigma\!}x_{\bar\eta}^2-\cir{x}_{\bar\mu}^2}{{}^{\sigma\!}x_{\bar\eta}^2-\cir{x}_N^2}\frac{\cir{U}_{\bar\mu}}{\cir{U}_N}\;,
\\
    \mathcal{W}_{\bar\mu\bar\nu}^p &=\frac{\cir{c}_{\bar\nu}}{\cir{c}_{\bar\mu}}\frac{{}^{\sigma\!}x_{\bar\eta}^2-\cir{x}_{\bar\mu}^2}{{}^{\sigma\!}x_{\bar\eta}^2-\cir{x}_{\bar\nu}^2}\frac{\cir{U}_{\bar\mu}}{\cir{U}_{\bar\nu}}\;,
    \quad \bar\nu\neq\bar\mu\;.
\end{aligned}
\end{equation}

The construction of the surface $\Upsilon_{\bar\mu}^p$ whose conicity we want to calculate requires a particular non-Killing radial vector field. This vector field is supposed to Lie-commute with the axial Killing vector $\tb{\pp}_{\phi_{\bar\mu}^p}$ and generate integral curves that are approximately geodesics near the axis. A vector field $\tb{\pp}_{x_{\bar\eta}}$ satisfies both conditions. The 2-surface $\Upsilon_{\bar\mu}^p$ generated by $\tb{\pp}_{x_{\bar\eta}}$ and $\tb{\pp}_{\phi_{\bar\mu}^p}$ is a surface of constant $x_\nu$, ${\nu\neq\bar\eta}$, ${\varphi_{\bar\kappa}+\mathcal{W}_{\bar\mu\bar\kappa}^p\varphi_{\bar\mu}}$, ${\bar\kappa\neq\bar\mu}$, and ${\varphi_N+\mathcal{T}_{\bar\mu}^p\varphi_{\bar\mu}}$. To calculate the conicity of $\Upsilon_{\bar\mu}^p$, we need the norms of these vector fields, which are easily obtained from \eqref{eq:KerrNUTAdSmetricALTBul} and \eqref{eq:xcbul},
\begin{equation}
    \tb{\pp}_{x_{\bar\eta}}^2=\frac{U_{\bar\eta}}{X_{\bar\eta}}\;,
    \quad
    \tb{\pp}_{\phi_{\bar\mu}^p}^2=4\sum_\nu\frac{X_{\nu}}{U_{\nu}}\bigg(\frac{{J_\nu({}^{\sigma\!}x_{\bar\eta}^2)}}{{X'_{\bar\eta}}\big|_{{}^{\sigma\!}x_{\bar\eta}}}\bigg)^{\!\!2}\;.
\end{equation}
Expanding these expressions near the particular part of the axes, ${|x_{\bar\eta}-{}^{\sigma\!}x_{\bar\eta}|\ll 1}$ (cf. \eqref{eq:metricfuncexp} with \eqref{eq:xizeta}), we get 
\begin{equation}
\begin{aligned}
	\tb{\pp}_{x_{\bar\eta}}^2 &= 
	\frac{U_{\bar\eta}}{X'_{\bar\eta}}\bigg|_{{}^{\sigma\!}x_{\bar\eta}}
	\frac{1}{x_{\bar\eta}{-}{}^{\sigma\!}x_{\bar\eta}}
	+\mathcal{O}\big((x_{\bar\eta}{-}{}^{\sigma\!}x_{\bar\eta})^0\big)\;,
	\\
	\tb{\pp}_{\phi_{\bar\mu}^p}^2 &= 
		4\frac{X'_{\bar\eta}}{U_{\bar\eta}}\bigg|_{{}^{\sigma\!}x_{\bar\eta}}
		(x_{\bar\eta}{-}{}^{\sigma\!}x_{\bar\eta})
	+\mathcal{O}\big((x_{\bar\eta}{-}{}^{\sigma\!}x_{\bar\eta})^2\big)\;.
\end{aligned}
\end{equation}
Following the definition provided in \eqref{eq:conicities}, we integrate the length of integral curves in direction $\tb{\pp}_{\phi_{\bar\mu}^p}$ that winds around the axis and divide it by the approximate geodesic distance in direction $\tb{\pp}_{x_{\bar\eta}^p}$. We find that the conicity of $\Upsilon_{\bar\mu}^p$ coincides with the value of the prefactor $\mathcal{K}_{\bar\mu}^p$,
\begin{equation}\label{eq:conicitiesKN}
\begin{aligned}
    \mathcal{C}_{\bar\mu}^p &=\lim_{x_{\bar\eta}\to {}^{\sigma\!}x_{\bar\eta}}\frac{\int_0^{2\pi\mathcal{K}_{\bar\mu}^p}|\tb{\pp}_{\phi_{\bar\mu}^p}|\,\dd\phi_{\bar\mu}^p}{2\pi\sigma\int_{x_{\bar\eta}}^{{}^{\sigma\!}x_{\bar\eta}}|\tb{\pp}_{x_{\bar\eta}}|\,\dd x_{\bar\eta}}=\mathcal{K}_{\bar\mu}^p
    \\
    &=\frac{|X'_{\bar\eta}|\big|_{{}^{\sigma\!}x_{\bar\eta}}\!\!}{2\cir{c}_{\bar\mu}}\frac{\cir{U}_{\bar\mu}}{\cir{J}_{\bar\mu}({}^{\sigma\!}x_{\bar\eta}^2)}\;.
\end{aligned}
\end{equation}

As mentioned above, the quantities \eqref{eq:tstwistKN} and \eqref{eq:conicitiesKN} are not uniquely determined. In Appx.~\ref{ap:TKTW}, we show that the ambiguities arise from two reasons: a possibility of multiple sets of cyclic Killing vectors \eqref{eq:phimubambi} and the freedom in the choice of the temporal Killing vector \eqref{eq:phiNambi}. They lead to the transformation \eqref{eq:KTWtransf} between various sets of Killing frames. The former is naturally fixed by requiring that the Kerr--NUT--(A)dS spacetimes are ${2\pi}$-periodic in coordinates ${\varphi_{\bar\mu}}$ with cyclic Killing vectors ${\tb{\pp}_{\varphi_{\bar\mu}}}$ so that the metrics in these coordinates take the form \eqref{eq:KerrNUTAdSmetricALT}. The latter freedom might be adjusted by setting parameters $\cir{x}_N$ and $\cir{c}_N$ to particular values. As discussed in the previous subsection, an example of a privileged temporal Killing vector for ${\lambda>0}$ can be obtained by choosing \eqref{eq:paramxcgauge} with $q$ being a cosmological horizon. In the limit ${\lambda\to0}$, this vector reduces to a Killing vector that is timelike far from the axes and normalized to ${-1}$.

With respect to this set of Killing vectors, the time shift $\mathcal{T}_{\bar\mu}^p$ and twists $\mathcal{W}_{\bar\mu\bar\nu}^p$ vanish only if the condition \eqref{eq:cirxcchoice1} is satisfied. In addition, \eqref{eq:cirxcchoice1} also implies that the conicity $\mathcal{C}_{\bar\mu}^p$ depends only on $\cir{c}_{\bar\mu}$ and is independent of $\cir{x}$'s. On the other hand, the condition \eqref{eq:cirxcchoice2} is equivalent to the vanishing conicity, ${\mathcal{C}_{\bar\mu}^p=1}$. This confirms the fact that all parameters $\cir{x}_{\bar\mu}$ and $\cir{c}_{\bar\mu}$ are connected with real geometric quantities, unlike the parameters $\cir{x}_N$ and~$\cir{c}_N$.

\subsection{Ergosurfaces of cyclic Killing vectors}

Similar to the temporal Killing vectors $\tb{\pp}_{\varphi_N^q}$, the cyclic Killing vectors $\tb{\pp}_{\varphi_{\bar\mu}}$ may also become null at some surfaces. Since such surfaces are not necessarily null like black hole horizons, we call them the ergosurfaces in analogy with the ergosurface of the spinning cosmic string spacetime introduced in Sec.~\ref{ssc:SCS}. They separate the regions where the cyclic Killing vectors are spacelike from the regions where they are timelike and generate closed timelike curves. The norm of the cyclic vector $\tb{\pp}_{\varphi_{\bar\mu}}$ is given by the expression
\begin{equation}
    \tb{\pp}_{\varphi_{\bar\mu}}^2=\frac{1}{\cir{c}_{\bar\mu}^2}\sum_\nu\frac{X_\nu}{U_\nu}\big(J_\nu(\cir{x}_{\bar\mu}^2)\big)^2\;.
\end{equation}
The general equation of the ergosurface, ${\tb{\pp}_{\varphi_{\bar\mu}}^2=0}$, is a complicated polynomial equation of degree $D$ in any $x_\nu$. (Its general closed-form solution may not even exist for ${D>4}$.) In what follows, we discuss only the basic properties of a few particular examples. The general behavior of these surfaces is certainly much more involved.

\begin{figure}
    \centering
	\includegraphics[width=\columnwidth/\real{2.03}]{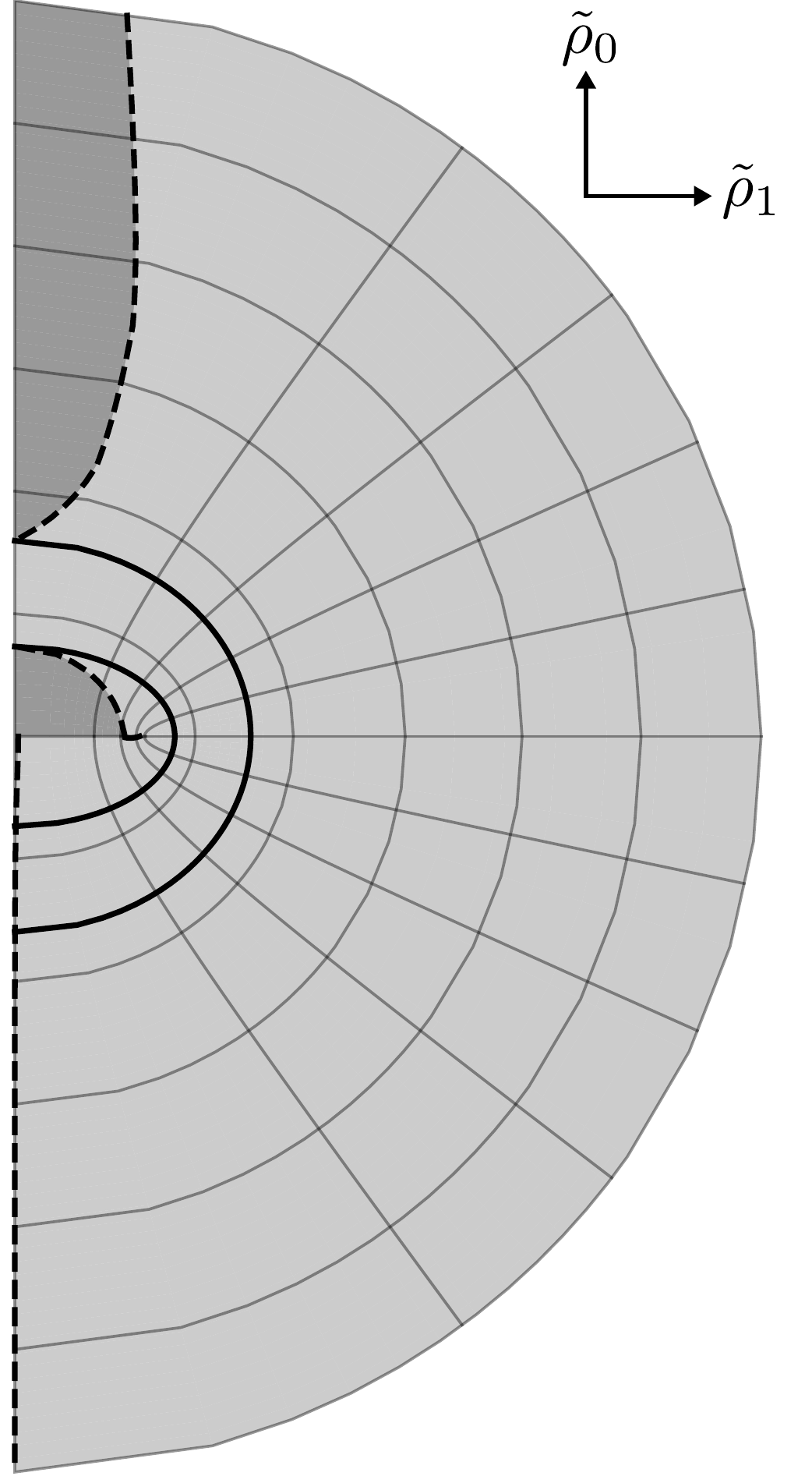}
	\includegraphics[width=\columnwidth/\real{2.03}]{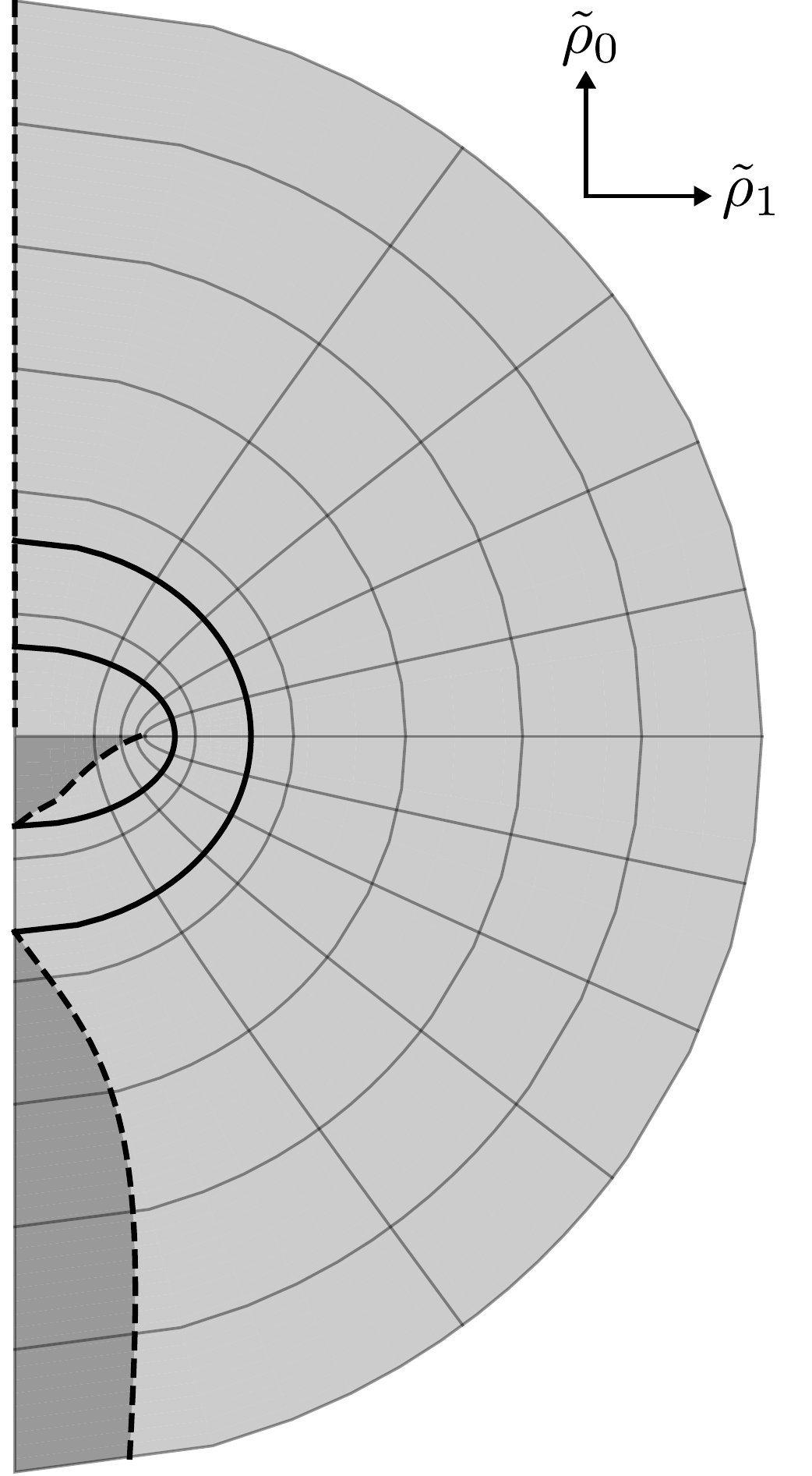}
    \caption{Ergosurfaces of the cyclic Killing vector $\tb{\pp}_{\varphi_1}$ for ${D=4}$, ${\lambda=0}$. The graphs show spacetimes with the regular part of the axis ${(\bar\mu,p)=(1,-1)}$ (left) and ${(\bar\mu,p)=(1,0)}$ (right), i.e., ${\cir{x}_1^2={}^{-\!}x_1^2}$ and ${\cir{x}_1^2={}^{+\!}x_1^2}$, respectively. The lighter gray area denote regions where $\tb{\pp}_{\varphi_1}$ is spacelike, ${\tb{\pp}_{\varphi_1}^2>0}$, while the darker gray area stands for regions where $\tb{\pp}_{\varphi_1}$ is timelike, ${\tb{\pp}_{\varphi_1}^2<0}$. Dashed lines represent ergosurfaces or fixed points of $\tb{\pp}_{\varphi_1}$, ${\tb{\pp}_{\varphi_1}^2=0}$. The highlighted coordinates ${x_1=x}$, ${x_2=\imag r}$ are drawn with respect to $\tilde{\rho}_0$, $\tilde{\rho}_1$ (cf. \eqref{eq:xxprimetrans}, \eqref{eq:rhoprime}). Outer and inner horizons are represented by solid lines.} \label{fig:ERd4}
\end{figure}

\begin{figure*}
    \centering
	\includegraphics[width=\columnwidth/\real{2.03}]{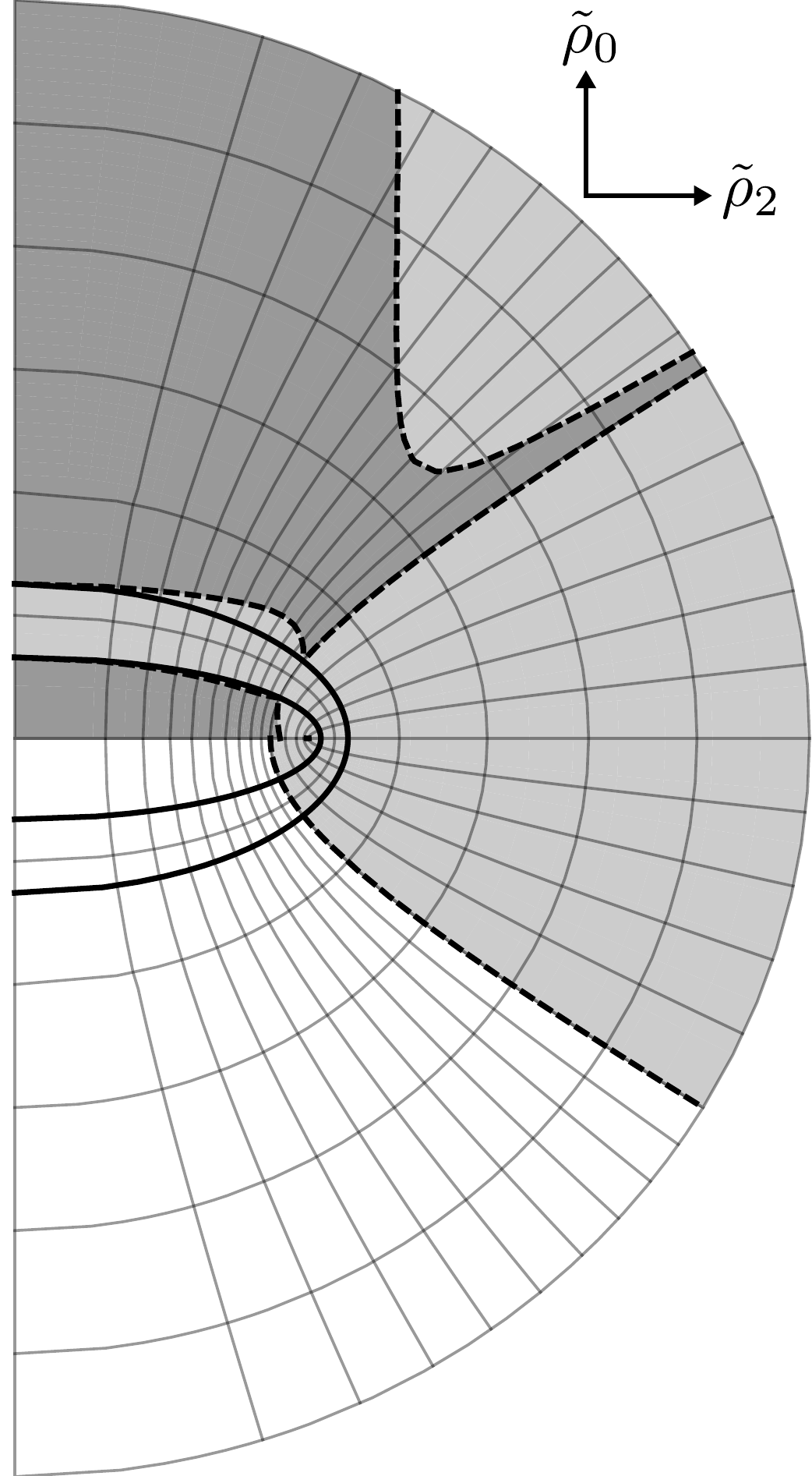}
	\includegraphics[width=\columnwidth/\real{2.03}]{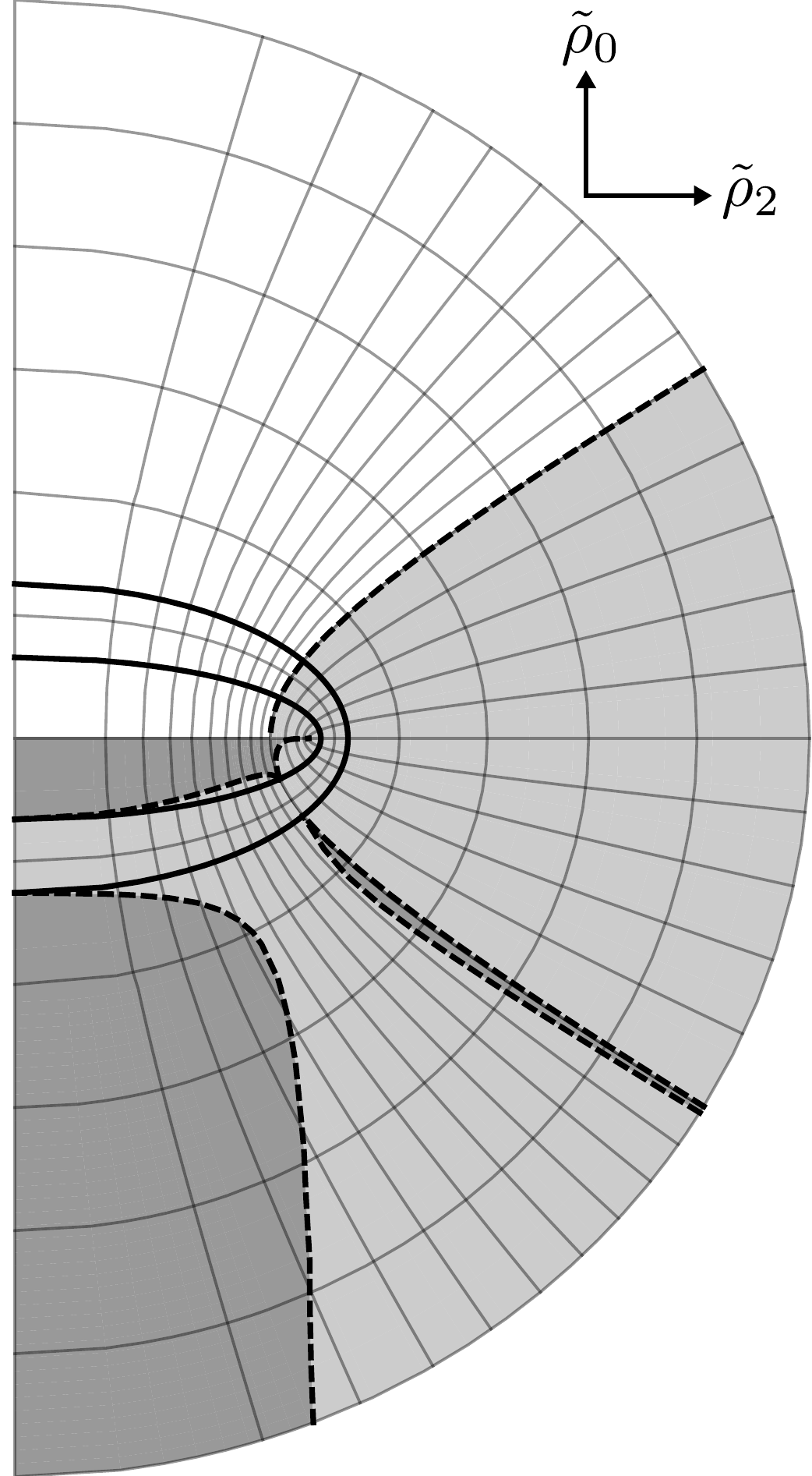}
	\includegraphics[width=\columnwidth/\real{2.03}]{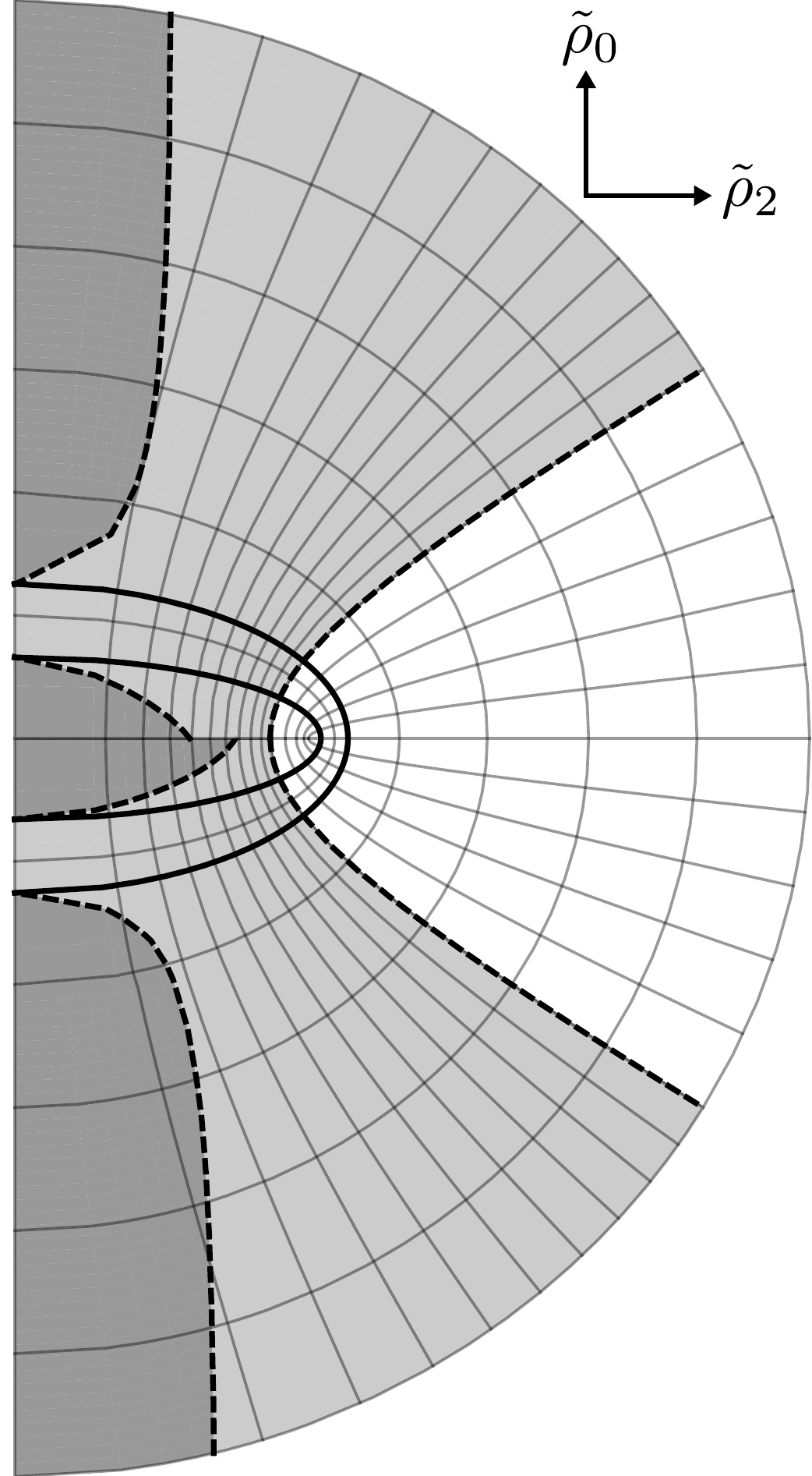}
	\hspace{1cm}
	\includegraphics[width=\columnwidth/\real{2.03}]{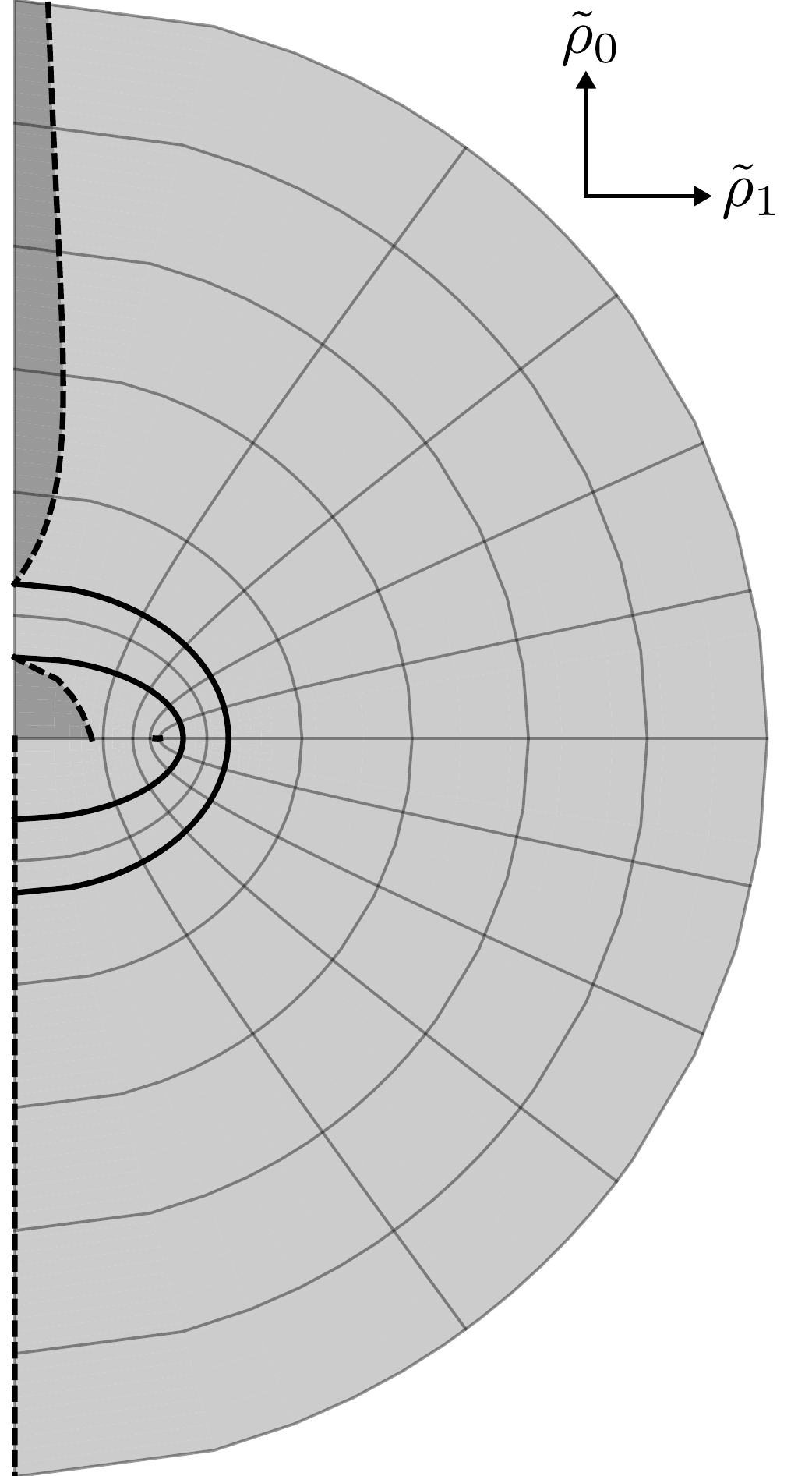}
	\includegraphics[width=\columnwidth/\real{2.03}]{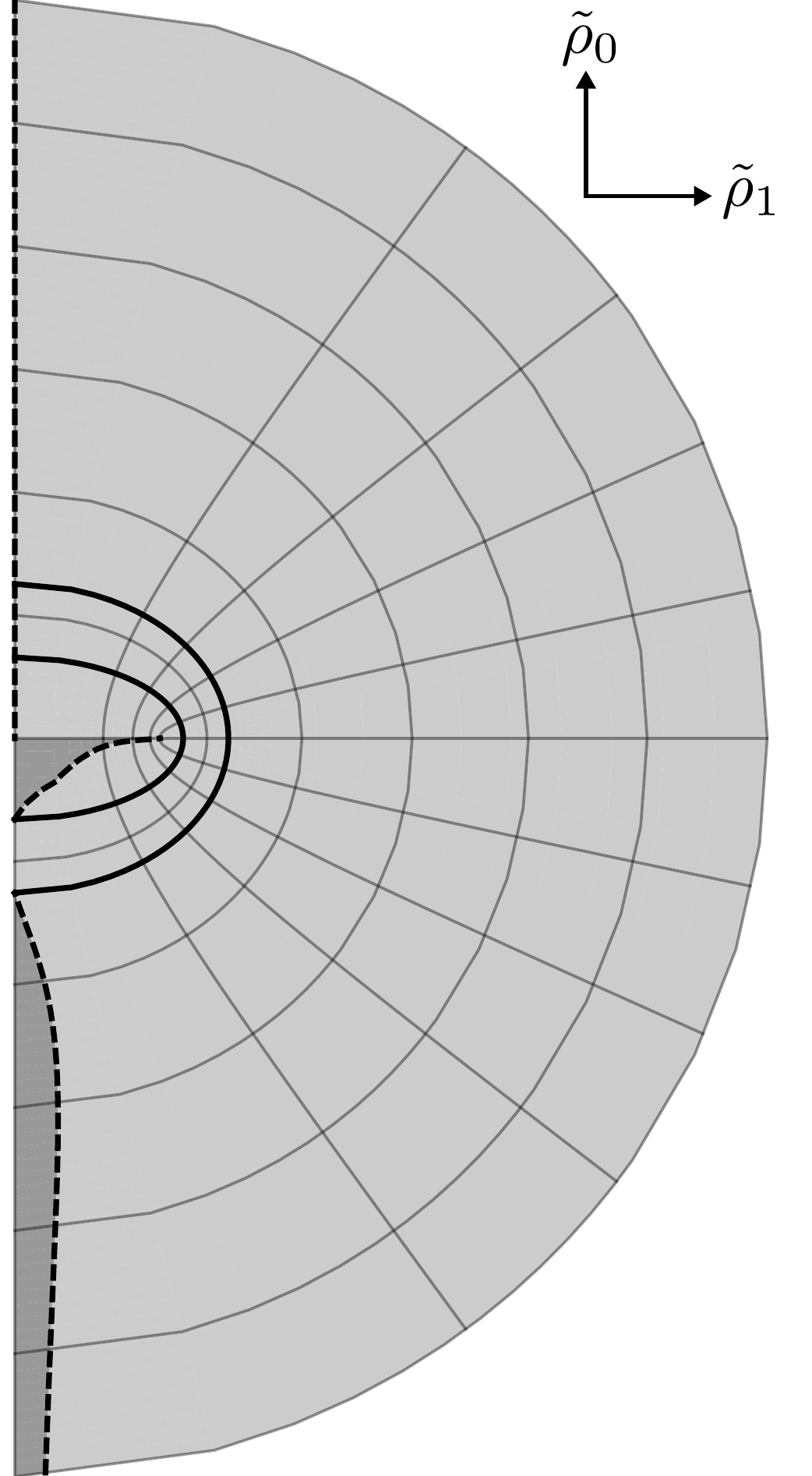}
	\includegraphics[width=\columnwidth/\real{2.03}]{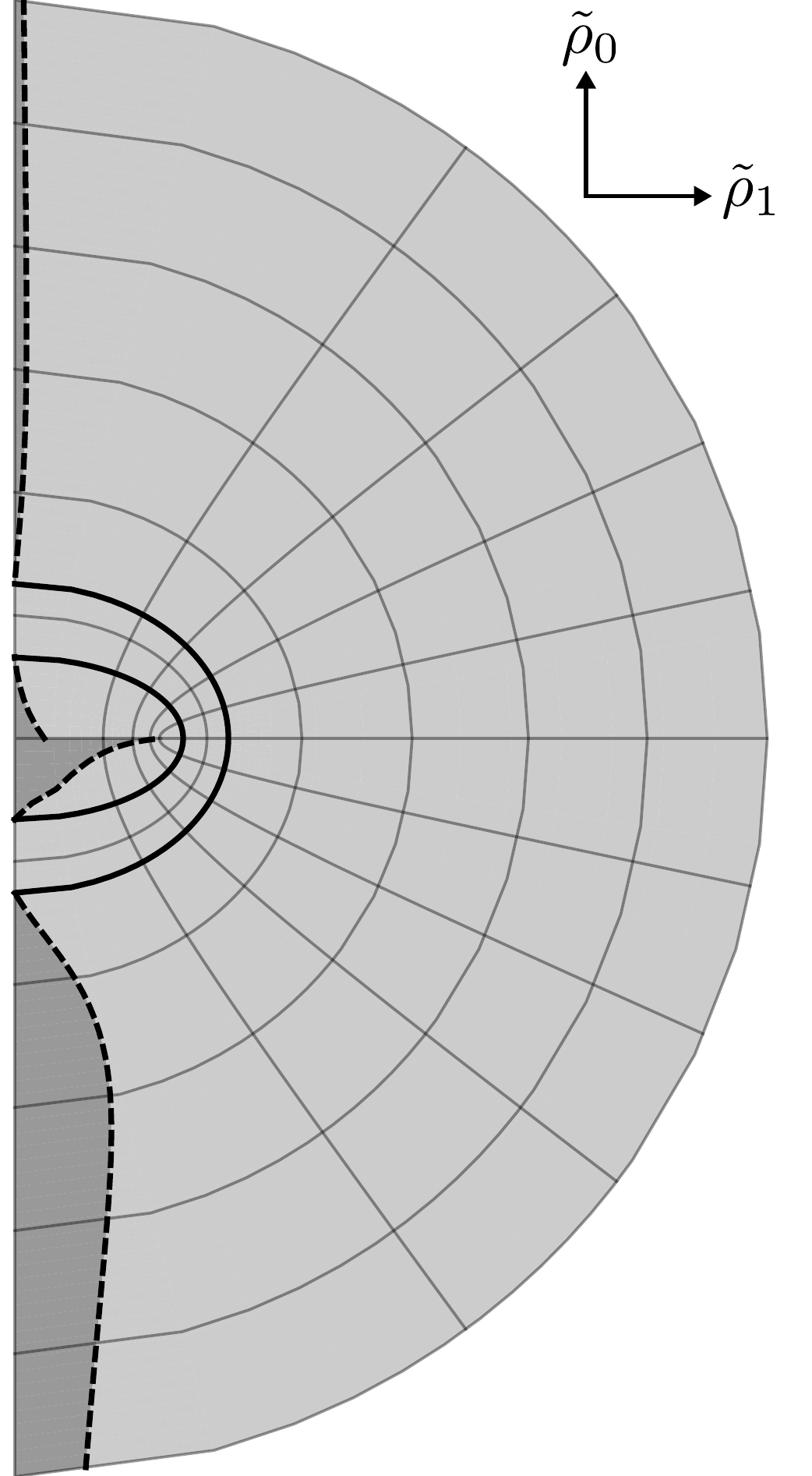}
    \caption{Ergosurfaces of the cyclic Killing vector $\tb{\pp}_{\varphi_1}$ for ${D=6}$, ${\lambda=0}$. The graphs show spacetimes with the regular part of the axis ${(\bar\mu,p)=(1,-1)}$ (left), ${(\bar\mu,p)=(1,0)}$ (center), and ${(\bar\mu,p)=(1,+1)}$ (right), i.e., ${\cir{x}_1^2={}^{-\!}x_1^2}$, ${\cir{x}_1^2={}^{+\!}x_1^2}$, and ${\cir{x}_1^2={}^{-\!}x_2^2}$, respectively. Top: the sections corresponding to the axis ${\tilde{\rho}_1=0}$. Bottom: the sections corresponding to the axis ${\tilde{\rho}_2=0}$. The lighter gray area denote regions where $\tb{\pp}_{\varphi_1}$ is spacelike, ${\tb{\pp}_{\varphi_1}^2>0}$, while the darker gray area stands for regions where $\tb{\pp}_{\varphi_1}$ is timelike, ${\tb{\pp}_{\varphi_1}^2<0}$. Dashed lines and white areas represent either ergosurfaces or fixed points of $\tb{\pp}_{\varphi_1}$, ${\tb{\pp}_{\varphi_1}^2=0}$. The highlighted coordinates ${x_1}$, ${x_2}$, ${x_3=\imag r}$ are drawn with respect to $\tilde{\rho}_0$, $\tilde{\rho}_1$, $\tilde{\rho}_2$ (cf. \eqref{eq:xxprimetrans}, \eqref{eq:rhoprime}). Outer and inner horizons are represented by solid lines.} \label{fig:ERd6_1v}
\end{figure*}

\begin{figure}
    \centering
    \includegraphics[width=\columnwidth/\real{2.03}]{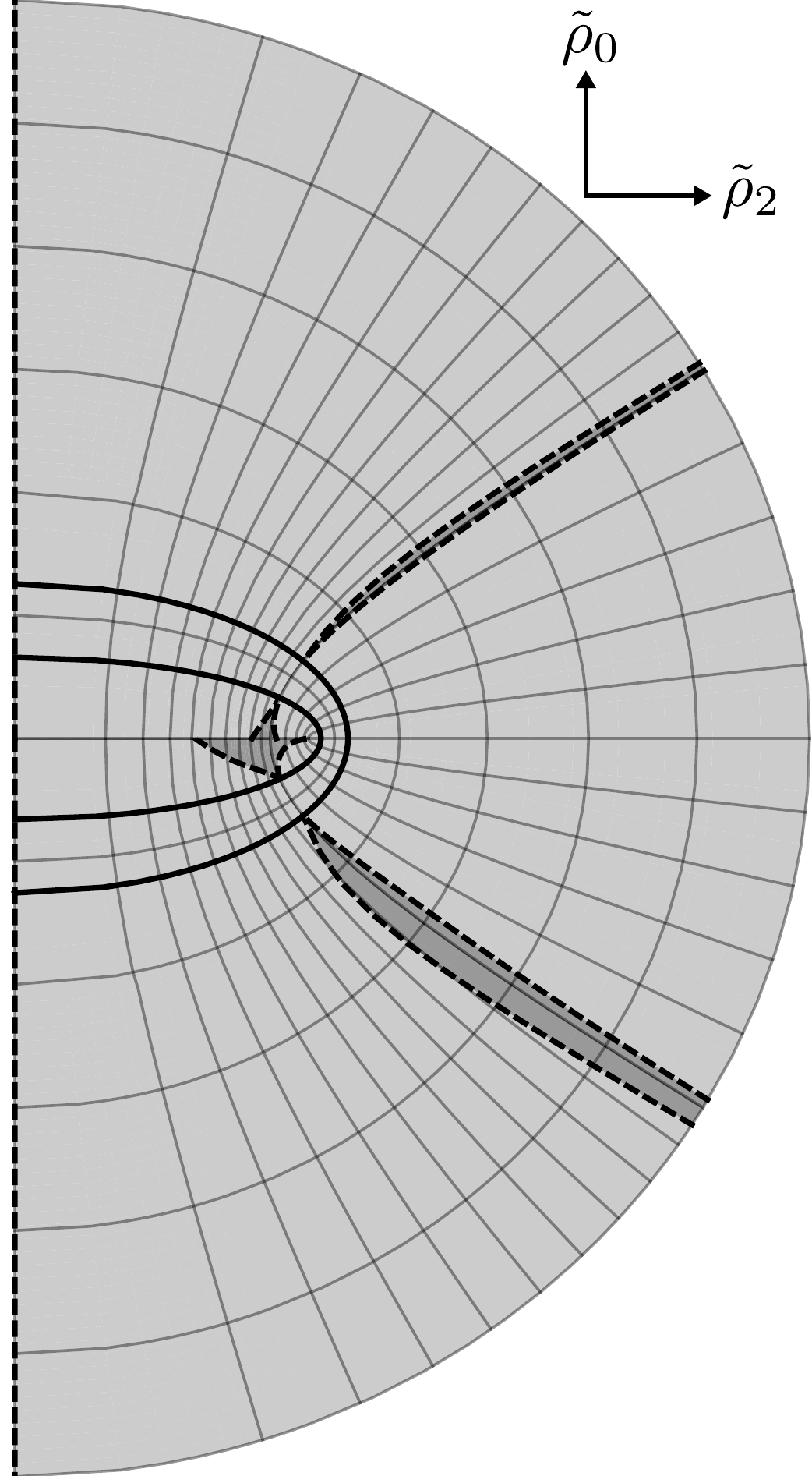}
	\includegraphics[width=\columnwidth/\real{2.03}]{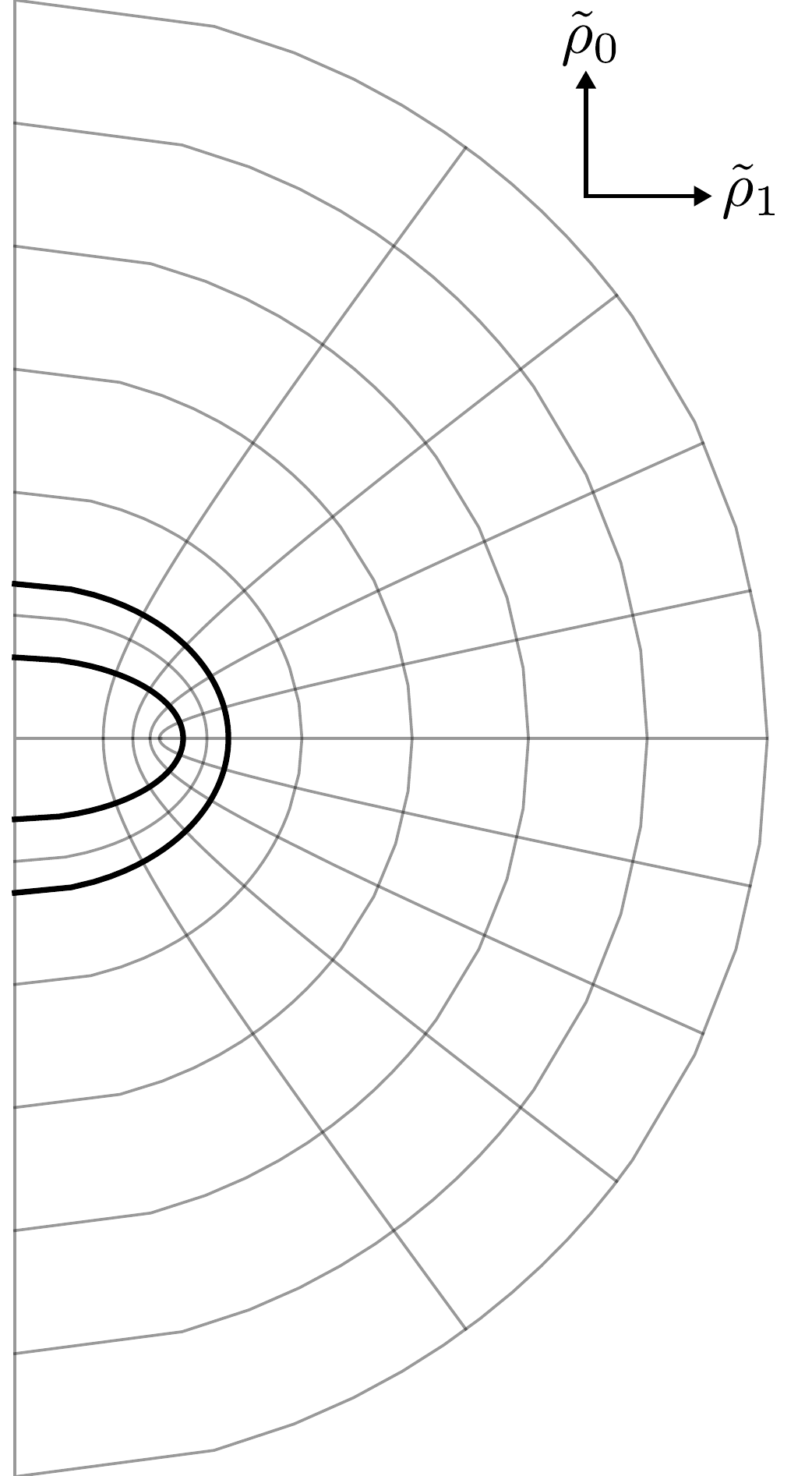}
    \caption{Ergosurfaces of the cyclic Killing vector $\tb{\pp}_{\varphi_2}$ for ${D=6}$, ${\lambda=0}$. Graphs show a spacetime with the regular part of the axis ${(\bar\mu,p)=(2,0)}$, i.e., ${\cir{x}_2^2={}^{+\!}x_2^2}$. Left: the section corresponding to the axis ${\tilde{\rho}_1=0}$. Right: the section corresponding to the axis ${\tilde{\rho}_2=0}$. The lighter gray area denote regions where $\tb{\pp}_{\varphi_2}$ is spacelike, ${\tb{\pp}_{\varphi_2}^2>0}$, while the darker gray area stands for regions where $\tb{\pp}_{\varphi_2}$ is spacelike, ${\tb{\pp}_{\varphi_2}^2<0}$. Dashed lines and white areas represent either ergosurfaces or fixed points of $\tb{\pp}_{\varphi_2}$, ${\tb{\pp}_{\varphi_2}^2=0}$. The highlighted coordinates ${x_1}$, ${x_2}$, ${x_3=\imag r}$ are drawn with respect to $\tilde{\rho}_0$, $\tilde{\rho}_1$, $\tilde{\rho}_2$ (cf. \eqref{eq:xxprimetrans}, \eqref{eq:rhoprime}). Outer and inner horizons are represented by solid lines.} \label{fig:ERd6_2v}
\end{figure}

Since it is impossible for the axis comprising multiple parts to be completely non-singular in the NUT-charged spacetimes, we consider the cases when at least one part of each axis is regular. The regularity can be adjusted by means of the parameter $\cir{x}_{\bar\mu}$, cf. \eqref{eq:cirxcchoice3}. The ergosurfaces of such spacetimes in four dimensions, ${D=4}$, are illustrated in Fig.~\ref{fig:ERd4}. The six-dimensional case, ${D=6}$, is depicted in Fig.~\ref{fig:ERd6_1v} and Fig.~\ref{fig:ERd6_2v}, where display the sections of individual axes only. For a given regular part of the axis, the regions with closed timelike curves of the associated axial/cyclic Killing vector typically appear near the intersections with the remaining parts of the same axis or the other axes. These regions get larger for larger values of NUT charges and smaller for smaller values until they finally disappear when the ergosurfaces merge with the symmetry axes for the vanishing NUT charges.

The terminology we use for the symmetry axes of the Kerr--NUT--(A)dS spacetimes is adopted from the weak-field limit. In particular, we say that the symmetry axes consist of multiple regular/singular parts associated with multiple axial Killing vectors. Alternatively, it would also be reasonable to refer to individual parts as the `axes' themselves. In this language, the whole axes are regular/singular and their intersections are surrounded by ergosurfaces as it is depicted in Fig.~\ref{fig:ERd6_1v} and Fig.~\ref{fig:ERd6_2v} for ${D=6}$. If we analytically extend the spacetimes to ${r<0}$, these intersections are separated from each other.\footnote{In ${D=4}$, the ergosurfaces occur around singular semi-axes. The semi-axes themselves are separated from each other when extended to ${r<0}$, see Fig.~\ref{fig:ERd4}.}


\section{Conclusions}
\label{sc:C}

The Kerr--NUT--(A)dS spacetimes have been studied extensively in the past decade. Their rich structure of explicit and hidden symetries led to a discovery of many remarkable properties such as the complete integrability of a particle motion and the separability of several fundamental equations. In order to appropriately utilize these mathematical results, one needs to have a thorough understanding of these geometries. The interpretation of the Kerr--NUT--(A)dS spacetimes is, however, quite problematic due to the presence of the NUT charges. Despite great effort the meaning of the NUT charges is still not very well understood. A significant observation is that the NUT-charged spacetimes usually suffer from the topological defects caused by the axial singularities. The symmetry axes of the Kerr--NUT--(A)dS spacetimes, therefore, deserve special attention.

In this paper we concentrated on the properties of the symmetry axes and their connection with the Kerr--NUT--(A)dS parameters. The motivation for the presented work was a similarity between the properties associated with the NUT charges and known properties of the spinning cosmic string spacetime, which can be constructed by a simple identification of the flat Minkowski spacetime. Similar to the NUT-charged spacetimes, this spacetime also features axial singularities surrounded by a region with closed timelike curves. Inspired by this geometry, we introduced the geometric quantities describing different types of axial singularities such as the conicities, time shifts, and twists in \eqref{eq:conicities}, \eqref{eq:timeshifttwist}. 

To distinguish physically meaningful cases of the Kerr--NUT--(A)dS spacetimes, we briefly reviewed some possible choices for the ranges of the non-Killing coordinates from \cite{KolarKrtous:2017}. Also, we extended this discussion by introducing Killing coordinates corresponding to the cyclic directions and rewrote the metrics in the form \eqref{eq:KerrNUTAdSmetricALT}.

The main results of this paper were presented in Sec.~\ref{sc:SAKNA}, where the properties of the symmetry axes of the Kerr--NUT--(A)dS spacetimes were investigated. Employing the weak-field limit we identified the natural candidates for the symmetry axes corresponding to the endpoints of the non-Killing coordinates and expanded the Kerr--NUT--(A)dS metrics in the vicinity of these singular points, see \eqref{eq:gaxis}. We found the Killing vectors that vanish there and give rise to the symmetry axes. In terms of the cyclic Killing vectors, these axial Killing vectors read \eqref{eq:axialKVKerrNUTAdS}. From this relation, we derived the expressions \eqref{eq:tstwistKN} and \eqref{eq:conicitiesKN} for the geometric quantities associated with topological defects mentioned above in terms of the Kerr--NUT--(A)dS parameters. It turned out that some of these parameters only describe transformation of coordinates and can be adjusted to match values corresponding to different horizons. In some cases, these coordinates may be regarded as Boyer--Lindquist-type coordinates. Finally, we discussed several four-dimensional and six-dimensional examples of the Kerr--NUT spacetimes whose symmetry axes are partially regular. We showed that the intersections of the parts of the axes are often surrounded by regions with closed timelike curves. 

As discussed in Appx.~\ref{ap:TKTW}, the values of the conicities, time shifts, and twists depend on the choice of the set of the cyclic Killing vectors as well as the temporal Killing vector. An appropriate identification of points of the Kerr--NUT--(A)dS spacetimes with respect to certain choices of cyclic Killing vectors may result in some nontrivial changes of spacetime that are worth further investigation. The ambiguity in the temporal Killing vector also poses an interesting open problem. For example, if ${\lambda<0}$ the spacetime admits no cosmological horizon and it is unclear how to fix the normalization of the temporal Killing vector.

Apart from the isometries whose fixed points form the generalized symmetry axes, there also exist isometries with fixed points forming the bifurcation surfaces. Similar to the axial Killing vectors, a construction of the corresponding Killing vectors vanishing at fixed points requires the specification of the (semi-)regular frames at the horizons. An appropriate expansion of the metric should lead to the approximate Kruskal--Szekeres-type coordinates and allow for an analytic extension across the horizons. The properties of the bifurcate Killing horizons of the Kerr--NUT--(A)dS are definitely worth a deeper investigation.


\section*{Acknowledgements}
I.K. acknowledges the support by Charles University Grants No. GAUK-196516 and No. SVV-260441. P.K. was supported by Czech Science Foundation Grant 17-01625S. The work was done under the auspices of the Albert Einstein Center for 
Gravitation and Astrophysics, Czech Republic. We are grateful for the hospitality of the Theoretical Physics Institute of the University of Alberta where this work began and the Perimeter Institute for Theoretical Physics where it was partially done. Authors would like to thank Valeri Frolov for valuable discussions in the early phase of research.

\appendix


\section{Transformations of $\mathcal{K}_a^p$, $\mathcal{T}_a^p$, $\mathcal{W}_{ab}^p$}
\label{ap:TKTW}

\begin{figure}
    \centering
	\includegraphics[width=\columnwidth/\real{1.3}]{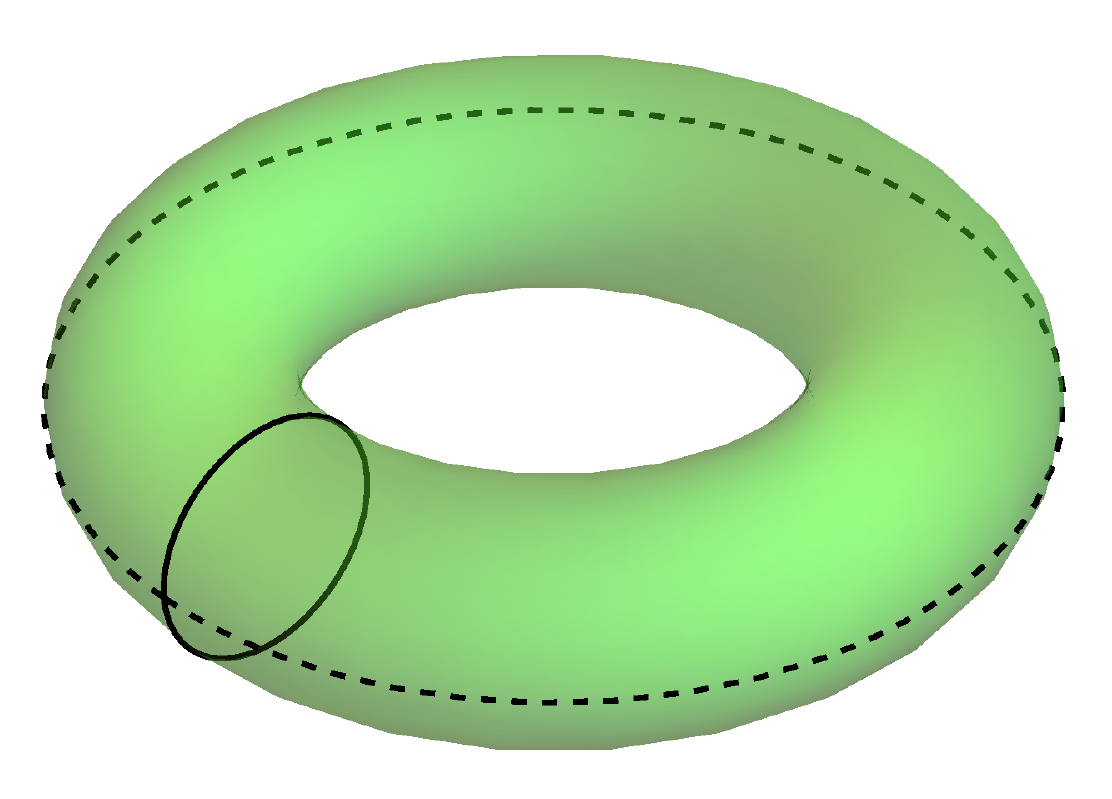}
	\includegraphics[width=\columnwidth/\real{1.3}]{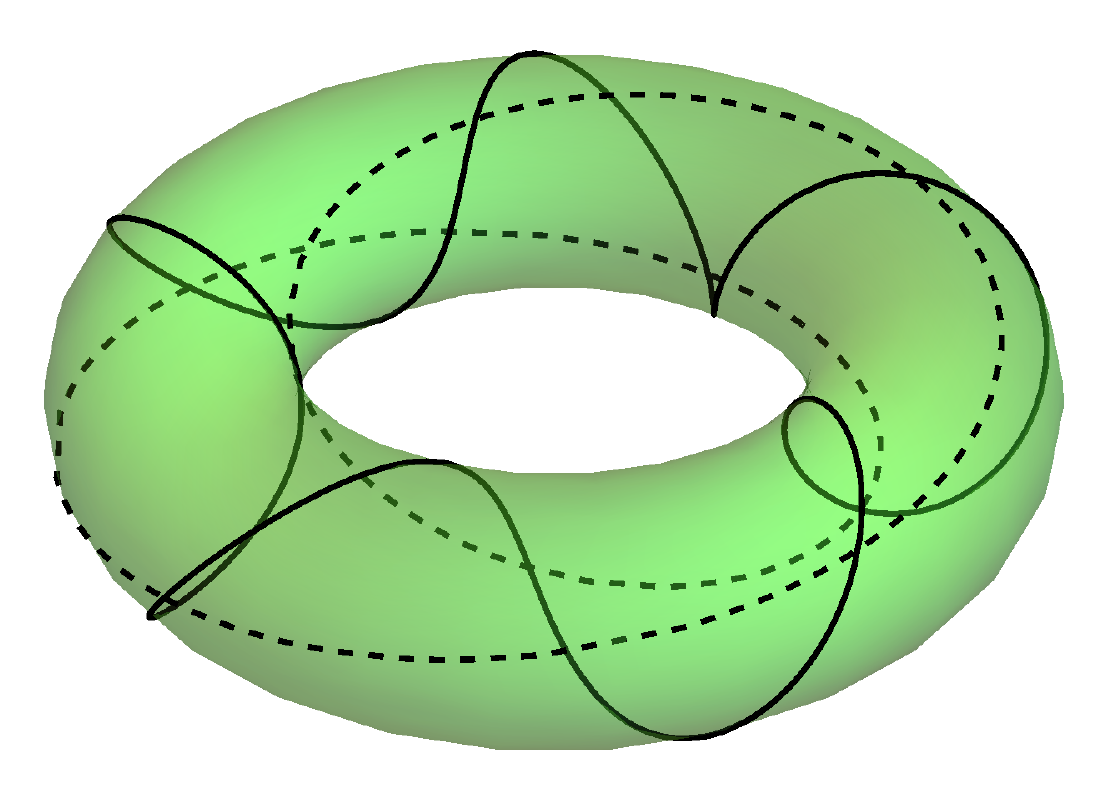}
    \caption{Examples of closed orbits on $\mathbb{T}^2$ corresponding to different winding numbers $n_a^b$. Top: orbits with winding numbers ${n_1^b=(1,0)}$ (solid line) and ${n_2^b=(0,1)}$ (dashed line). Bottom:
    orbits with winding numbers ${n_1^b=(5,1)}$ (solid line)
    and ${n_2^b=(1,2)}$ (dashed line).}
    \label{fig:tor}
\end{figure}

As in Sec.~\ref{ssc:CTST}, we consider a torus ${\mathbb{T}^n}$ generated by a set of cyclic Killing vectors $\tb{u}_a$ of the spacetime ${(\mathcal{M},g)}$. These Killing vectors Lie-commute and generate $2\pi$-periodic coordinates smoothly covering the whole ${\mathbb{T}^n}$. Each other set of cyclic Killing vectors is given by a linear combination of $\tb{u}_a$ with rational coefficients $p_a^b$,
\begin{equation}\label{eq:windnumb}
    \tilde{\tb{u}}_a=\sum_b p_a^b\tb{u}_b\;, \quad p_a^b\in\mathbb{Q}\;.
\end{equation}
This is because the integral curves associated with such linear combinations are closed. By contrast, the irrational combinations of Killing vectors generate open orbits which wind around infinitely many times and never return to the original point. The linear combination \eqref{eq:windnumb} can be rewritten in terms of the integers $n_a^b$ and rational prefactors ${k_a>0}$, such that ${p_a^b=k_a n_a^b}$,
\begin{equation}\label{eq:winding}
    \tilde{\tb{u}}_a=k_a\sum_b n_a^b\tb{u}_b\;, \quad k_a\in\mathbb{Q}\;,\quad n_a^b\in\mathbb{Z}\;.
\end{equation}
The numbers $n_a^b$, sometimes called the \textit{winding numbers}, are assumed to form coprime sets for each index $a$. The examples of closed orbits on $\mathbb{T}^2$ corresponding to various winding numbers are depicted in Fig.~\ref{fig:tor}. 

One may also introduce $2\pi$-periodic coordinates whose tangent vectors are $\tilde{\tb{u}}_a$. In general, such coordinates cover only certain parts of $\mathbb{T}^n$, or they cover the same points several times depending on the choice of the prefactors $k_a$. Another peculiarity, which occurs for ${|\det{n_a^b}|\neq1}$, is that the coordinate lines connect at the endpoints with a certain shift given by a fraction of $2\pi$. All these features can be avoided by assuming that ${\det{n_a^b}=\pm 1}$ and ${k_a=1}$. We refer to such coordinates as the \textit{simple coordinates},
\begin{equation}\label{eq:windingtrans}
    \tilde{\tb{u}}_a=\sum_b n_a^b\tb{u}_b\;, \quad \det{n_a^b}=\pm 1\;,\quad n_a^b\in\mathbb{Z}\;.
\end{equation}

The $n$-dimensional torus ${\mathbb{T}^n}$ can be represented by means of its covering space ${\mathbb{R}^n}$. In this representation, different coordinate patches correspond to different parallelotopes (parallelograms in ${n=2}$, parallelepipeds in ${n=3}$) generated by the Killing vectors ${\tilde{\tb{u}}_a}$. The simple coordinates are singled out by two conditions: 1)~The volume of the parallelotopes is the same as the volume of the original parallelotope generated by $\tb{u}_a$. 2)~All opposite sides of the parallelotopes are can be joined without any shift to completely cover the torus.

In order for the torus to be completely covered by non-simple coordinates, the latter condition cannot be satisfied. However, we can construct a new space by identifying the opposite sides in a standard way without any shift. This results in a new torus which is related to the original one in a nontrivial manner. The reason is that the non-simple coordinates become simple with respect to the new torus, and vice versa. A similar construction employing the simple coordinates leads to the original space. This means that the freedom in the choice of the cyclic Killing vectors \eqref{eq:windnumb} only comes from the transformations \eqref{eq:windingtrans} for a fixed spacetime ${(\mathcal{M},\tb{g})}$.

The spacetime ${(\mathcal{M},\tb{g})}$ possesses not only $n$ cyclic Killing vectors $\tb{u}_a$, but also a temporal Killing vector $\tb{t}$. Altogether, they generate the surfaces $\Gamma$ which have the topology ${\mathbb{T}^n\times\mathbb{R}}$. The Killing vector $\tb{t}$ is restricted only by the condition of being timelike in some region near the axes. Therefore, any linear combination of $\tb{t}$ and $\tb{u}_a$ with real constant coefficients $q^0$ and $q^a$,
\begin{equation}\label{eq:ttrans}
    \tilde{\tb{t}}=q^0\tb{t}+\sum_a q^a \tb{u}_a\;, \quad q^0>0\;, \quad q^a\in\mathbb{R}\;,
\end{equation}
satisfying such a condition is also permitted.

The conicities, time shifts, and twists, introduced in \eqref{eq:conicities} and \eqref{eq:timeshifttwist}, are defined with respect to a chosen Killing frame $\tb{u}_a$, $\tb{t}$. Under the transformations \eqref{eq:windingtrans} (involving simple coordinates only) and \eqref{eq:ttrans}, the quantities $\mathcal{K}_{a}^p$, $\mathcal{T}_{a}^p$, and $\mathcal{W}_{ab}^p$ change to
\begin{equation}\label{eq:KTWtransf}
\tilde{\mathcal{K}}_{a}^p = \frac{\mathcal{K}_{a}^p}{\mathcal{Q}_{aa}^p}\;,
\quad
\tilde{\mathcal{T}}_{a}^p = \frac{\mathcal{T}_{a}^p}{q^0\mathcal{Q}_{aa}^p}\;,
\quad
\tilde{\mathcal{W}}_{ab}^p = \frac{\mathcal{Q}_{ab}^p}{\mathcal{Q}_{aa}^p}\;,
\end{equation}
where $\mathcal{Q}_{ab}^p$ is defined by
\begin{equation}\label{eq:KTWtransfQ}
    \mathcal{Q}_{ab}^p= \bigg(1-q^a\frac{\mathcal{T}_a^p}{q^0}\bigg){n^{-1}}_a^b+\sum_{\substack{c \\ c\neq a}}\bigg(\mathcal{W}_{ac}^p-q^c\frac{\mathcal{T}_a^p}{q^0}\bigg){n^{-1}}_c^b\;,
\end{equation}
The symbol ${n^{-1}}_a^b$ denotes the inverse matrix of $n_a^b$.

In the formula for the axial Killing vector $\tb{v}_a^p$ in \eqref{eq:timeshifttwist}, we separated a cyclic Killing vector $\tb{u}_a$ so that these two Killing vectors share the same index $a$ as is in the case of the spinning cosmic string spacetime. The reason for this reparametrization becomes clearer when we also consider an additional natural condition on the Killing frame. Let us assume that each axial Killing vector $\tb{v}_a^p$ associated with the respective part of the axis approaches the cyclic Killing vector $\tb{u}_a$ in the limit when the corresponding part of the axis becomes regular,
\begin{equation}\label{eq:KTWlim}
    \mathcal{K}_a^p \to 1\;,
    \quad
    \mathcal{T}_a^p \to 0\;,
    \quad
    \mathcal{W}_{ab}^p \to 0\;.
\end{equation}
In other words, we match the axial Killing vectors with the cyclic Killing vectors so that they coincide in the limits of vanishing axial singularities. If the axial Killing vectors $\tb{v}_a^p$ were approaching different cyclic Killing vectors, we would need to consider a different set $\tilde{\tb{u}}_a$ so that \eqref{eq:KTWlim} holds true for tilded quantities. Although this is not a full specification of the Killing frame, it may restrict the sets of Killing vectors considerably. As an illustration, let us determine which Killing frames can be eliminated by this condition if the values of the coefficients $q^0$, $q^a$, and $n_a^b$ are independent of the limiting parameters. Starting with a Killing frame $\tb{u}_a$ satisfying \eqref{eq:KTWlim} and transforming by means of \eqref{eq:KTWtransf}, we find the limits of $\tilde{\mathcal{K}}_a^p$, $\tilde{\mathcal{T}}_a^p$, and $\tilde{\mathcal{W}}_{ab}^p$,
\begin{equation}\label{eq:KTWtilde}
    \tilde{\mathcal{K}}_a^p\to\frac{1}{{n^{-1}}_a^a}\;,
    \quad
    \tilde{\mathcal{T}}_a^p\to 0\;,
    \quad
    \tilde{\mathcal{W}}_{ab}^p\to\frac{{n^{-1}}_a^b}{{n^{-1}}_a^a}\;.
\end{equation}
Together with \eqref{eq:KTWlim} for tilded quantities, this implies that ${n_a^b=\delta_a^b}$. Thus, the cyclic Killing vectors ${\tilde{\tb{u}}_a=\tb{u}_a}$ are uniquely determined in this particular case. The condition \eqref{eq:KTWlim}, however, cannot fix the temporal Killing vector~$\tb{t}$, because \eqref{eq:KTWtilde} are independent of $q^0$ and $q^a$ in this example. Therefore, the Killing vector $\tb{t}$ should be specified by other means.

In Sec.~\ref{ssc:BH}, we introduced new Killing coordinates~${\varphi_\mu}$ by means of the transformation \eqref{eq:psiphitran} involving new parameters $\cir{x}_\mu$, $\cir{c}_\mu$. We assumed that the Kerr--NUT--(A)dS spacetimes are $2\pi$-periodic in $\varphi_{\bar\mu}$ and the tangent vectors ${\tb{\pp}_{\varphi_{\bar\mu}}}$ are cyclic Killing vector. The coordinates $\varphi_{N}$ along with the temporal Killing vector $\tb{\pp}_{\varphi_{N}}$ remained unspecified. In accordance with \eqref{eq:windingtrans}, there exist multiple sets of cyclic Killing vectors (generating $2\pi$-periodic coordinates) that are related by the transformation with integer coefficients $n_{\bar\mu}^{\bar\nu}$,
\begin{equation}\label{eq:phimubambi}
    \tb{\pp}_{{\tilde\varphi}_{\bar\mu}}=\sum_{\bar\nu}n_{\bar\mu}^{\bar\nu}\tb{\pp}_{\varphi_{\bar\nu}}\;, \quad\det{n_{\bar\mu}^{\bar\nu}}=\pm 1\;, \quad n_{\bar\mu}^{\bar\nu}\in\mathbb{Z}\;.
\end{equation}
They give rise to different closed curves on the $\bar{N}$-torus. Similarly, the freedom \eqref{eq:ttrans} leads to the transformation of the timelike Killing vector that contains arbitrary real coefficients $q^{\mu}$,
\begin{equation}\label{eq:phiNambi}
    \tb{\pp}_{{\tilde\varphi}_N} = \sum_\mu q^{\mu}\tb{\pp}_{\varphi_\mu}\;, \quad q^{\bar\mu}\in\mathbb{R}\;, \quad q^{N}>0\;.
\end{equation}

Let us assume that the coordinates ${\tilde\varphi}_{\mu}$ are also defined by means of the parameters $\tilde{\cir{x}}_\mu$, $\tilde{\cir{c}}_\mu$, cf. \eqref{eq:psiphitran}. By analogy with a similar coordinates \eqref{eq:varphiphidotbul}, the transformation between these two systems is
\begin{equation}
    \varphi_\mu=\sum_\nu\frac{\cir{c}_\mu}{\tilde{\cir{c}}_\nu}\frac{\cir{J}_\mu(\tilde{\cir{x}}_\nu^2)}{\cir{U}_\mu}\tilde{\varphi}_\nu\;,
    \quad
    \tilde{\varphi}_\mu=\sum_\nu\frac{\tilde{\cir{c}}_\mu}{\cir{c}_\nu}\frac{\tilde{\cir{J}}_\mu(\cir{x}_\nu^2)}{\tilde{\cir{U}}_\mu}\varphi_\nu\;.
\end{equation}
Therefore, it preserves the form of the metric \eqref{eq:KerrNUTAdSmetricALT} and implies the following transformation of the Killing vectors, cf. \eqref{eq:varphiphidotbulKV}:
\begin{equation}\label{eq:KVtildevarphi}
    \tb{\pp}_{\tilde{\varphi}_\mu}=\sum_\nu\frac{\cir{c}_\nu}{\tilde{\cir{c}}_\mu}\frac{\cir{J}_\nu(\tilde{\cir{x}}_\mu^2)}{\cir{U}_\nu}\tb{\pp}_{\varphi_\nu}\;.
\end{equation}
By comparing \eqref{eq:phimubambi} with \eqref{eq:KVtildevarphi} for ${\mu=\bar\mu}$ and using the condition \eqref{eq:xcircond1}, we find that ${\tilde{\cir{x}}_{\bar\mu}=\cir{x}_{\bar\mu}}$ and ${\tilde{\cir{c}}_{\bar\mu}=\cir{c}_{\bar\mu}}$. Therefore, the set of cyclic Killing vectors ${\tb{\pp}_{\tilde{\varphi}_\mu}=\tb{\pp}_{\varphi_\nu}}$ is uniquely fixed by requiring that the metric takes the form \eqref{eq:KerrNUTAdSmetricALT} in $2\pi$-periodic coordinates with tangent vectors being the cyclic Killing vectors. Furthermore, this Killing frame also satisfies \eqref{eq:KTWlim} as shown in Sec.~\ref{ssc:AS}.

The comparison of \eqref{eq:phiNambi} with \eqref{eq:KVtildevarphi} for ${\mu=N}$ does not lead to any further restriction on parameters $\cir{x}_N$, $\cir{c}_N$, which means that they are only related to a choice of coordinates. As it was discussed in Sec.~\ref{ssc:HAC}, we can use this gauge freedom to introduce the horizon-adjusted coordinates. An example of such coordinates are the Boyer--Lindquist-type coordinates that arise from choosing the cosmological horizon and taking the limit ${\lambda\to0}$.


\section{Functions $J$, $A$, $U$}
\label{ap:FJAU}

Throughout the paper we employ many symbols, such as $J(x^2)$, $A^{(k)}$, $U_\mu$, and their generalizations. Even though they are just polynomial expressions, they turned out to be very useful in many calculations regarding the Kerr--NUT--(A)dS spacetimes. We use the notation which was established mainly in \cite{ChenLuPope:2006,KrtousEtal:2016}. The definitions come in two variants -- polynomials of variables $x_\mu$ and $a_\mu$, but we often use their analogies with different variables too. The most important functions are the functions $J$, which give rise to both~$A$ and~$U$. Their simplest variants are defined by
\begin{equation}\label{eq:J}
\begin{gathered}
J(x^2) =\prod_\mu(x_\mu^2-x^2)=\sum_{k=0}^N A^{(k)}{(-x^2)}^{N-k}\;,
\\
\mathcal{J}(a^2) =\prod_\mu(a_\mu^2-a^2)=\sum_{k=0}^N \mathcal{A}^{(k)}{(-a^2)}^{N-k}\;,
\end{gathered}
\end{equation}
where the coefficients are
\begin{equation}\label{eq:A}
A^{(k)}=\sum_{\mathclap{\substack{\mu_1,\,\dots,\,\mu_k\\ \mu_1<\dots<\mu_k}}}x_{\mu_1}^2\dots x_{\mu_k}^2\;,
\quad
\mathcal{A}^{(k)}=\sum_{\mathclap{\substack{\mu_1,\,\dots,\,\mu_k\\ \mu_1<\dots<\mu_k}}}a_{\mu_1}^2\dots a_{\mu_k}^2\;.
\end{equation}
We can generalize \eqref{eq:J} by omitting an index~$\mu$,
\begin{equation}\label{eq:Jmu}
\begin{gathered}
J_\mu(x^2) =\prod_{\substack{\nu\\ \nu\neq\mu}}(x_\nu^2-x^2)=\sum_k A_{\mu}^{(k)}{(-x^2)}^{N-k-1}\;,
\\
\mathcal{J}_\mu(a^2) =\prod_{\substack{\nu\\ \nu\neq\mu}}(a_\nu^2-a^2)=\sum_k \mathcal{A}_{\mu}^{(k)}{(-a^2)}^{N-k-1}\;,
\end{gathered}
\end{equation}
which generate the functions
\begin{equation}\label{eq:Amu}
A_\mu^{(k)}=\sum_{\mathclap{\substack{\nu_1,\,\dots,\,\nu_k\\ \nu_1<\dots<\nu_k\\ \nu_j\neq\mu}}}x_{\nu_1}^2\dots x_{\nu_k}^2\;,
\quad
\mathcal{A}_\mu^{(k)}=\sum_{\mathclap{\substack{\nu_1,\,\dots,\,\nu_k\\ \nu_1<\dots<\nu_k\\ \nu_j\neq\mu}}}a_{\nu_1}^2\dots a_{\nu_k}^2\;.
\end{equation}
Similarly, we could also define the functions
$J_{\mu\nu}(x^2)$, $\mathcal{J}_{\mu\nu}(x^2)$, $A_{\mu\nu}^{(k)}$, $\mathcal{A}_{\mu\nu}^{(k)}$ by skipping the indices $\mu$, $\nu$. Besides this we set
\begin{equation}
\begin{gathered}
	A^{(0)}=A_\mu^{(0)}=A_{\mu\nu}^{(0)}=\dots=1\;,
	\\
	\mathcal{A}^{(0)}=\mathcal{A}_\mu^{(0)}=\mathcal{A}_{\mu\nu}^{(0)}=\dots=1\;,
\end{gathered}
\end{equation}
and
\begin{equation}
\begin{gathered}
	J_\mu(x^2)\big|_{N=1}=J_{\mu\nu}(x^2)\big|_{N=2}=\dots=1\;,
	\\
	\mathcal{J}_\mu(a^2)\big|_{N=1}=\mathcal{J}_{\mu\nu}(a^2)\big|_{N=2}=\dots=1\;.
\end{gathered}
\end{equation}
We also assume that the functions $J$ and $A$ are zero if the indices ${\mu,\nu}$ overflow. Finally, the special case of \eqref{eq:Jmu} are the functions
\begin{equation}\label{eq:Umu}
\begin{gathered}
	U_\mu=J_\mu(x_\mu^2)=\prod_{\substack{\nu\\ \nu\neq\mu}}(x_\nu^2-x_\mu^2)\;,
	\\
	\mathcal{U}_\mu=\mathcal{J}_\mu(a_\mu^2)=\prod_{\substack{\nu\\ \nu\neq\mu}}(a_\nu^2-a_\mu^2)\;.
\end{gathered}
\end{equation}

These functions satisfy the following known identities:
\begin{equation}\label{eq:AAmuAmu}
	A^{(k)} =A_{\mu}^{(k)}+x_\mu^2 A_{\mu}^{(k-1)}\;,
\end{equation}

\begin{equation}\label{eq:AmuAmuAmunu}
	A_\mu^{(k)} =A_{\mu\nu}^{(k)}+x_\nu^2 A_{\mu\nu}^{(k-1)}\;,
\end{equation}
\begin{equation}\label{eq:AmuU}
\begin{gathered}
\sum_\mu A_\mu^{(l)}\frac{{(-x_\mu^2)}^{N-k-1}}{U_\mu}=\delta_k^l\;,
\\
\sum_k A_\mu^{(k)}\frac{{(-x_\nu^2)}^{N-k-1}}{U_\nu}=\delta_\mu^\nu\;,
\end{gathered}
\end{equation}	
\begin{equation}\label{eq:AN1}
	\sum_\mu\frac{J(a_\mu^2)}{-a_\mu^2\mathcal{U}_\mu}=1-\frac{A^N}{\mathcal{A}^N}\;,
\end{equation}
\begin{equation}\label{eq:JJ}
\sum_\kappa\frac{J_\nu(a_{\kappa}^2)}{\mathcal{U}_{\kappa}}\frac{\mathcal{J}_\kappa(x_{\mu}^2)}{U_{\mu}}=\delta^\mu_\nu\;,
\end{equation}
\begin{equation}\label{eq:idJJdJ}
\sum_\kappa\frac{J_\mu(a_\kappa^2)J_\nu(a_\kappa^2)}{J(a_\kappa^2)\mathcal{U}_\kappa}=-\frac{U_\mu}{\mathcal{J}(x_\mu^2)}\delta_{\mu\nu}\;,
\end{equation}
\begin{equation}\label{eq:idJJJ}
\sum_\mu\frac{\mathcal{J}(x_\mu^2)}{U_\mu}J_\mu(a_\nu^2)J_\mu(a_\kappa^2)=-J(a_\nu^2)\mathcal{U}_\nu\delta^{\nu\kappa}\;.
\end{equation}
\begin{equation}\label{eq:Ader}
 	A_{\mu,\nu}^{(k)}=\frac{2x_\nu}{x_\nu^2-x_\mu^2}\big(A_{\mu}^{(k)}-A_{\nu}^{(k)}\big)\;,
\end{equation}
\begin{equation}\label{eq:Jder}
 	\big(J_\mu(x^2)\big)_{,\nu}=(1{-}\delta_{\mu\nu})\,2x_\mu J_{\mu\nu}(x^2)\;,
\end{equation}
\begin{equation}\label{eq:Jderx}
 	\frac{\pp}{\pp x}\big(J_\mu(x^2)\big)=2x\sum_{\substack{\nu \\ \nu\neq\mu}} J_{\mu\nu}(x^2)\;,
\end{equation}
\begin{equation}\label{eq:Uder}
 	U_{\mu,\nu} =\delta_{\mu\nu}\sum_{\substack{\rho\\ \rho\neq\mu}}\frac{2x_\mu}{x_\mu^2{-}x_\rho^2}U_{\mu}+(1{-}\delta_{\mu\nu})\frac{2x_\nu}{x_\nu^2{-}x_\mu^2}U_{\mu}\;.
\end{equation}


%

\end{document}